\documentclass[aps,reprint,prc,showpacs,amsmath,amssymb,nofootinbib]{revtex4-1}
\usepackage{graphicx}
\usepackage{color}
\usepackage{times}
\usepackage{bm}
\usepackage{multirow}
\usepackage{float}
\usepackage{url}
\usepackage{natbib}
\usepackage{tensor}
\usepackage{float}
\usepackage[colorlinks=true,citecolor=blue,urlcolor=blue,linkcolor=red]{hyperref}

\begin{document}

\title{Rotating neutron stars with quark cores}%
\author{\large Ishfaq A. Rather$^1$}
\email{ishfaqrather81@gmail.com}
\author{\large Usuf Rahaman$^1$}
\author{\large M. Imran$^1$}
\author{H. C. Das$^{2,3}$}
\author{\large A. A. Usmani$^1$}
\author{\large S. K. Patra$^{2,3}$}
\affiliation{$^1$Department of Physics, Aligarh Muslim University, Aligarh 202002, India}
\affiliation{$^2$Institute of Physics, Bhubaneswar 751005, India}
\affiliation{$^3$ Homi Bhabha National Institute, Training School Complex, Anushakti Nagar, Mumbai 400094, India}
\begin{abstract}
 The rotating neutron-star properties are studied to investigate a phase transition to quark matter. The density-dependent relativistic mean-field model (DD-RMF) is employed to study the hadron matter, while the vector-enhanced bag (vBag) model is used to study the quark matter. The star matter properties such as mass, radius, the moment of inertia, rotational frequency, Kerr parameter, and other important quantities are studied to see their effect on quark matter. The maximum mass of a rotating neutron star with the DD-LZ1 and DD-MEX parameter sets is found to be around 3$M_{\odot}$ for pure hadronic phase and decreases to around 2.6$M_{\odot}$ upon phase transitioning to quark matter, which satisfies the recent GW190814 possible maximum mass constraint, implying that the secondary component of GW190814 could be a fast-rotating hybrid star. For DDV, DDVT, and DDVTD parameter sets, the maximum mass decreases to satisfy 2$M_{\odot}$. The moment of inertia calculated for various DD-RMF parameter sets decreases with the increasing mass satisfying constraints from various measurements. Other important quantities calculated also vary with the bag constant and hence show that the presence of quarks inside neutron stars can also allow us to constraint these quantities to determine a proper equation of state. Also, the theoretical study along with the accurate measurement of uniformly rotating neutron-star properties may offer some valuable information concerning the high-density part of the equation of state.
\end{abstract}
\maketitle

\section{Introduction}
Compact objects such as neutron star (NSs) in the known universe are  ideal sources to study the properties and composition of highly dense matter. The measurement of mass and the radius for spherically symmetric and static stars impose constraints on the properties of matter at high density. The study of rotating NS (RNS) properties may lead to significant new constraints. From the past decade, the successful discoveries of various gravitational waves by LIGO and Virgo collaborations (LVC) have allowed us to study the dense matter properties with more constraints imposed on the NS equation of state (EoS). The measurement of tidal deformability for static NSs ruled out many EoSs with either too large or too small maximum mass.\par 
 The binary NS (BNS) merger event GW170817 \cite{PhysRevLett.119.161101,PhysRevLett.121.161101} constrained the maximum mass and the tidal deformability of NSs and hence the EoS. The total mass of the GW170817 event was around 2.7$M_{\odot}$ with the heavier component mass 1.16$M_{\odot}$-1.60$M_{\odot}$ for low spin priors. The maximum mass approached 1.9$M_{\odot}$ for high spin priors \cite{PhysRevX.9.011001}. 
The tidal deformability dependence on the NS radius $\Lambda \propto R^5$ provided a more strong constraint on the high dense nuclear EoS. A new gravitational wave event (GW190814) was observed recently by LVC  with a black hole merger of mass 22.2$M_{\odot}$-24.3$M_{\odot}$ and a massive secondary component of mass 2.50$M_{\odot}$-2.67$M_{\odot}$ \cite{Abbott_2020a}. The secondary component of GW190814 gained a lot of attention about its nature whether it is a black hole, a NS, or some other exotic object \cite{dexheimer2020gw190814,tan2020neutron,Fishbach_2020,rather2020hadronquark,godzieba2020maximum,10.1093/mnrasl/slaa168,tan2020neutron,Zhang_2020,tsokaros2020gw190814,fattoyev2020gw190814,lim2020revisiting,tews2020nature}.\par
 
 A proper knowledge of a NS maximum mass is assumed to be the most important parameter determining the possible outcome of a BNS merger \cite{PhysRevD.73.064027,PhysRevLett.107.051102,PhysRevD.88.044026,PhysRevLett.111.131101,PhysRevD.92.044045,PhysRevD.94.024023,Lehner_2016,Radice_2018,K_ppel_2019}. The constraints on the EoS at high density are imposed with accurate information of a NS maximum mass and radius \cite{PhysRevLett.105.161102,Hebeler_2013,doi:10.1146/annurev-nucl-102711-095018,Miller_2019,Annala2020}. The precise measurement of masses of millisecond pulsars such as PSR J1614-2230 (1.928$\pm$0.017)$M_{\odot}$ \cite{Demorest2010},PSR J0348+0432(2.01$\pm$0.04)$M_{\odot}$ \cite{Antoniadis1233232}, and PSR J0740+6620 (2.04$^{+0.10}_{-0.09}$)$M_{\odot}$ \cite{Cromartie2020} show that the theoretical maximum mass of a NS should be around 2$M_{\odot}$. Combining the GW observations of BNS with quasi-universal relations, a maximum mass of $M_{max} \lesssim 2.17_{-0.15}^{+0.17}$ $M_{\odot}$ is attained for nonrotating NSs \cite{Rezzolla_2018}. By combining the total binary mass of GW170817 inferred from GW signal with electromagnetic (EM) observations, an upper limit of $M_{max}\lesssim$ 2.17 $M_{\odot}$ is predicted \cite{Margalit_2017}. Further analysis employing both energy and momentum conservations along with the numerical-relativity simulations show that the maximum mass of cold NSs is weakly constrained as $M_{max}\lesssim$ 2.3 $M_{\odot}$ \cite{PhysRevD.100.023015}. However, with the discovery of the recent secondary component of GW190814 predicting a maximum mass around 2.5$M_{\odot}$-2.67$M_{\odot}$, the maximum mass limit for a NS seems to be weekly constrained.\par 
  The effect of the EoS on the properties of a RNS has been studied since the late 90s by various groups \cite{1994ApJ...424..823C,Margalit_2017,Stergioulas1998,Paschalidis2017}. To investigate the NS structure and its properties, the choice of the EoS becomes the starting point. There proper choice of EoS for NS matter invites theoretical discussions. Every single EoS produces a NS with different properties. Despite predicting several NS properties, the composition at several times the normal nuclear density is still not known properly. The core of a NS is considered to be a nuclear matter in $\beta$-equilibrium and charge-neutral conditions. Neutron, proton, electron, and muon are the basic components of the core of a NS. The NS structure with several exotic degrees of freedom like quarks, kaons, and hyperons is also studied \cite{Rathernm, Ratherbag, rather2020hadronquark,PhysRevD.100.103017,doi:10.1142/S021830131550007X}. The presence of such exotic phases significantly affects the NS properties. \par 
   NS matter containing only hadrons are studied by employing different model parameters at high densities. Density functional theories (DFTs) have been widely used to determine the saturation properties of high dense nuclear matter (NM) \cite{PhysRevC.5.626,SHEN1998435,PhysRevC.65.035802,refId0,PhysRevC.89.045807,PhysRevC.90.045802}. At saturation density, the NM EoS is well constrained and its corresponding properties are determined with less uncertainty. These EoSs at several times the normal nuclear density describe the NS properties. The relativistic mean-field (RMF) model  has been very successful in describing both finite and infinite NM \cite{Walecka:1974qa}. The basic mechanism involves the interaction of nucleons via mesons. Different mesons like $\rho$, $\sigma$, $\omega$, and $\delta$ have reduced the large uncertainties present in the NM properties and constrained the properties to well within the limits \cite{PhysRevLett.86.5647,SUGAHARA1994557,BOGUTA1977413,SEROT1979146,PhysRevC.89.044001,Kumara:2017bti,PhysRevC.97.045806}. The RMF EoS like NL3 \cite{PhysRevC.55.540} and BigApple \cite{fattoyev2020gw190814,das2020bigapple} determine NS with a maximum mass around 2.7$M_{\odot}$. The density-dependent RMF (DD-RMF) model contains the density-dependent coupling constants replacing the self- and cross-coupling of various mesons in the basic RMF model \cite{PhysRevLett.68.3408}. DD-RMF parameters like DD-ME1 \cite{PhysRevC.66.024306}, DD-ME2 \cite{PhysRevC.71.024312} generate very massive NSs with a 2.3$M_{\odot}$-2.5$M_{\odot}$ maximum mass. Several new DD-RMF parameter sets were proposed recently such as DD-LZ1 \cite{ddmex}, DD-MEX \cite{TANINAH2020135065}, DDV, DDVT, and DDVTD \cite{typel}. These recently proposed parameter sets are divided into two categories. The DD-LZ1 and DD-MEX parameter sets produce very stiff EoSs and hence a large NS maximum mass and belong to the stiff EoS group. Parameter sets such as DDV, DDVT, and DDVTD produce soft EoSs and hence lie in the softer EoS group. Both the stiff and soft EoS groups are used in the current study to determine the NS properties for the static and rotating case.\par 
    Exotic degrees of freedom like quarks have been studied over the past decades. The presence of quarks in the core of NSs at very high densities has been proposed \cite{Annala2020}. Thus the phase transition to quark matter (QM) inside NSs is possible at very high density  \cite{PhysRevD.30.272,PhysRevD.30.2379}. A NS with hadrons in the core followed by a phase transition to the QM at several times the normal nuclear density is termed the hybrid star (HS)  \cite{PhysRevD.46.1274,zel_2010,PhysRevD.88.085001,refId0,Bombaci2016}. \par 
    The MIT bag model \cite{PhysRevD.30.2379,PhysRevD.9.3471,PhysRevD.17.1109} was first proposed to study  strange and hybrid stars. The  Nambu-Jona-Lasinio (NJL) model \cite{KUBIS1997191,PhysRev.122.345,PhysRev.124.246,RevModPhys.64.649,BUBALLA2005205} was later introduced and explained the QM more precisely than the bag model. The modified NJL models have been  very successful in explaining the stable HSs and also satisfying the recent GW170817 constraints \cite{PhysRevD.95.056018,PhysRevD.97.103013}.    
    The modified bag model, termed the vector-enhanced bag model (vBag) \cite{Kl_hn_2015} was introduced as an effective model to study the astrophysical processes. The vBag model is favored over the simple bag model and NJL model because it accounts for the repulsive vector interactions along with the dynamic chiral symmetry breaking (D$\chi$SB). The repulsive vector interaction and the deconfinement for the construction of a mixed-phase allowed it to describe the strange or hybrid stars which attain the  2$M_{\odot}$ limit. Recent work by Roupas show the secondary component of GW190814 to be a strange star in the color-flavor locked (CFL) phase \cite{roupas2020qcd}.\par 
     In the present work, we study the properties of a RNS by considering a phase transition from hadron matter (HM) to QM. The star matter properties such as mass, radius, moment of inertia, and Kerr parameter are studied along with some other important properties. The dependence of these quantities on the NS mass is discussed. Several properties of a static star such as mass, radius, and tidal deformability are also discussed.
     
This article is organized as follows: the DD-RMF model for the HM and vBag model for QM and the phase-transition properties are discussed in Sec. (\ref{sec:headings}. The static and rotating NS structure and various properties associated with the star matter are discussed in Sec. (\ref{nsprop}). In Sec. (\ref{results}), the parameter sets for the NM and the saturation density properties are defined. The EoS for the hadronic and hybrid star configurations are explained. The static and RNS properties like mass, radius, and moment of inertia are discussed in Sec. (\ref{sub2}). Finally, the summary and concluding remarks are given in Sec. (\ref{summary}). 
\section{Theory and Formalism}
\label{sec:headings}

The RMF Lagrangian involves the interaction between the nucleons through various mesons defined as Dirac particles. The most basic and simplest RMF Lagrangian involves the scalar-isoscalar sigma $\sigma$ and vector-isoscalar $\omega$ mesons without any interaction among themselves \cite{walecka}, which results in large NM incompressibility $K_0$ \cite{Walecka:1974qa}. Boguta and Bodmer included a nonlinear self-coupling of the $\sigma$ field which lowered the value of NM incompressibility to reasonable values \cite{BOGUTA1977413}. Apart from $\sigma$, $\omega$, and $\rho$ mesons, the addition of the scalar-isoscalar $\delta$ meson is included to study the isovector effect on the scalar potential of the nucleon. Both NM and NS matter properties are obtained which lie well within the limits \cite{KUBIS1997191,PhysRevC.89.044001}. The effective field theory motivated RMF (E-RMF) is the extended RMF model which includes all possible self- and cross-couplings between the mesons \cite{FURNSTAHL1996539,FURNSTAHL1997441,Kumara:2017bti}. The RMF model gained a lot of success in investigating both finite and infinite NM properties. The various nonlinear meson coupling terms can be replaced by the density-dependent nucleon-meson coupling constants in the density-dependent relativistic Hartree-Fock (DD-RHF) \cite{PhysRevC.36.380,PhysRevC.18.1510,LONG2006150} and DD-RMF \cite{PhysRevLett.68.3408}. The density-dependent models take into account the nuclear medium effect caused by the relativistic Brueckner-Hartree-Fock mode \cite{PhysRevLett.68.3408}. Unlike the RMF model, the coupling constants in the DD-RMF are density-dependent i.e., they vary with density. The DD-RMF coupling constants depend either on the scalar density $\rho_s$ or the vector density $\rho_B$, but the vector density parametrizations are usually considered which does not influence the total energy of the system.\par
The DD-RMF Lagrangian density is

\begin{widetext}
\begin{equation}
\begin{split}
\mathcal{L}  =\sum_{\alpha=n,p} \bar{\psi}_{\alpha} \Biggl\{\gamma^{\mu}\Bigg(i\partial_{\mu}-g_{\omega}(\rho_B)\omega_{\mu}-\frac{1}{2}g_{\rho}(\rho_B)\gamma^{\mu}\rho_{\mu}\tau\Bigg)-\Bigg(M-g_{\sigma}(\rho_B)\sigma-g_{\delta}(\rho_B)\delta\tau\Bigg)\Biggr\} \psi_{\alpha} \\
 +\frac{1}{2}\Bigg(\partial^{\mu}\sigma \partial_{\mu}\sigma-m_{\sigma}^2 \sigma^2\Bigg) +\frac{1}{2}\Bigg(\partial^{\mu}\delta \partial_{\mu}\delta-m_{\delta}^2 \delta^2\Bigg)
-\frac{1}{4}W^{\mu \nu}W_{\mu \nu}\\
+\frac{1}{2}m_{\omega}^2 \omega_{\mu} \omega^{\mu}-\frac{1}{4}R^{\mu \nu} R_{\mu \nu}+\frac{1}{2}m_{\rho}^2 \rho_{\mu} \rho^{\mu},
\end{split}
\end{equation}
\end{widetext}

where $\psi_{\alpha}, (\alpha=n,p)$ denotes the neutron and proton wave-function. $g_{\sigma},g_{\omega},g_{\rho}$, and $g_{\delta}$ are the meson coupling constants which are density-dependent, and $m_{\sigma},m_{\omega},m_{\rho}$ and $m_{\delta}$ are the masses for  $\sigma, \omega, \rho$ and $\delta$ mesons respectively. The tensor fields $W^{\mu \nu}$ and $R^{\mu \nu}$ are defined as 
	\begin{equation}
	\begin{gathered}
	W^{\mu \nu}=\partial^{\mu}W^{\nu}-\partial^{\nu}W^{\mu}, \\
R^{\mu \nu}=\partial^{\mu}R^{\nu}-\partial^{\nu}R^{\mu}\\
	\end{gathered}
	\end{equation} 
The coupling constants of $\sigma$ and $\omega$ mesons for the DD-MEX, DDV, DDVT, and DDVTD parameter sets are expressed as a fraction of the vector density. The density-dependent coupling constants for various parametrizations are given as
\begin{equation}
g_i(\rho_B) = g_i(\rho_0) f_i(x),
\end{equation}
where the function $f_i(x)$ is given by
\begin{equation}\label{eq5}
f_i(x) = a_i \frac{1+b_i (x+d_i)^2}{1+c_i(x+d_i)^2},i=\sigma,\omega
\end{equation}
as a function of $ x=\rho_B/\rho_{0}$, where $\rho_0$ is the nuclear matter saturation density.\par 
For the function $f_i(x)$, the number of constraint conditions defined as $f_i(1)=1$,$f^{''}_{\sigma}(1)=f^{''}_{\omega}(1)$, $f^{''}_i(0)=0$ reduce the number of free parameters from eight to three in  Eq. (\ref{eq5}). Out of them, the first two constraints are
\begin{equation}
a_i=\frac{1+c_i(1+d_i)^2}{1+b_i(1+d_i)^2},
3c_id_i^2=1
\end{equation} 
For the isovector $\rho$ and $\delta$ mesons, the coupling constants are given by an exponential dependence as
\begin{equation}
g_i(\rho_B)=g_i(\rho_0)exp[-a_i(x-1)]
\end{equation}
For the DD-LZ1 parameter set, the coefficient $g_i$ is fixed at $\rho_B$=0 for $i=\sigma,\omega$;
\begin{equation}
g_i(\rho_B)=g_i(0)f_i(x).
\end{equation}
There are only four constraint conditions for $\sigma$ and $\omega$ in the DD-LZ1 parameter set. The constraint $f^{''}_{\sigma}(1)=f^{''}_{\omega}(1)$ is removed, which changes the coupling constant of $\rho$ meson as
\begin{equation}
g_{\rho}(\rho_B)=g_{\rho}(0)exp(-a_i x).
\end{equation}
Following the Euler-Lagrange equation, we obtain the equations of motion for nucleons and mesons.

The scalar density $\rho_s$, baryon density $\rho_B$,  isovector densities $\rho_{s3}$, and $\rho_3$ are defined  as
\begin{equation}
\rho_s = \sum_{\alpha=n,p}\bar{\psi}\psi =\rho_{sp} +\rho_{sn}=\sum_{\alpha}\frac{2}{(2\pi)^3}\int_{0}^{k_{\alpha}}d^3k \frac{M_{\alpha}^*}{E_{\alpha}^*},
\end{equation}
\begin{equation}
\rho_B = \sum_{\alpha=n,p}\psi^{\dagger}\psi =\rho_{p} +\rho_{n}=\sum_{\alpha}\frac{2}{(2\pi)^3}\int_{0}^{k_{\alpha}}d^3k,
\end{equation}
\begin{equation}
\rho_{s3} = \sum_{\alpha}\bar{\psi}\tau_3\psi =\rho_{sp} -\rho_{sn},
\end{equation}
\begin{equation}
\rho_3 = \sum_{\alpha}\psi^{\dagger}\tau_3\psi =\rho_p -\rho_n.
\end{equation}
The effective masses of nucleons are given as
\begin{equation}
M_p^* =M-g_{\sigma}(\rho_B)\sigma -g_{\delta}(\rho_B)\delta,
\end{equation},
and
\begin{equation}
M_n^* =M-g_{\sigma}(\rho_B)\sigma +g_{\delta}(\rho_B)\delta
\end{equation}
Also,
\begin{equation}
E_{\alpha}^*=\sqrt{k_{\alpha}^2+M_{\alpha}^{*2}},
\end{equation}
is the effective energy of nucleon with nucleon momentum $k_{\alpha}$.
The energy-momentum tensor determines the total energy density and the pressure for the NM as
\begin{equation}
\begin{gathered}
\mathcal{E}_{DD}= \mathcal{E}_H +\mathcal{E}_{kin},\\
P_{DD}=P_H +P_{kin}\\
\end{gathered}
\end{equation}
where $\mathcal{E}_H $ and $P_H$ are the energy density and the pressure of hadronic matter, which are given as
\begin{widetext}
\begin{equation} \label{eq18}
\begin{split}
\mathcal{E}_H = \frac{1}{2}m_{\sigma}^2 \sigma^2-\frac{1}{2}m_{\omega}^2 \omega^2-\frac{1}{2}m_{\rho}^2 \rho^2+\frac{1}{2}m_{\delta}^2 \delta^2+g_{\omega}(\rho_B)\omega \rho_B+\frac{g_{\rho}(\rho_B)}{2}\rho \rho_3,\\
P_H = -\frac{1}{2}m_{\sigma}^2 \sigma^2+\frac{1}{2}m_{\omega}^2 \omega^2+\frac{1}{2}m_{\rho}^2 \rho^2-\frac{1}{2}m_{\delta}^2 \delta^2-\rho_B \sum_R (\rho_B),\\
\end{split}
\end{equation}
\end{widetext}
and $\mathcal{E}_{kin}$ and $P_{kin}$ are the energy density and pressure from the kinetic part,
	\begin{equation} \label{eq20}
	\begin{split}
	\mathcal{E}_{kin} =\frac{1}{8\pi^2}\Big[k_{\alpha}E_{\alpha}^* \big(2k_{\alpha}^2+M_{\alpha}^2\big)+M_{\alpha}^4 ln\frac{M_{\alpha}}{k_{\alpha}+E_{\alpha}^*}\Big],\\
	P_{kin} = \frac{1}{24\pi^2}\Big[k_{\alpha} E_{\alpha}^*\big(2k_{\alpha}^2-3M_{\alpha}^2\big)+3M_{\alpha}^4 ln\frac{k_{\alpha}+E_{\alpha}^*}{M_{\alpha}^*}\Big].
	\end{split}
	\end{equation}

 For NS matter, the $\beta$-equilibrium condition is
\begin{equation}\label{c1}
\mu_e =\mu_\mu =  \mu_n - \mu_p. 
\end{equation}
where \\
\begin{equation}
\begin{gathered}
\mu_{\alpha=n,p} =\sqrt{k_{\alpha}^2 +M_{\alpha}^{*2}}\\
+\Big[g_{\omega}(\rho_B)\omega +\frac{g_{\rho}(\rho_B)}{2}\rho \tau_3 +\sum_R (\rho_B)\Big], \\
\mu_{l=\mu,e} = \sqrt{k_l^2 +m_l^2}.
\end{gathered}
\end{equation}
  The charge neutrality condition implies
\begin{equation}
q_{total} = \sum_{i=n,p} q_i k_i^3/(3\pi^2)+\sum_l q_l k_l^3/(3\pi^2)=0.
\end{equation}
To study the phase transition from HM to QM, the vBag \cite{Kl_hn_2015} is employed  which is an extension of the simple bag model \cite{PhysRevD.9.3471,PhysRevD.17.1109,PhysRevD.30.2379}. The vBag model accounts for D$\chi$SB and also the additional repulsive vector interactions which allow the strange stars to achieve the 2$M_{\odot}$ limit on the maximum mass and hence satisfy the constraints from recently measured masses of pulsars such as PSR J1614-2230 \cite{Demorest2010}, PSR 0348+0432 \cite{Antoniadis1233232}, and MSP J0740+6620 \cite{Cromartie2020}. \\

The energy density and pressure in the vBag model follow as \cite{vbageos} 
\begin{equation}
\mathcal{E}_Q = \sum_{f=u,d,s} \mathcal{E}_{vBag,f}-B_{dc},
\end{equation}
\begin{equation}
P_Q = \sum_{f=u,d,s} P_{vBag,f}+B_{dc},
\end{equation}
where $B_{dc}$ represents the deconfined bag constant introduced which lowers the energy per particle, thus favoring stable strange matter. The energy density and pressure of a single quark flavor are defined as
 \begin{equation}
 \mathcal{E}_{vBag,f}(\mu_f) = \mathcal{E}_{FG,f}(\mu_f^*)+\frac{1}{2}K_{\nu}n_{FG,f}^2 (\mu_f^*)+B_{\chi,f},
 \end{equation}   
 \begin{equation}
 P_{vBag,f}(\mu_f) = P_{FG,f}(\mu_f^*)+\frac{1}{2}K_{\nu}n_{FG,f}^2 (\mu_f^*)-B_{\chi,f},
 \end{equation}
 where FG denotes the zero-temperature Fermi gas. The coupling constant parameter $K_{\nu}$ results from the vector interactions and controls the stiffness of the star matter curve \cite{Wei_2019}. The bag constant for a single quark flavor is denoted $B_{\chi,f}$ . The chemical potential $\mu_f^*$ of the system is
 \begin{equation}
 \mu_f =\mu_f^* +K_{\nu}n_{FG,f}(\mu_F^*).
 \end{equation}  
 An effective bag constant is defined in the vBag model so that the phase transition to QM occurs at the same chemical potential
 \begin{equation}
 B_{eff}=\sum_{f=u,d,s}B_{\chi,f}-B_{dc}.
 \end{equation}
 The effective bag constant $B_{eff}$ is an extension to the deconfined bag constant to allow the phase transition to occur at the same chemical potential.
 This also illustrates how  $B_{eff}$ can be used in two and three flavor QM.\\
 
  The charge neutrality and $\beta$-equilibrium conditions for the QM are
  \begin{equation}
  \frac{2}{3}\rho_u -\frac{1}{2}(\rho_d+\rho_s)-\rho_e-\rho_u =0,
  \end{equation}
\begin{equation}
\mu_s=\mu_d=\mu_u+\mu_e;
\mu_u=\mu_e.
\end{equation}

The density range over which a phase transition exists between HM and QM is determined by beta-equilibrium and charge-neutral conditions \cite{PhysRevD.46.1274,PhysRevC.60.025801,PhysRevC.75.035808,PhysRevC.66.025802,PhysRevC.89.015806}. The phase transition can be either by a local charge condition (Maxwell construction) \cite{PhysRevD.88.063001} or global charge  neutrality condition (Gibbs construction) \cite{PhysRevD.46.1274}. The global charge neutrality condition allows the HM and QM to be separately charged, unlike the local charge-neutrality condition. In this study, we used the Gibbs method to construct the hadron-quark phase transition. 
The global charge neutrality condition follows as
\begin{equation}
\chi \rho_{Q}+(1-\chi) \rho_{H}+\rho_l=0,
\end{equation} 
where the quark volume fraction in the mixed-phase is given by $\chi= V_Q/(V_T)$ which varies from $\chi=0$ to $\chi=1$ in the pure hadron and pure quark phases respectively. The charge densities of quarks, hadrons, and leptons are represented by $\rho_Q$, $\rho_H$, and $\rho_l$ respectively.\par 
 The equations governing the mixed-phase chemical potential, pressure, energy, and baryon density are defined as:
\begin{equation}\label{e1}
\mu_{B,H} = \mu_{B,Q}; \mu_{l,H} = \mu_{l,Q},
\end{equation}
and
\begin{equation}\label{e2}
P_{H}(\mu_B,\mu_l) = P_{Q}(\mu_B,\mu_l) = P_{MP}.
\end{equation}
\begin{equation}\label{e3}
\varepsilon_{MP} = \chi \varepsilon_{Q} +(1-\chi)\varepsilon_{H} +\varepsilon_l,
\end{equation}
and
\begin{equation}\label{e4}
\rho_{MP} = \chi \rho_{Q} +(1-\chi)\rho_{H}.
\end{equation}
  The above equations determine the properties of the mixed-phase and combined with the hadron equations generate the overall properties of the star. \par

\section{Neutron star structure and properties}
\label{nsprop}
\subsection{Static neutron star}
For a spherically symmetric, static NS (SNS), the metric element has the Schwarzschild form ($G$ = $c$ = 1)
\begin{equation}
ds^2=-e^{2\phi(r)}dt^2+e^{2\Lambda(r)}dr^2+r^2(d\theta^2+sin^2\theta d\phi^2),
\end{equation}
where the metric functions $e^{-2\phi(r)}$ and $e^{2\Lambda(r)}$ are defined as
\begin{equation}
e^{-2\phi(r)} =(1-\gamma(r))^{-1},
\end{equation}
\begin{equation}
e^{2\Lambda(r)}=(1-\gamma(r)),
\end{equation}
with
\begin{equation}
\gamma(r)=2M(r)/r
\end{equation}
 The energy-momentum tensor reduces the Einstein field equations to well-known 
 Tolman-Oppenheimer-Volkoff coupled differential equations given by\cite{PhysRev.55.364,PhysRev.55.374}
\begin{equation}\label{tov1}
\frac{dP(r)}{dr}= -\frac{[\mathcal{E}(r) +P(r)][M(r)+4\pi r^3 P(r)]}{r^2(1-2M(r)/r) }
\end{equation}
and
\begin{equation}\label{tov2}
\frac{dM(r)}{dr}= 4\pi r^2 \mathcal{E}(r)
\end{equation}
where $M(r)$ represents the gravitational mass at radius $r$ with fixed central density. The boundary conditions  $P(0)=P_c$, $M(0)=0$ allows one to solve the above differential equations and determine the properties of a NS. \par
 The tidal deformability $\lambda$ is defined as the ratio of the induced quadrupole mass $Q_{ij}$ to the external tidal field $\mathcal{E}_{ij}$ as \cite{PhysRevD.81.123016,PhysRevC.95.015801}
\begin{equation}\label{l1}
\lambda=-\frac{Q_{ij}}{\mathcal{E}_{ij}} = \frac{2}{3}k_2 R^5
\end{equation}
The dimensionless tidal deformability $\Lambda$ is defined as
\begin{equation}\label{l2}
\Lambda=\frac{\lambda}{M^5}=\frac{2k_2}{3C^5}
\end{equation}
where $C=M/R$ is the compactness parameter and $k_2$ is the second Love number. The expression for the Love number is written as \cite{PhysRevD.81.123016}
\begin{equation}\label{l3}
\begin{split}
k_2=\frac{8}{5}(1-2C)^2 [2C(y-1)]\Bigl\{2C(4(y+1)C^4
+(6y-4)C^3\\
+(26-22y)C^2+3(5y-8)C-3y+6)\\
-3(1-2C)^2(2C(y-1)-y+2)log\Big(\frac{1}{1-2C}\Big)\Bigr\}^{-1}.
\end{split}
\end{equation}
The function $y=y(R)$ can be computed by solving the differential equation \cite{PhysRevC.95.015801,Hinderer_2008}
\begin{equation}\label{l4}
r\frac{dy(r)}{dr}+y(r)^2+y(r)F(r)+r^2 Q(r)=0,
\end{equation}
where
\begin{equation}\label{l5}
F(r)=\frac{r-4\pi r^3 [\mathcal{E}(r)-P(r)]}{r-2M(r)},
\end{equation}
\begin{equation}\label{l6}
\begin{split}
Q(r)=\frac{4\pi r\Big(5\mathcal{E}(r)+9P(r)+\frac{\mathcal{E}(r)+P(r)}{\partial P(r)/\partial\mathcal{E}(r)}-\frac{6}{4\pi r^2}\Big)}{r-2M(r)}\\
-4\Bigg[\frac{M(r)+4\pi r^3 P(r)}{r^2 (1-2M(r)/r)}\Bigg]^2.
\end{split}
\end{equation}
The above equations are solved for spherically symmetric and static NS to determine the properties like mass, radii, and tidal deformability.
\subsection{Rotating neutron star}
For a rapidly rotating NSs with a nonaxisymmetric configuration, they would emit gravitational waves until they achieve axisymmetric configuration. The rotation deforms the NS. Here we study the rapidly rotating NS assuming a stationary, axisymmetric space-time. The energy-momentum tensor for such a perfect fluid describing the matter is given by
\begin{equation}
T^{\mu \nu} = (\mathcal{E}+P)u^{\mu}u^{\nu}+Pg^{\mu \nu},
\end{equation}
where the first term represents the contribution from matter. $u^{\mu}$ denotes the fluid-four-velocity, $\mathcal{E}$ is the energy density, and $P$ is the pressure. For RNS, the metric tensor is given by \cite{1976ApJ...204..200B,1986ApJ...304..115F,PhysRevLett.62.3015}
\begin{equation}
\begin{gathered}
ds^2=-e^{2\nu(r,\theta)}dt^2+e^{2\psi(r,\theta)}(d\phi-\omega(r)dt)^2\\ +e^{2\mu(r,\theta)}d\theta^2 +e^{2\lambda(r,\theta)}dr^2,\\
\end{gathered}
\end{equation}
where the gravitational potentials $\nu$, $\mu$, $\psi$, and $\lambda$ are the functions of $r$ and $\theta$ only. The Einstein's field equations are solved for the given potential to determine the physical properties that govern the structure of the RNS. Global properties like gravitational mass, equitorial radius, moment of inertia, angular momentum and quadrupole moment are calculated.\par 
 For a RNS, the angular momentum $J$ is easy to calculate. By defining the angular velocity of the fluid relative to a local inertial frame, $\bar{\omega}(r) =\Omega-\omega(r)$, $\bar{\omega}$ satisfies the following differential equation
 \begin{equation}
 \frac{1}{r^4}\frac{d}{dr}\Bigg(r^4 j \frac{d\bar{\omega}}{dr}\Bigg)+\frac{4}{r}\frac{dj}{r}\bar{\omega}=0,
 \end{equation}
where $j=j(r)=e^{-(\nu+\lambda)/2}$.\par 
The angular momentum of the star is then given by the relation
\begin{equation}
J=\frac{1}{6}R^4 \Bigg(\frac{d\bar{\omega}}{dr}\Bigg)_{r=R},
\end{equation}
which relates the angular velocity as
\begin{equation}
\Omega=\bar{\omega}(R)+\frac{2J}{R^3}.
\end{equation}
The moment of inertia defined by $I=J/\Omega$, is given by \cite{LATTIMER2000121,Worley_2008}
\begin{equation}\label{e5}
I\approx\frac{8\pi}{3}\int_{0}^{R} (\mathcal{E}+P)e^{-\phi(r)} \Big[1-\frac{2m(r)}{r}\Big]^{-1}\frac{\bar{\omega}}{\Omega}r^4 dr,
\end{equation}
The properties of a RNS are calculated by using the RNS code \cite{Stergioulas2003,Stergioulas_1995,rnscode,refId0}.

\section{Results and Discussions}
\label{results}
\subsection{Parameter sets}
To determine the properties of SNSs and RNSs, we used several recent DD-RMF parametrizations such as DD-MEX \cite{TANINAH2020135065}, DD-LZ1 \cite{ddmex}, and DDV, DDVT, DDVTD \cite{typel}. Apart from the basic DD-MEX and DD-LZ1 parameter sets, the DDV, DDVT, and DDVTD sets include the necessary tensor couplings of the vector mesons to nucleons.\par

Table \ref{tab1} shows the necessary nucleon masses, meson masses, and the coupling constants of the parameter sets used. The meson coupling constants in the DD-LZ1 parameter set are the values at zero density while for the other parameter sets, these coupling constants are obtained at the nuclear matter saturation density $\rho_{0}$.\par

\begin{table}[ht]
	\centering
	\caption{Nucleon and meson masses and different coupling constants for various DD-RMF parameter sets. }
	\begin{tabular}{ cccccc }
	    \hline
		\hline
		&DD-LZ1&DD-MEX&DDV&DDVT&DDVTD \\
		\hline
		$m_n$ & 938.9000&939.0000&939.5654&939.5654&939.5654\\
		$m_p$&938.9000&939.0000&938.2721&938.2721&938.2721\\
		$m_{\sigma}$&538.6192&547.3327&537.6001&502.5986&502.6198\\
		$m_{\omega}$&783.0000&783.0000&783.0000&783.0000&783.0000\\
		$m_{\rho}$&769.0000&763.0000&763.0000&763.0000&763.0000\\
		$g_{\sigma}(\rho_0)$&12.0014&10.7067&10.1369&8.3829&8.3793\\
		$g_{\omega}(\rho_0)$&14.2925&13.3388&12.7704&10.9871&10.9804\\
		$g_{\rho}(\rho_0)$&15.1509&7.2380&7.8483&7.6971&8.0604\\
		\hline
		$a_{\sigma}$&1.0627&1.3970&1.2099&1.2040&1.1964\\
		$b_{\sigma}$&1.7636&1.3350&0.2129&0.1921&0.1917\\
		$c_{\sigma}$&2.3089&2.0671&0.3080&0.2777&0.2738\\
		$d_{\sigma}$&0.3799&0.4016&1.0403&1.0955&1.1034\\
		$a_{\omega}$&1.0592&1.3926&1.2375&1.1608&1.1693\\
		$b_{\omega}$&0.4183&1.0191&0.0391&0.04460&0.0264\\
		$c_{\omega}$&0.5386&1.6060&0.0724&0.0672&0.0423\\
		$d_{\omega}$&0.7866&0.4556&2.1457&2.2269&2.8062\\
		$a_{\rho}$&0.7761&0.6202&0.3326&0.5487&0.5579\\
		\hline
		\hline 
	\end{tabular}
	\label{tab1}
\end{table}

Table \ref{tab2} displays the NM properties such as symmetry energy, incompressibility, and slope parameter at saturation density for various DD-RMF parameter sets. The $E/A$ for all the parameter sets lies well around -16 MeV. The value of $J$ lies in the range $\approx$ 31-34 MeV which is compatible with the measurement from various astrophysical observations, $J=31.6\pm 2.66$ MeV \cite{LI2013276}. The $L$ value lies outside the constraints $L=59.57\pm10.06$MeV for the DD-LZ1 parameter set, while others satisfy this constraint properly \cite{PhysRevC.101.034303,DANIELEWICZ20141}.  The proton and neutron effective masses are very large for the DDVT and DDVTD parameter sets as compared with the DDV set.\\

\begin{table}
	\centering
	\caption{NM properties binding energy ($E/A$), incompressibility ($K_0$), symmetry energy ($J$), slope parameter ($L$) at saturation density for various DD-RMF parameter sets. }
	\begin{tabular}{ cccccc }
		\hline
		\hline
		Parameter&DD-LZ1&DD-MEX&DDV&DDVT&DDVTD \\
        \hline
		$\rho_0(fm^{-3})$ &0.158&0.152&0.151&0.154&0.154\\
		$E/A$(MeV)&-16.126&-16.140&-16.097&-16.924&-16.915\\
		$K_0$(MeV)&231.237&267.059&239.499&239.999&239.914\\
		$J$(MeV)&32.016&32.269&33.589&31.558&31.817\\
		$L$(MeV)&42.467&49.692&69.646&42.348&42.583\\
		$M_n^*/M$&0.558&0.556&0.586&0.667&0.667\\
		$M_p^*/M$&0.5582&0.556&0.585&0.666&0.666\\
		\hline
		\hline
	\end{tabular}
	\label{tab2}
\end{table}

\subsection{Equation of State}
\begin{figure}[ht]
	\includegraphics[scale=0.35]{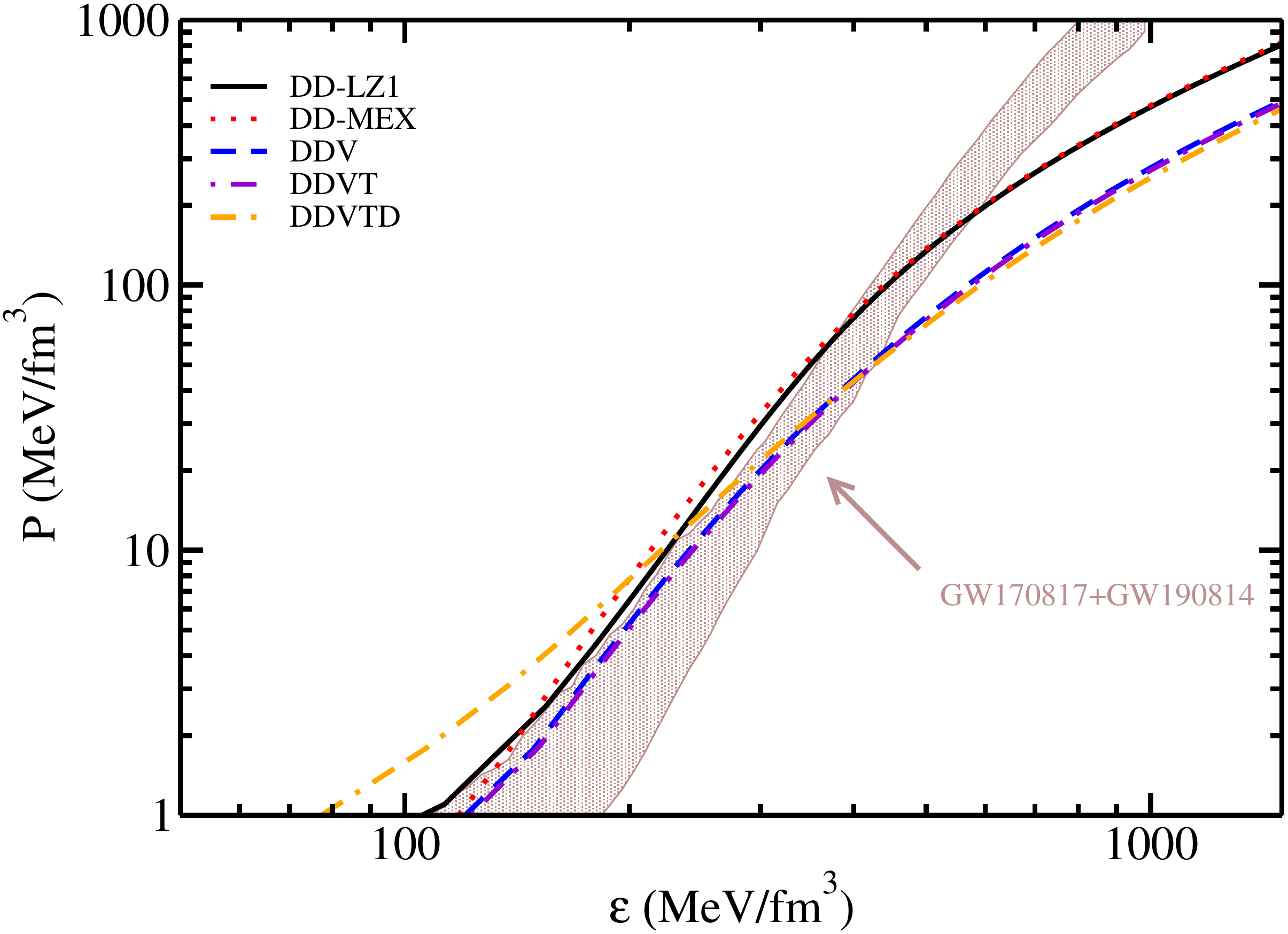}
	\caption{(color online) EoS profile for  DD-LZ1, DD-MEX, DDV, DDVT, and DDVTD  parameter sets. The recent combined constraints from GW170817 \cite{PhysRevLett.121.161101} and GW190814 \cite{Abbott_2020a} are also shown \cite{Abbott_eos}. }
	\label{fig1} 
\end{figure}

Figure \ref{fig1} displays the various EoS for various DD-RMF parameter sets for a NS in beta-equilibrium and charge-neutrality conditions. The DDVTD parameter set produces the stiffest EoS at low densities and softest EoS at high density as compared with other parameter sets. DDV and DDVT sets produce soft EoS at high densities which represent a NS with small maximum mass. The DD-LZ1 and DD-MEX parameter sets produce stiff EoSs at high densities and hence larger NS maximum masses. The recently combined constraints from the gravitational wave data GW170817 and GW190814 in the shaded region are adopted from Ref. \cite{Abbott_2020a}. This joint constraint was introduced by considering the GW190814 event as neutron star-black hole (NSBH) merger, with its secondary component assumed to be a NS. For this scenario, the maximum mass was assumed to be not less than secondary component of GW190814, which constraints the distribution of EoSs compatible with astrophysical data. For a unified EoS, the Baym-Pethick-Sutherland (BPS) EoS \cite{Baym:1971pw} is used for the outer crust part which lies in the density region 10$^4$-10$^{11}$ g/cm$^{3}$. Since the outer crust EoS does not effect the NS maximum and the radius, therefore the it is chosen for the outer crust part of the NS for all parameter sets. The inner crust EoS has a high impact on the NS radius, $R_{1.4M_{\odot}}$ at the canonical mass,  while a small change is seen in the maximum mass and radius \cite{rather2020effect}. For the parameter sets used in this work, the inner crust EoS is not available. Thus, we have employed the DD-ME2 inner crust EoS \cite{PhysRevC.71.024312} for all the parameter sets but with matching symmetry energy and slope parameter \cite{PhysRevC.94.015808,PhysRevC.90.045803}.\par

For the mixed-phase HM and QM, the Gibbs construction method, which corresponds to the global charge neutrality between two different phases, has been employed. The effective bag model with an effective bag constant $B^{1/4}$ is used to study the QM. The coupling constant parameter $K_{\nu}$ is fixed at 6 GeV$^{-2}$ for the three flavor configuration. Three different values of effective bag constant are used $B_{eff}^{1/4}$=130, 145, and 160 MeV.
  
  \begin{figure}[h]
  	\includegraphics[scale=0.35]{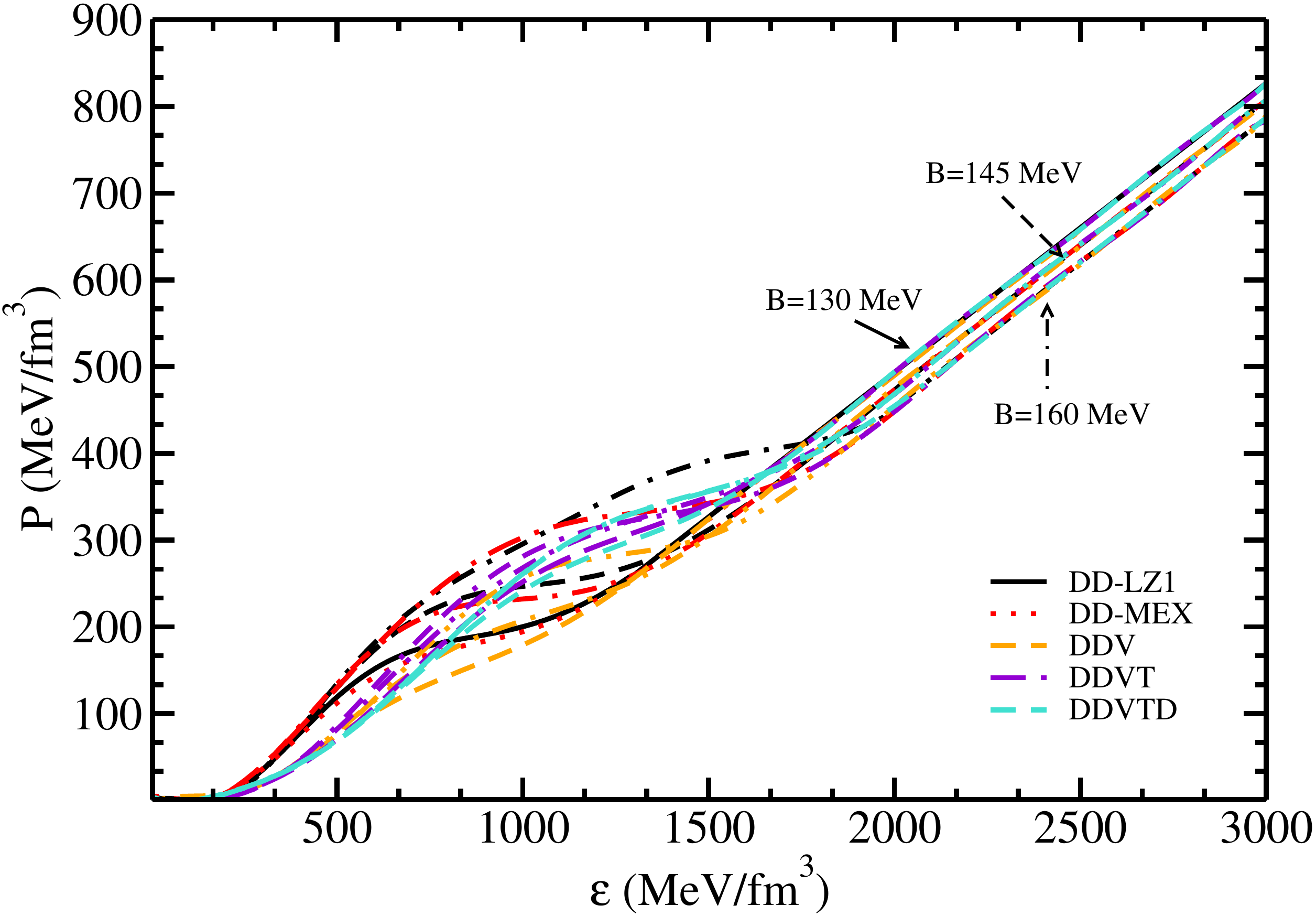}
  	\caption{(color online) Equation of state for the hadron-quark phase transition for DDV, DDVT, DDVTD, DD-LZ1, and DD-MEX hadronic parameter sets and vBag QM at different effective bag constants. The solid lines represents the hybrid EoS at $B_{eff}^{1/4}$=130 MeV while dashed and dot-dashed lines represent the hybrid EoS at $B_{eff}^{1/4}$=145 \& 160 MeV, respectively. }
  	\label{fig2} 
  \end{figure}
  
  Figure \ref{fig2} shows the hadron-quark phase transition
  with DD-RMF parameter sets for hadronic matter and vBag model for QM using the Gibbs method for constructing mixed-phase which ensures a smooth transition between the two different phases. With the increasing effective bag constant $B_{eff}^{1/4}$, the phase transition density increases, and the mixed-phase region also expands. For bag constant $B_{eff}^{1/4}$=130 MeV, the mixed-phase region starts from $\rho=2.47 \rho_0$ and extends up to 4.03 $\rho_0$. For $B_{eff}^{1/4}$=145 and 160 MeV, the mixed-phase region lies in the density range (3.03-4.82) $\rho_{0}$ and (3.69-5.31)$\rho_0$, respectively. DD-LZ1 and DD-MEX parameter sets produce stiff EoS and thus the mixed-phase region lies in a higher pressure region than the DDV, DDVT, and DDVTD parameter sets. The mixed-phase region in the DD-LZ1 parameter sets lies in the density range (2.56-4.23)$\rho_{0}$ for 130 MeV, (2.73-4.95)$\rho_{0}$ for 145 MeV ,and (3.04-5.43)$\rho_{0}$ for 160 bag constants. Thus, DD-LZ1 and DD-MEX sets predict a large mixed-phase region as compared with the other parameter sets. \par 
 
\subsection{Neutron star properties}
\label{sub2}
\begin{figure}
	\includegraphics[width=0.48\textwidth]{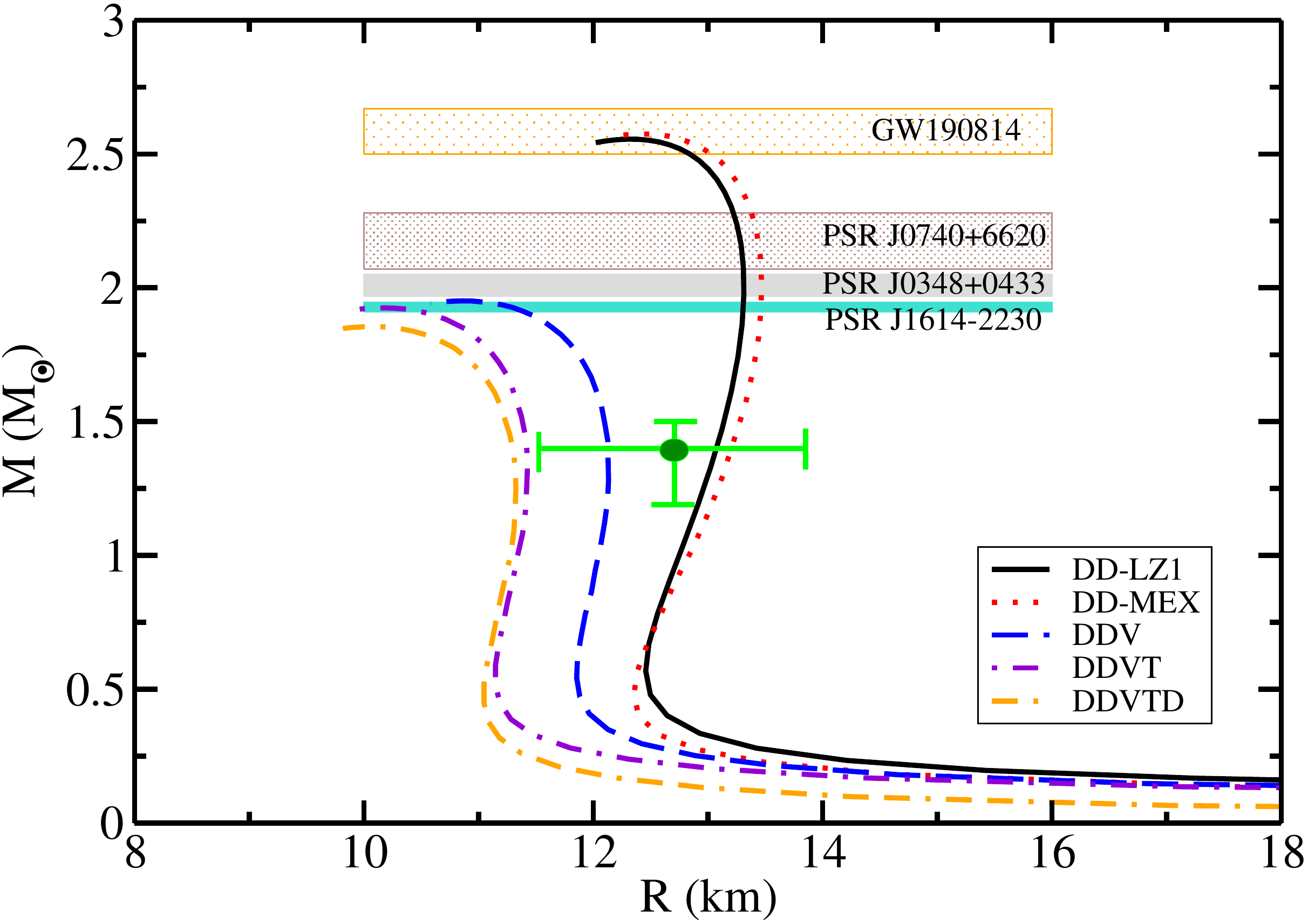}
	\caption{(color online) Mass vs Radius profiles for pure DD-LZ1, DD-MEX, DDV, DDVT, and DDVTD parameters for a static NS. The recent constraints on mass from the various gravitational wave data and the pulsars (shaded region) \cite{Abbott_2020a,Demorest2010,Antoniadis1233232,Cromartie2020} and radii \cite{Miller_2019,Riley_2019} are also shown.}
	\label{fig3}
\end{figure}

Figure \ref{fig3} displays the hadronic mass vs radius curves for DD-LZ1, DD-MEX, DDV, DDVT, and DDVTD parameter sets. The DD-LZ1 set produces a NS with a maximum mass of 2.55$M_{\odot}$  and with a radius of 12.30 km. DD-MEX set produces a 2.57$M_{\odot}$ NS with a 12.46 km radius. Both these parameter sets satisfy the constraints from recent gravitational wave data GW190814 and recently measured mass and radius of PSR J0030+0451, $M=1.34_{-0.16}^{+0.15}M_{\odot}$ and $R=12.71_{-1.19}^{+1.14}$ km by NICER \cite{Miller_2019,Riley_2019}. The DDV, DDVT, and DDVTD predict a maximum mass of 1.95$M_{\odot}$, 1.93$M_{\odot}$, and 1.85$M_{\odot}$ for a static NS with 12.11, 11.40, and 11.33 km radius at canonical mass, $R_{1.4}$, respectively. DDV and DDVT satisfy the mass constraint from PSR J1614-2230 and radius constraint from PSR J0030+0451. The DDVTD parameter set produces a NS with a slightly lower maximum mass than PSR J1614-2230. The shaded regions display the constraints on the maximum mass of a NS from PSR J1614-2230 (1.928$\pm$0.017$M_{\odot}$)\cite{Demorest2010}, PSR J0348+0432 (2.01$\pm$0.04$M_{\odot}$) \cite{Antoniadis1233232}, MSP J0740+6620 (2.14$^{+0.10}_{-0.09}$$M_{\odot}$) \cite{Cromartie2020}, and GW190814 (2.50-2.67 $M_{\odot}$) \cite{Abbott_2020a}.\par 

\begin{figure}
	\includegraphics[width=0.48\textwidth]{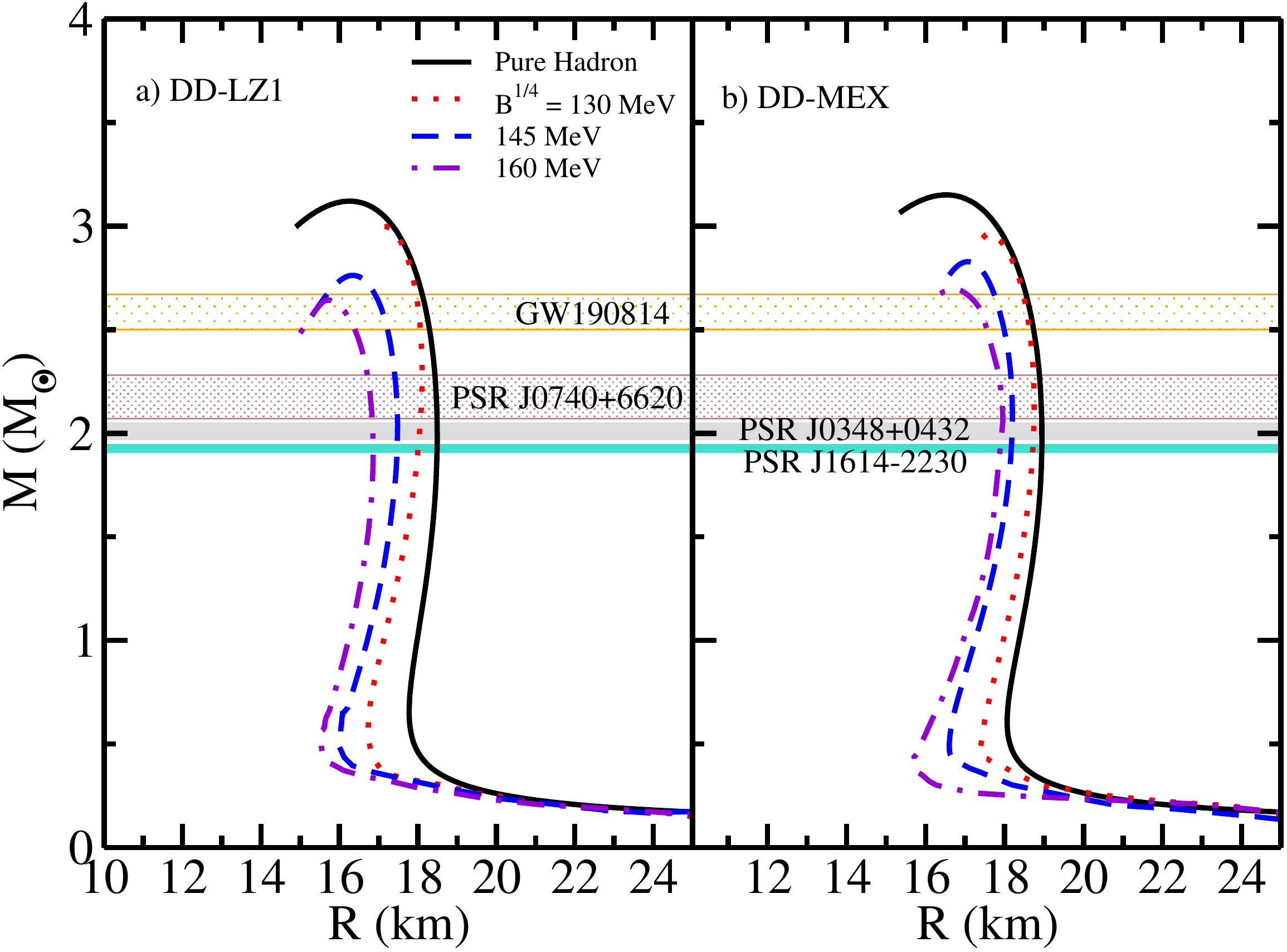}
	\caption{(color online) Mass-Radius profile for pure hadronic and hybrid rotating NSs for (a) DD-LZ1 and (b) DD-MEX parameter sets at bag values $B_{eff}^{1/4}$=130, 145 \& 160 MeV. The shaded regions represent recent constraints on the mass from various measured astronomical observables.  }
	\label{fig4}
\end{figure}

The RNS mass-radius profile for DD-LZ1 and DD-MEX parameter sets are shown in Fig. \ref{fig4}. The solid lines represent the pure hadronic star while the dashed lines represent the HS at different bag constants. The effective bag constant $B_{eff}^{1/4}$ is written as $B^{1/4}$ for convenience. The DD-LZ1 EoS produces a pure hadronic RNS with a maximum mass of 3.11$M_{\odot}$ with a radius of 18.23 km. With the phase transition from HM to QM, the maximum mass and the corresponding radius decrease with the increase in the bag constant. For the DD-LZ1 set, the maximum mass decreases from 3.11$M_{\odot}$ to 2.98$M_{\odot}$ for $B^{1/4}$=130 MeV, and to 2.75$M_{\odot}$ and 2.64$M_{\odot}$ for $B^{1/4}$=145 and 160 MeV, respectively. The radius $R_{1.4}$ decreases from 18.32 km for pure HM to 16.64 km for hybrid star matter at 160 MeV bag value. Similarly, for the DD-MEX parameter set, the maximum mass for pure hadronic matter is 3.15$M_{\odot}$ at radius 16.53 km which reduces to 2.69$M_{\odot}$ at 16.63 km for bag constant of 160 MeV. Thus, while the pure hadronic RNS predict a large maximum mass, the phase transition to QM lowers the maximum mass and the radius thereby satisfying the maximum mass constraint from GW190814. These results imply that the secondary component of GW190814 could be a possible fast-rotating hybrid star.

\begin{figure}
	\includegraphics[width=0.48\textwidth]{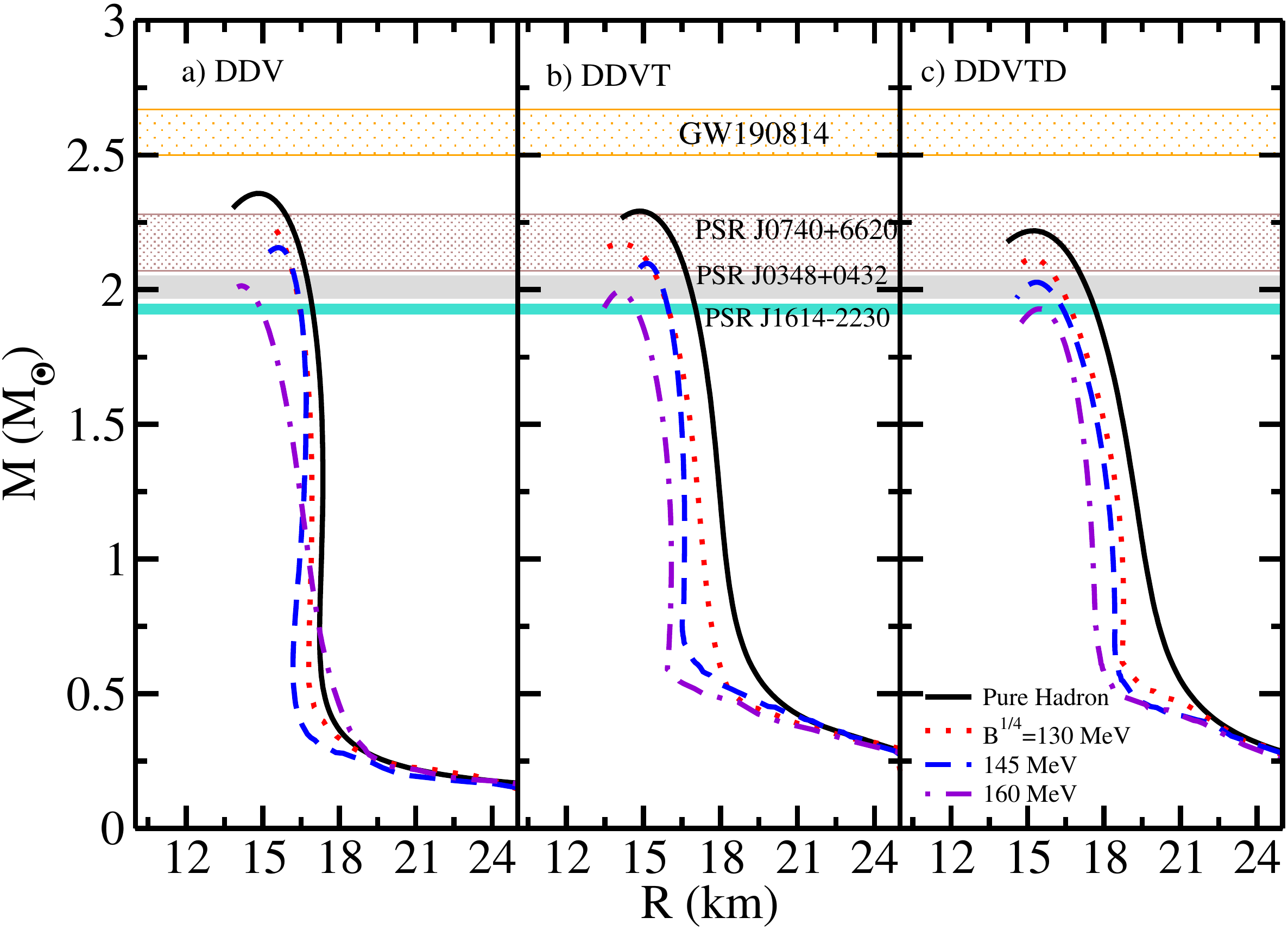}
	\caption{(color online) Same as Fig. \ref{fig4}, but for (a) DDV, (b) DDVT, and (c) DDVTD EoSs. }
	\label{fig5}
\end{figure}
Figure \ref{fig5} displays the mass-radius relation for hadronic and hybrid rotating NS with DDV, DDVT, and DDVTD EoSs. The maximum mass for a RNS with DDV EoS is 2.37$M_{\odot}$ with a 17.41 km radius at the canonical mass. Both the maximum mass and the radius decrease to 2.23$M_{\odot}$, 2.13$M_{\odot}$, 2.01$M_{\odot}$ and 16.91, 16.68, 16.13 km for bag constants $B^{1/4}$=130, 145, and 160 MeV, respectively, thereby satisfying the 2$M_{\odot}$ constraint. For DDVT, the maximum mass reduces from 2.28$M_{\odot}$ to 1.99$M_{\odot}$. $R_{1.4}$ also decreases from 17.82 km to 16.01 km. Similarly for the DDVTD EoS, the RNS maximum mass reduces to 1.93$M_{\odot}$ from 2.21$M_{\odot}$ at $B^{1/4}$=160 MeV. For all the parameter sets, the phase transition to QM lowers the maximum mass which satisfies the 2$M_{\odot}$ limit.\par
\begin{figure}
	\includegraphics[width=0.48\textwidth]{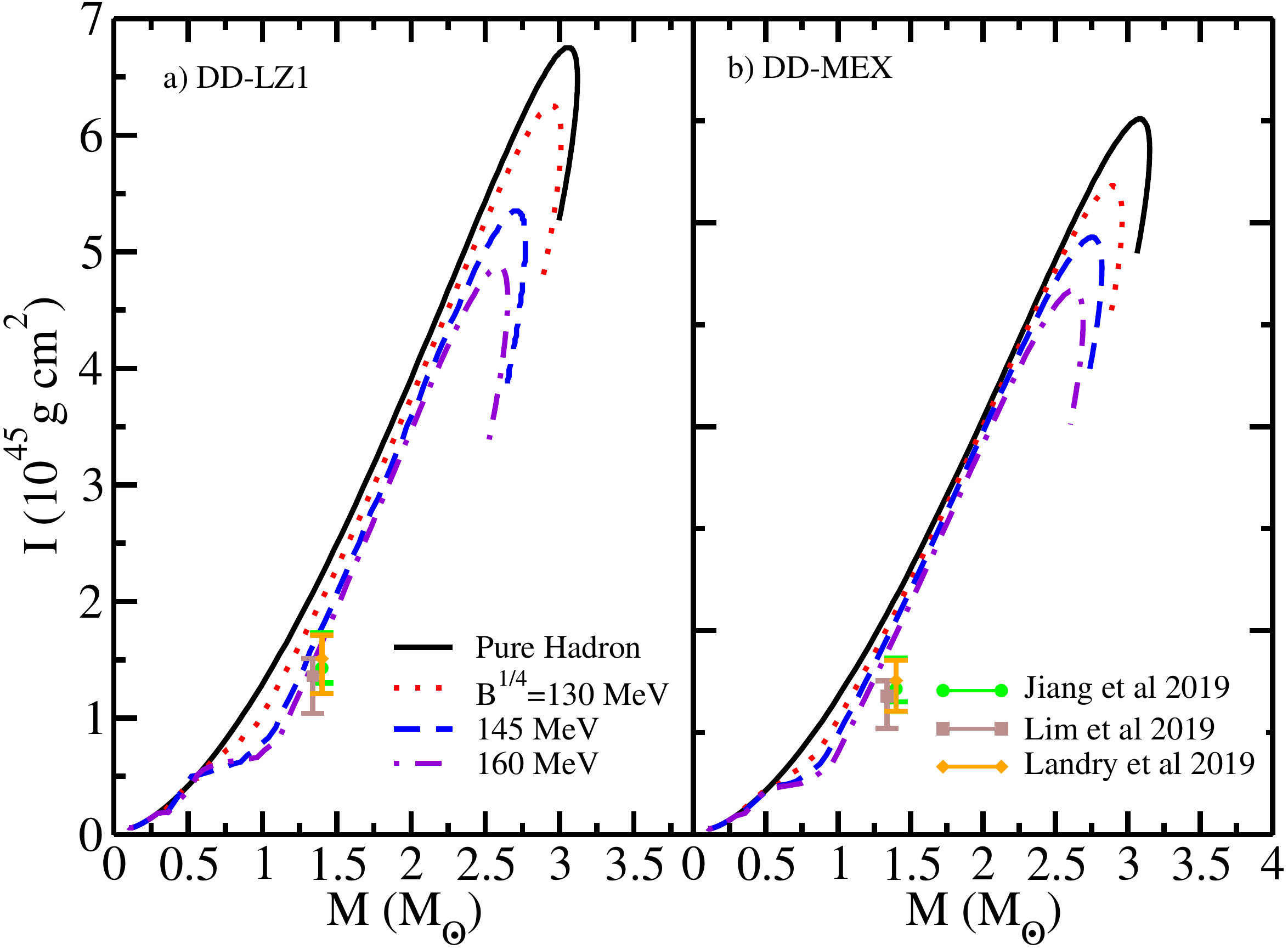}
	\caption{(color online) Moment of inertia variation with the gravitational mass for (a) DD-LZ1 and (b) DD-MEX EoSs. The constraints on canonical moment of inertia are also shown \cite{PhysRevC.100.035802}. The constraint from joint PSR J0030+0451, GW170817, and the nuclear data analysis are shown by green bar \cite{Jiang_2020}. The predicted moment of inertia of pulsar J0737-3039A using Bayesian analysis of nuclear EoS is shown by brown bar \cite{PhysRevD.101.123007}. }
	\label{fig6}
\end{figure}

 
The measurement of the NS moment of inertia is important because it follows a universal relation with the tidal deformability and the compactness of a NS.
The moment of inertia as a function of gravitational mass for the RNS is displayed in Fig. \ref{fig6}. The constraint on the moment of inertia obtained from the joint PSR J0030+0451, GW170817, and the nuclear data analysis predicting $I_{1.4}=1.43_{-0.13}^{+0.30}\times 10^{38}$ kg.m$^2$ is given in Ref. \cite{Jiang_2020}. The predicted moment of inertia of pulsar PSR J0737-3093A,  $I_{1.338}=1.36_{-0.32}^{+0.15}\times 10^{45}$ g.cm$^2$ is also given \cite{PhysRevD.101.123007}. For pure hadronic matter, DD-LZ1 and DD-MEX EoSs predicts an NS with a moment of inertia 2.22 and 2.35 $\times$ 10$^{45}$ g.cm$^2$, respectively. The phase transition to the QM reduces the moment of inertia to a value 1.65 and 1.93$\times$ 10$^{45}$ g.cm$^2$ for DD-LZ1 and DD-MEX parameter sets at bag constant $B^{1/4}$=160 MeV, which satisfies the constraint from Refs. \cite{Jiang_2020,PhysRevD.101.123007,PhysRevC.100.035802}.

\begin{figure}
	\includegraphics[width=0.48\textwidth]{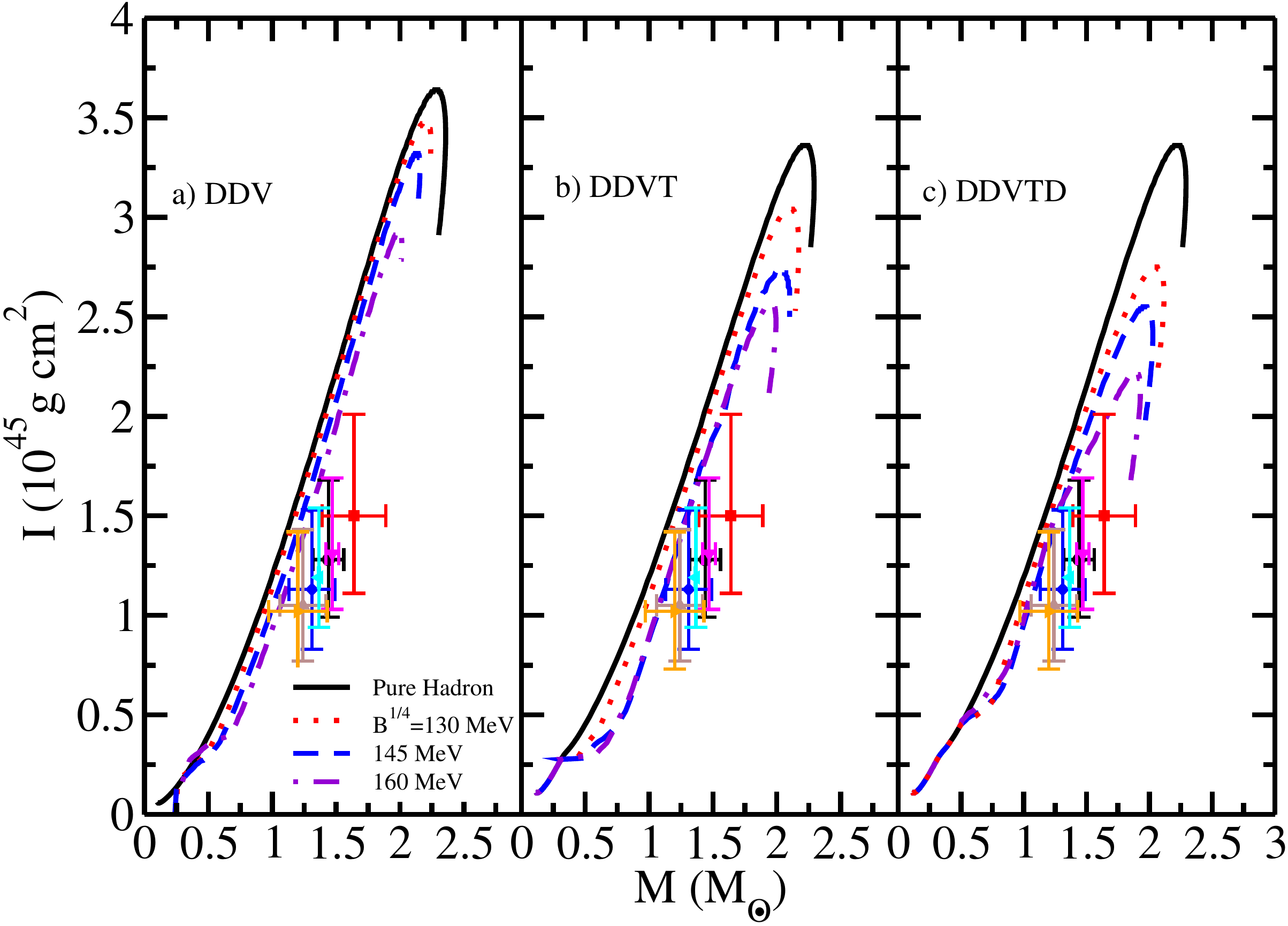}
	\caption{(color online) Same as Fig. \ref{fig6} but for (a) DDV, (b) DDVT, and (c) DDVTD parameter sets. The constraints on the moment of inertia of MSPs obtained from universal relations with GW170817 are shown \cite{PhysRevD.99.123026}.}
	\label{fig7}
\end{figure}

Figure \ref{fig7} displays the moment of inertia variation with the gravitational mass for DDV, DDVT, and DDVTD parameter sets. The solid lines represent the pure hadronic matter, while the dashed lines represent the hadron-quark mixed phase at bag constants $B^{1/4}$=130, 145, and 160 MeV. The constraints on the moment of inertia obtained from millisecond pulsars (MSPs) with GW170817 universal relations are shown in Ref. \cite{PhysRevD.99.123026}. For the DDV EoS, the moment of inertia of a pure hadronic star is found to be 2.01$\times$ 10$^{45}$ g.cm$^2$ while for the DDVT and DDVTD EoSs, the value is found to be 1.95 and 1.88 $\times$ 10$^{45}$ g.cm$^2$, respectively.  For the hybrid EoS, the moment of inertia is lowered to a value of 1.71$\times$ 10$^{45}$ g.cm$^2$ for the DDV set at bag constant 160 MeV. For DDVT and DDVTD sets, this value reduces to 1.68 and 1.64$\times$ 10$^{45}$ g.cm$^2$ respectively for a 160 MeV bag constant. The phase transition to QM produces a NS with the moment of inertia that satisfies the constraints from various measurements. \par 
For a static NS, the maximum mass is usually determined as the first maximum of a $M$-$\varepsilon_c$ curve, i.e., $\partial M/\partial \varepsilon_c$=0, where $\varepsilon_c$ is the central energy density. For RNSs, the situation becomes complicated. To determine the axisymmetric instability points, several methods have been used in the literature. Friedman et al. \cite{1988ApJ...325..722F} described a method to determine the points at which instability is reached in rotating NSs \cite{Sorkin:1981jc,Sorkin:1982ut}.
\begin{equation}
\Bigg|\frac{\partial M (\varepsilon_c,J)}{\partial \varepsilon_c}\Bigg|_{J=constant}=0,
\end{equation}  
where $J$ is the angular momentum of the star, which is obtained self-consistently in the solution of the Einstein's equation for a rotating NS. Once the secular instability is initiated, the star evolves until it reaches a point of dynamical instability where the gravitational collapse starts \cite{Stergioulas2003}. The maximum mass of the rotating star lies at the termination point of uniformly rotating star. \par 
The above equation defining an upper limit on the mass at a given angular momentum is sufficient but not a necessary condition for the instability. The limit on the dynamic instability is shown in Ref. \cite{takami}.\par 
\begin{figure}
	\includegraphics[width=0.48\textwidth]{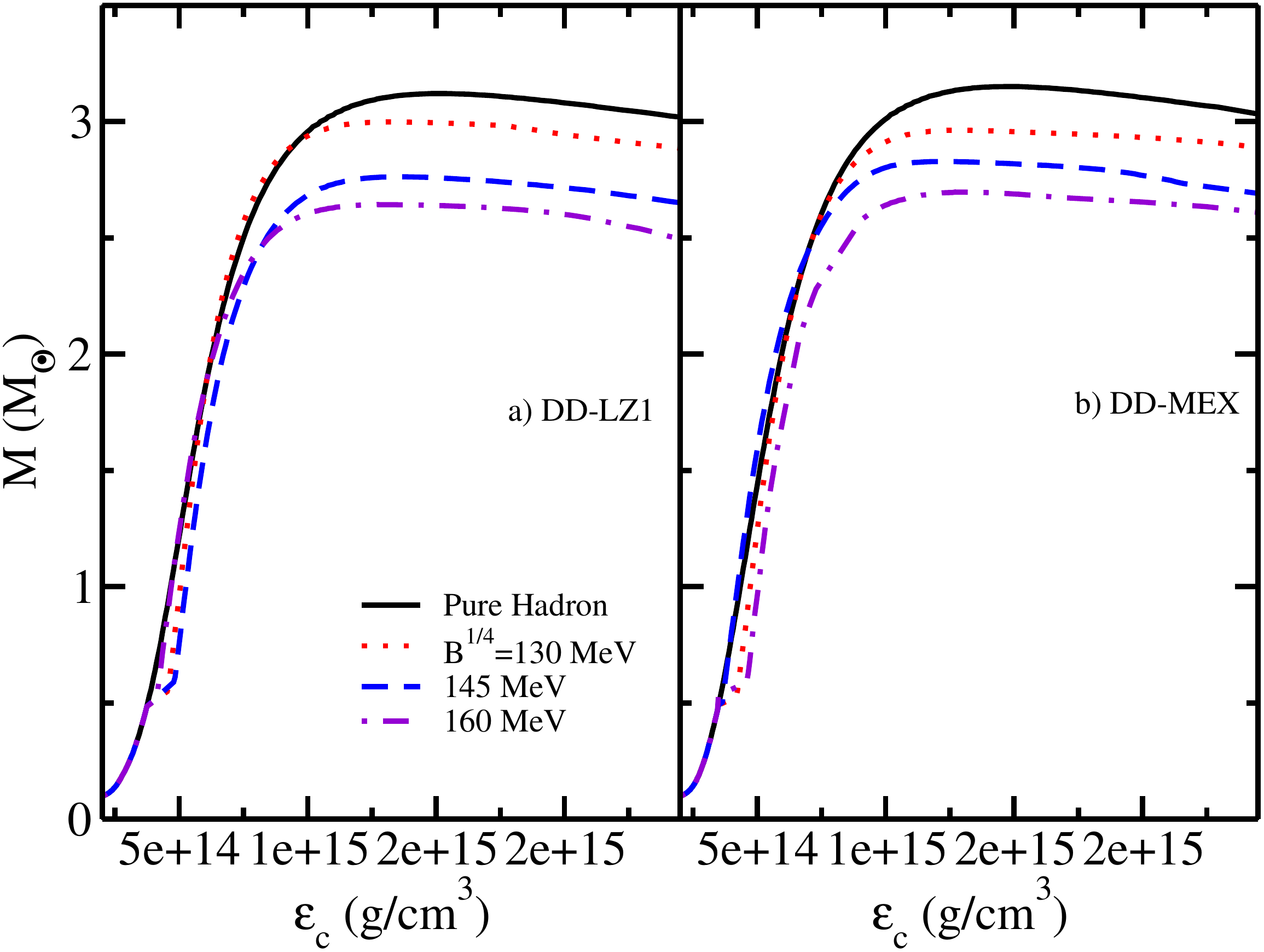}
	\caption{(color online) Gravitational mass versus central density for (a) DD-LZ1 and (b) DD-MEX EoSs. The solid lines represent pure hadronic rotating stars while dashed lines represent the hybrid stars at different bag constants.}
	\label{fig8}
\end{figure}

\begin{figure}
	\includegraphics[width=0.48\textwidth]{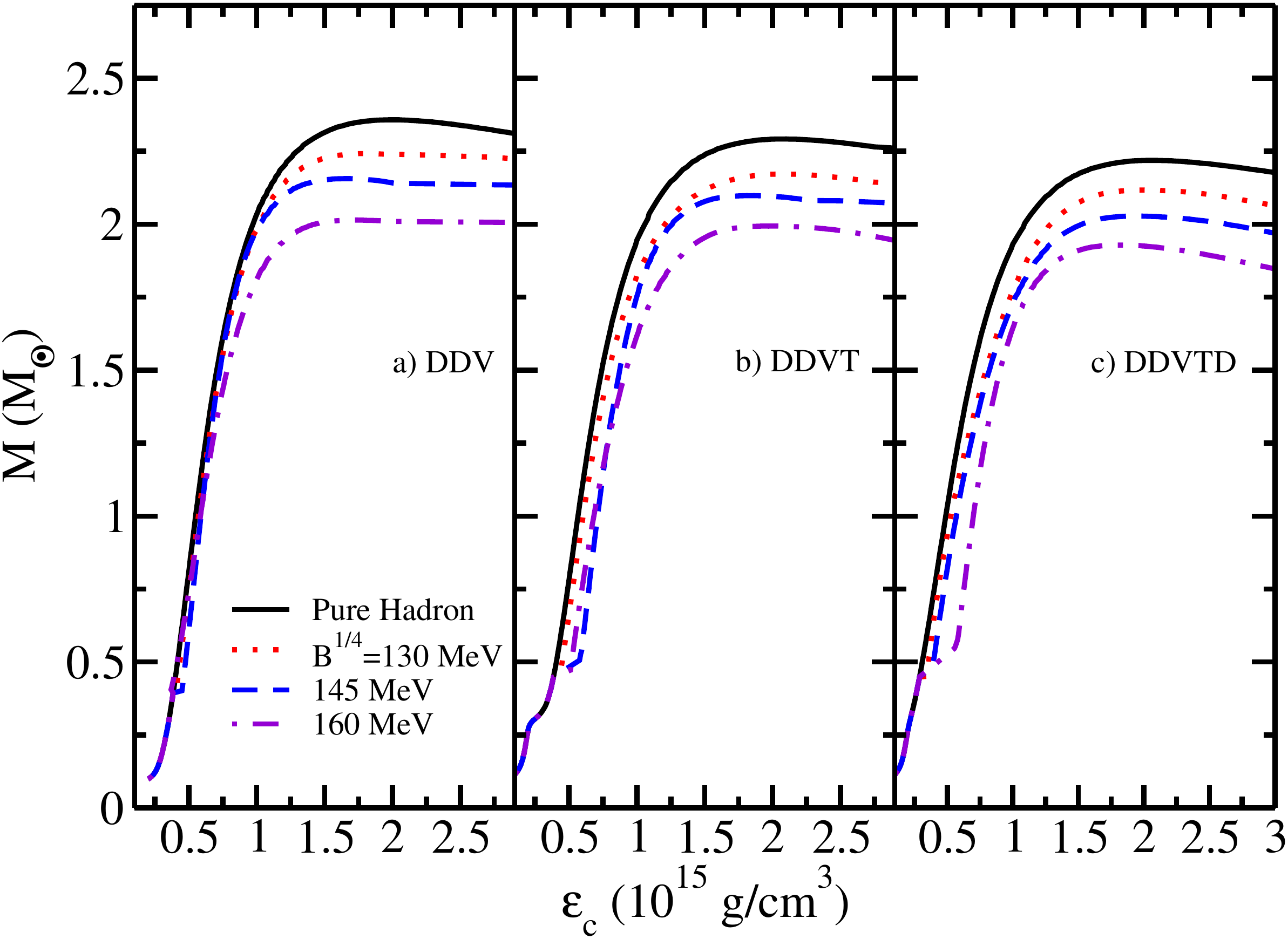}
	\caption{(color online) Same as Fig. \ref{fig8} but for (a) DDV, (b) DDVT, and (c) DDVTD EoSs.}
	\label{fig9}
\end{figure}

Figure \ref{fig8} shows the variation in the gravitational mass of a rotating NS with the central density for DD-LZ1 and DD-MEX parameter sets. Figure \ref{fig9} repreents the same for DDV, DDVT, and DDVTD parameter sets. The maximum mass of 3.11$M_{\odot}$ for DD-LZ1 EoS is produced at a density of 1.40$\times$10$^{15}$ g/cm$^3$. The phase transition to QM at bag constant $B^{1/4}$=160 MeV reduces the maximum mass to 2.64$M_{\odot}$ at 1.17$\times$10$^{15}$ g/cm$^3$ energy density. For the DD-MEX parameter set, the maximum mass of 3.15$M_{\odot}$ occurs at 1.47$\times$10$^{15}$ g/cm$^3$ reduces to 2.69$M_{\odot}$ at 1.25$\times$10$^{15}$ g/cm$^3$. \par 
A star rotating at a Keplerian rate becomes unstable due to the loss of mass from its surface. The mass shedding limit angular velocity which is the maximum angular velocity of a rotating star is the Keplerian angular velocity evaluated at the equatorial radius $R_e$, i.e., $\Omega_K^{J\ne0}=\Omega_{orb}(r=R_e).$  \par 
\begin{figure}
\includegraphics[width=0.48\textwidth]{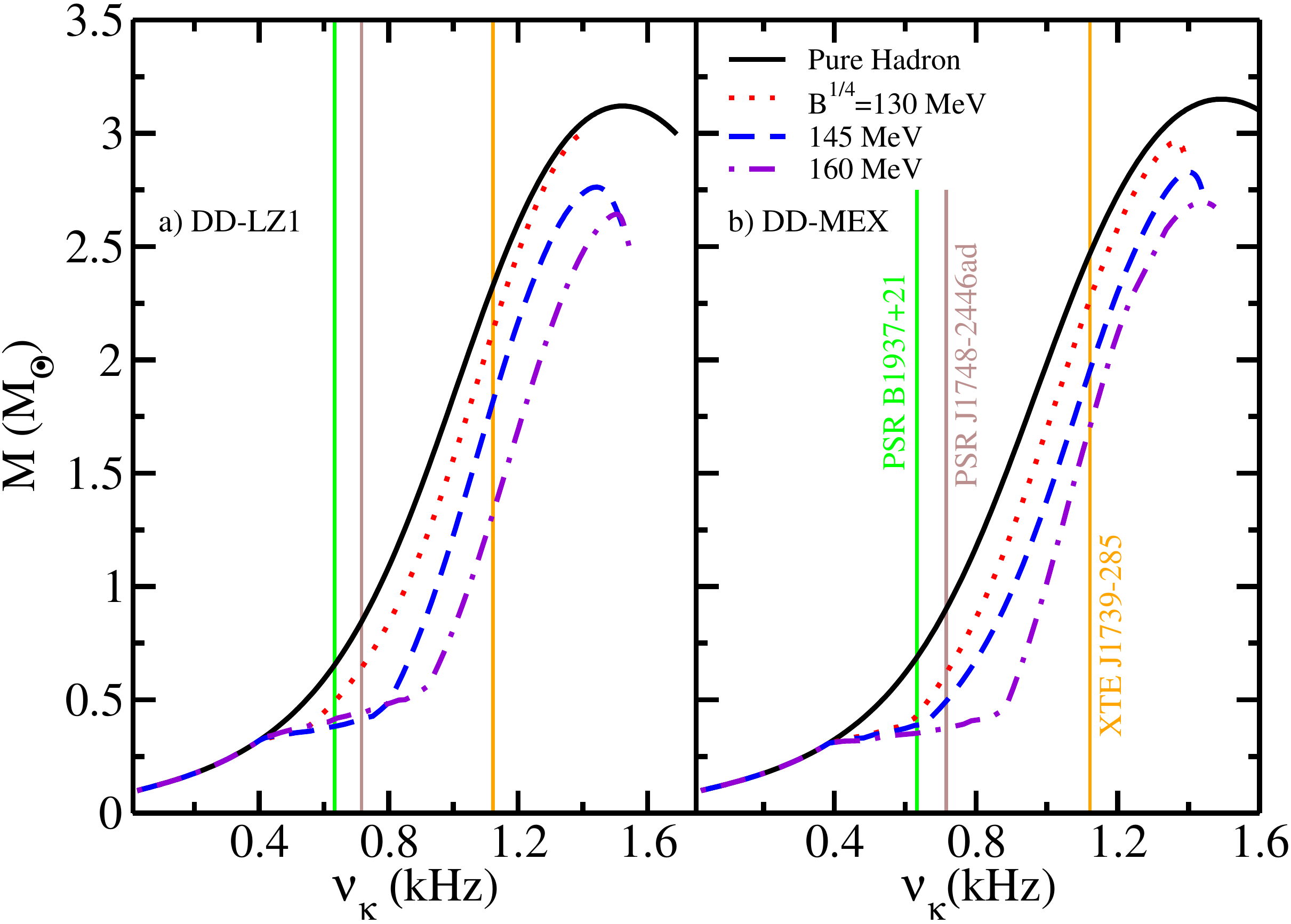}
\caption{(color online) Variation of rotational frequency with the NS gravitational mass at Keplerian velocity for (a) DD-LZ1 and (b) DD-MEX EoSs. Solid lines represent a pure hadronic star while the dashed lines represent a hybrid star at bag constants $B^{1/4}$=130, 145, and 160 MeV. The vertical lines represent the observational limits imposed on the frequency from rapidly rotating pulsars like PSR B1937+21 ($\nu_k$=633 Hz) \cite{Backer1982}, PSR  J1748-2446ad ($\nu_k$=716 Hz) \cite{Hessels1901}, and XTE J1739-285($\nu_k$=1122 Hz) \cite{Kaaret_2007}.}
\label{fig10}
\end{figure}

Figures \ref{fig10}a and \ref{fig10}b display the NS gravitational mass as a function of the Kepler frequency $\nu_k$ for the DD-LZ1 and DD-MEX EoSs, respectively. The limits imposed on the rotational frequency by various pulsars such as PSR B1937+21 ($\nu_k$=633 Hz) \cite{Backer1982}, PSR  J1748-2446ad ($\nu_k$=716 Hz) \cite{Hessels1901}, and XTE J1739-285 ($\nu_k$=1122 Hz) \cite{Kaaret_2007} are also shown. For the DD-LZ1 EoS, the pure hadronic star rotates with a maximum frequency of 1525 Hz. For a HS at bag value $B^{1/4}$=130 MeV, the star rotates with a frequency of 1405 Hz. For bag values of 145 and 160, the frequency obtained is 1431 and 1497 Hz, respectively. Similarly for DD-MEX EoS, the maximum rotational frequency for a pure hadronic star is found to be 1503 Hz, which changes to 1361 Hz at a bag constant of 130 MeV, 1408 and 1438 Hz for the HS at 145 and 160 MeV bag values. Both pure hadronic and HSs rotate at a frequency greater than $\nu_k$=1122 Hz. Also, the hybrid star $M$-$\nu_k$ curves coincide with the pure hadronic curves upto $\nu_k \approx$ 400 Hz, which then show a transition towards higher frequency depending upon the bag constant. \par

\begin{figure}
	\includegraphics[width=0.48\textwidth]{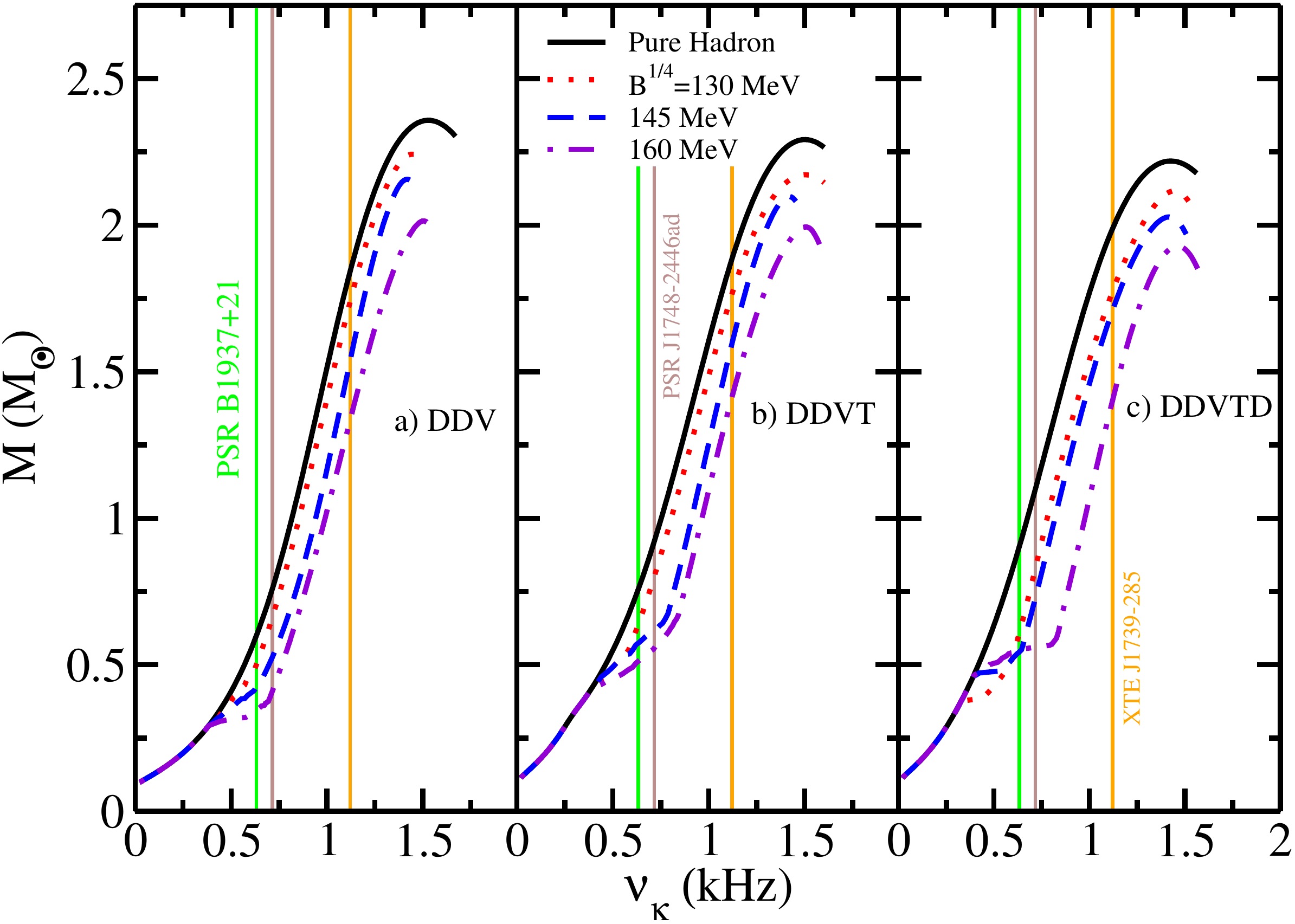}
	\caption{(color online) Same as Fig. \ref{fig10} but for (a) DDV, (b) DDVT, and (c) DDVTD EoSs.}
	\label{fig11}
\end{figure}
 Figure \ref{fig11} displays the same gravitational mass variation with the Kepler frequency for DDV, DDVT, and DDVTD parameter sets. For the DDV set, the pure hadronic star rotates with a rotational frequency of 1498 Hz. The HSs produced with bag constants $B^{1/4}$=130, 145, and 160 MeV have a rotational frequency of 1454, 1446, and 1520 Hz, respectively. Similarly for DDVT and DDVTD EoSs, the pure hadronic star has a rotational frequency of 1473 and 1418 Hz respectively which then changes to 1503 and 1456 Hz respectively for HS at 160 MeV bag constant. Thus it is seen that HSs with a hadron-quark phase transition initially produce a low mass NS with a low rotating frequency than the pure hadronic star at low bag constant ($B^{1/4}$=130 MeV). Thus the HSs can withstand higher rotation as the star is denser and has low maximum mass as compared with the pure hadronic star. \par 
 A useful parameter to characterize the rotation of a star is the ratio of rotational kinetic energy $T$ to the gravitational potential energy $W$, $\beta=T/W$. For a RNS, if $\beta > \beta_d$, where $\beta_d$ is the critical value, the star will be dynamically unstable. The critical value $\beta_d$ for a rigidly rotating star is found to be 0.27 \cite{10.1093/gji/21.1.103-a,1985ApJ...298..220T}. However, for different angular-momentum distributions, the value lies in the range 0.14 to 0.27 \cite{1996ApJ...458..714P,1995ApJ...444..363I,2001ApJ...550L.193C}. \par 
 
\begin{figure}
	\includegraphics[width=0.48\textwidth]{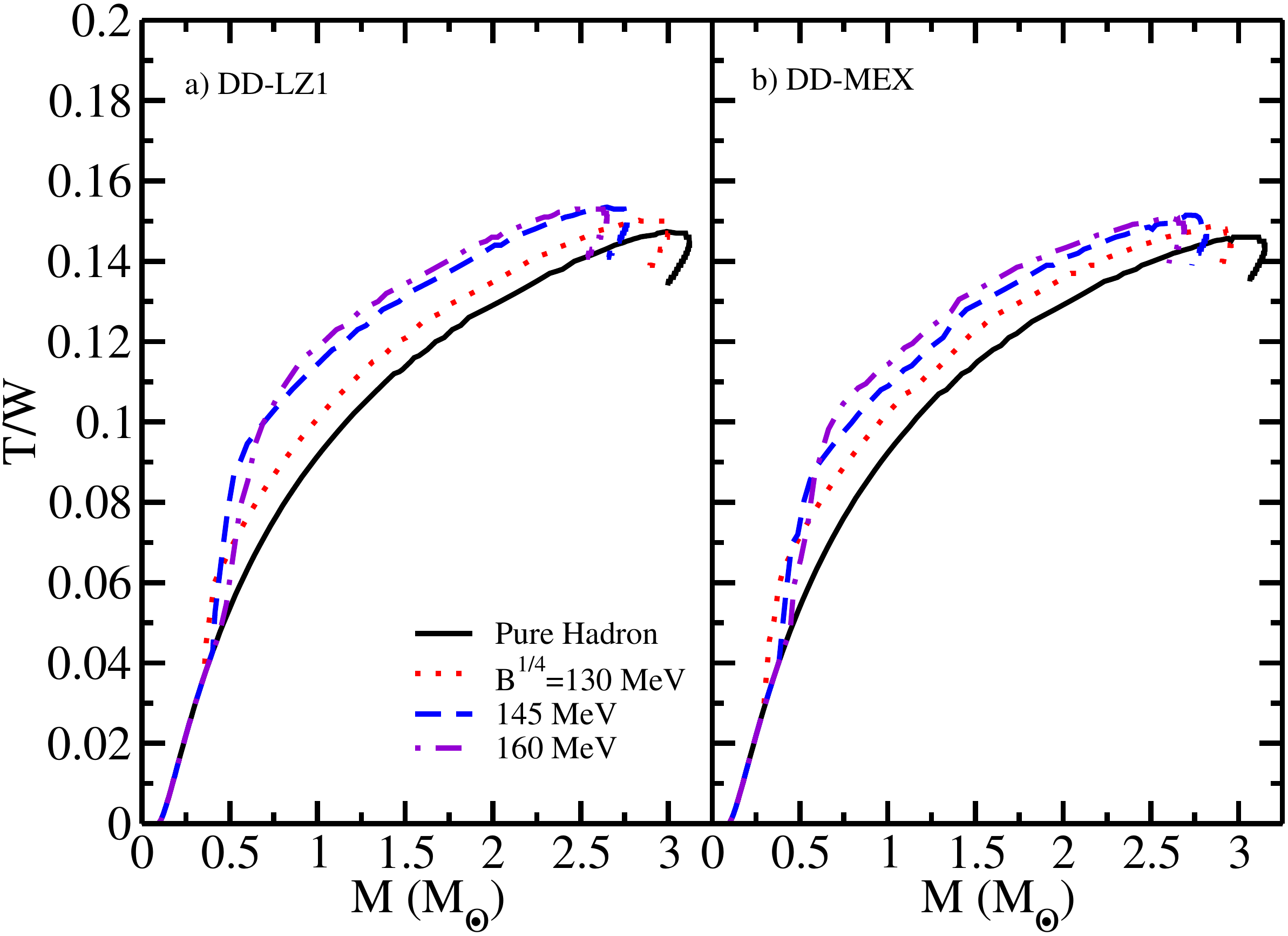}
	\caption{(color online) Variation in the ratio of rotational kinetic energy to the gravitational potential energy $T/W$ with gravitational mass for (a) DD-LZ1 and (b) DD-MEX EoSs. Solid lines represent pure hadronic stars while the dashed lines represent hybrid stars at bag constants $B^{1/4}$=130, 145, and 160 MeV.}
	\label{fig12}
\end{figure}

 The variation in the $T/W$ ratio of the pure hadron and HS with the gravitational mass is shown in Fig. (\ref{fig12}). The $T/W$ ratio for pure hadronic stars is 0.147 and 0.145 for DD-LZ1 and DD-MEX parameter sets, respectively. The HSs have large $T/W$ ratio and increase with bag constant. For DD-LZ1 set, the ratio increases from 0.150 at $B^{1/4}$=130 MeV to 0.153 at $B^{1/4}$=160 MeV. For the DD-MEX set, the ratio increases to 0.149 and 0.151 for bag values 130 and 160 MeV, respectively. The large value of the $T/W$ ratio in HSs is since the quark stars being bound by the strong interaction, unlike hadron stars which are bound by gravity.\par 

\begin{figure}
	\includegraphics[width=0.48\textwidth]{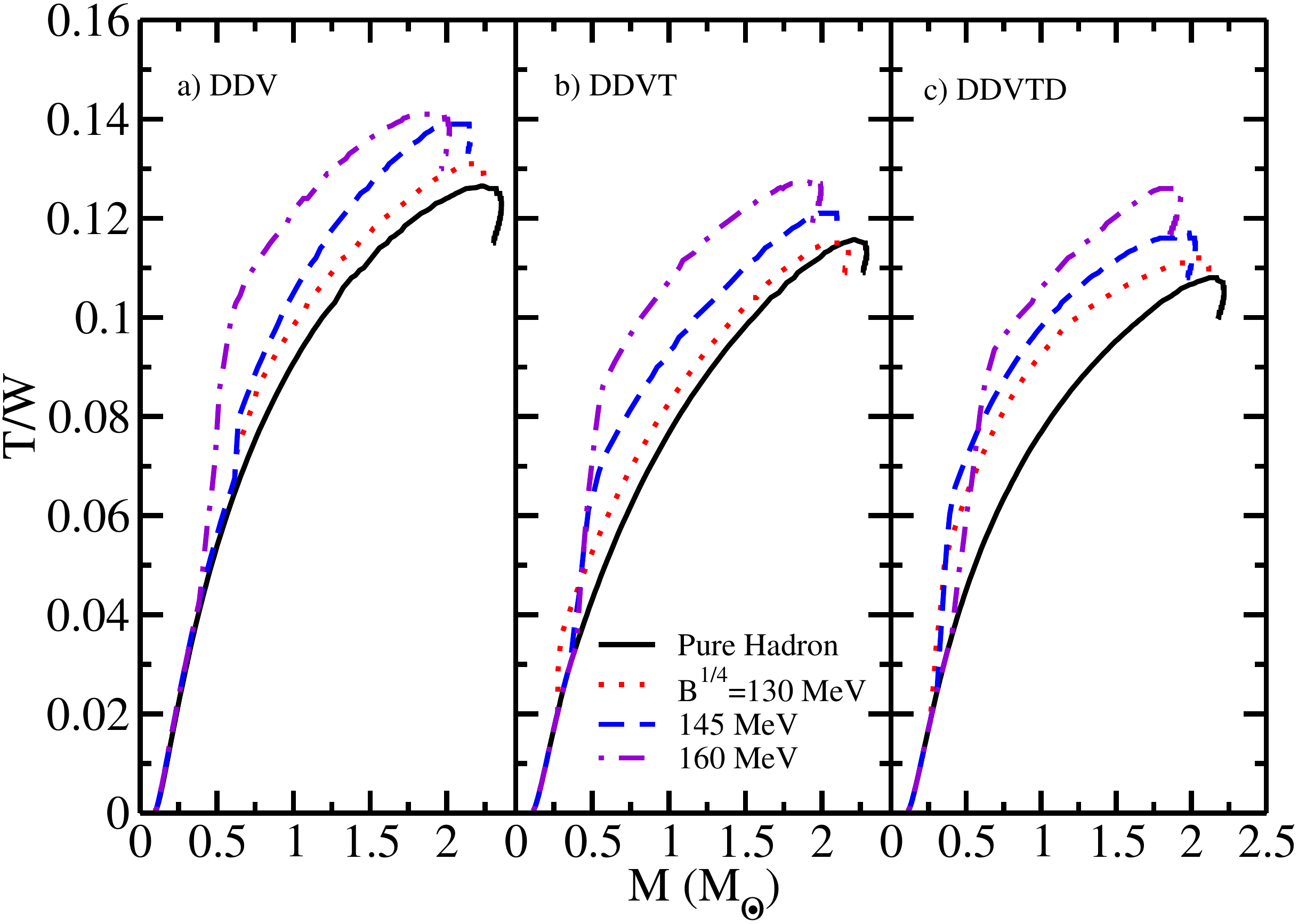}
\caption{(color online) Same as Fig. \ref{fig12} but for (a) DDV, (b) DDVT, and (c) DDVTD EoSs.}
\label{fig13}
\end{figure}
Figure \ref{fig13} depicts the $T/W$ variation with the gravitational mass for DDV, DDVT, and DDVTD parameter sets. For the DDV EoS, the pure hadronic star predicts a $T/W$ ratio of 0.127, which lies below the critical value $\beta_d$. For hybrid stars, this ratio increases 0.142 for a bag constant of 160 MeV thereby satisfying the critical $\beta_d$ limit and hence becomes dynamically unstable and emits gravitational waves. Similarly, for DDVT and DDVTD EoS, the pure hadron star produces a ratio of 0.115 and 0.108 while the HS at $B^{1/4}$=160 MeV gives a value of 0.127 and 0.125, respectively.\par
The Einstein's field equations provide Kerr space-time for so-called Kerr black holes which can be fully described by the angular momentum $J$ and  gravitational mass $M$ of rotating black holes \cite{spin.parameter, PhysRevD.92.023007}. The condition $J\ge GM^2/c$ must be satisfied to define a stable Kerr black hole. The gravitational collapse of a massive RNS constrained to angular-momentum conservation creates a black hole with mass and angular momentum resembling that of a NS. Thus, it is an important quantity used in the study of black holes as well as RNSs. The Kerr parameter leads to the possible limits on the compactness of a NS and also can be an important criterion for determining the final fate of the collapse of a rotating compact star \cite{PhysRevC.101.015805,spin.parameter}. The Kerr parameter is described by the relation
\begin{equation}
\kappa=\frac{cJ}{GM^2}
\end{equation}

where $J$ is the angular momentum and $M$ is the gravitational mass of the rotating NS. The Kerr parameter for black holes is an important and fundamental quantity with a maximum value of 1, but it is important for other compact stars as well.\par 
\begin{figure}
	\includegraphics[width=0.48\textwidth]{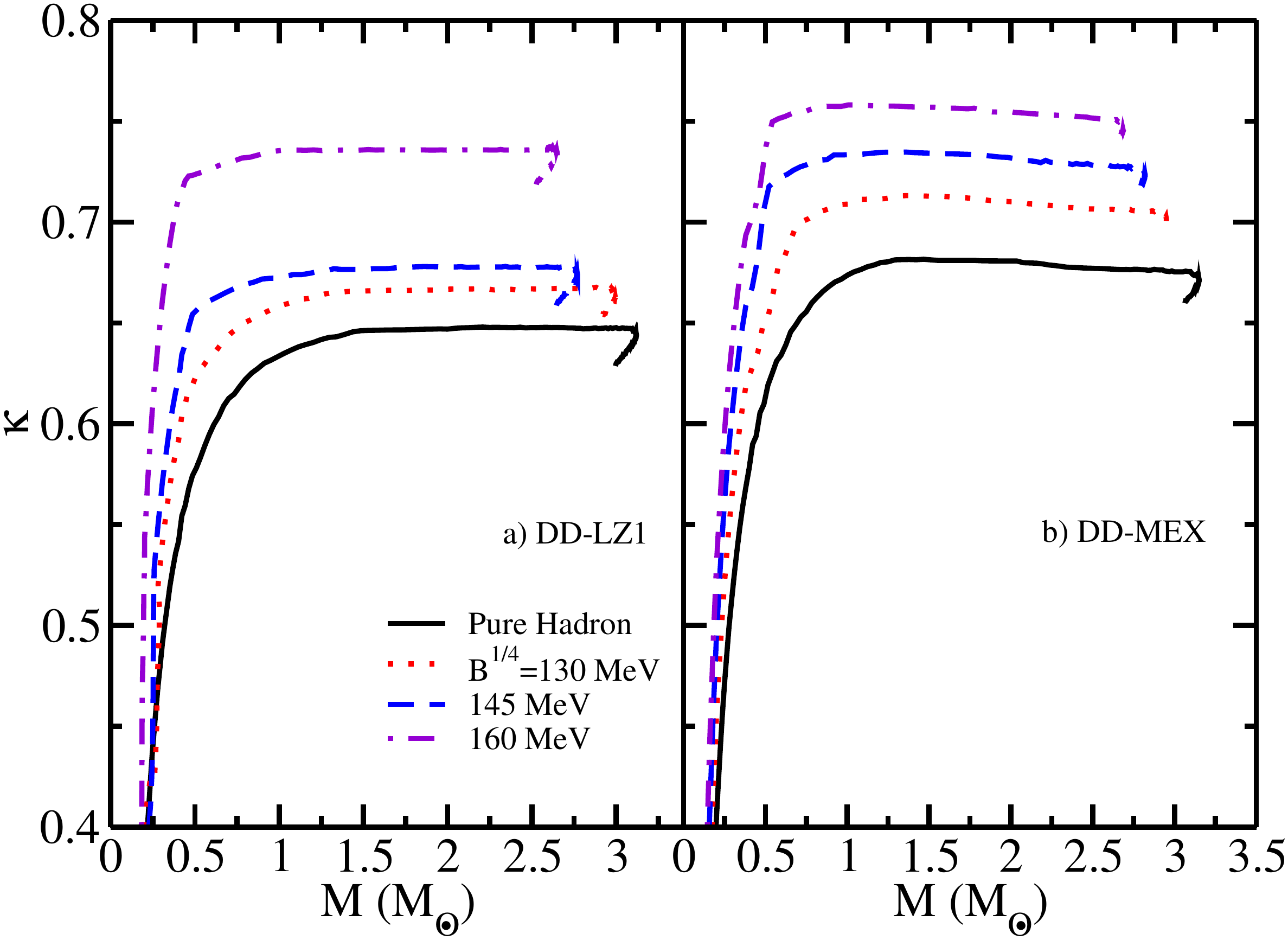}
	\caption{(color online) Kerr parameter $\kappa$ as a function of gravitational mass for (a) DD-LZ1 and (b) DD-MEX EoSs. The plot shows both pure hadronic stars (solid lines) and hybrid stars (dashed lines) at different bag constants.}
	\label{fig14}
\end{figure}

\begin{figure}
	\includegraphics[width=0.48\textwidth]{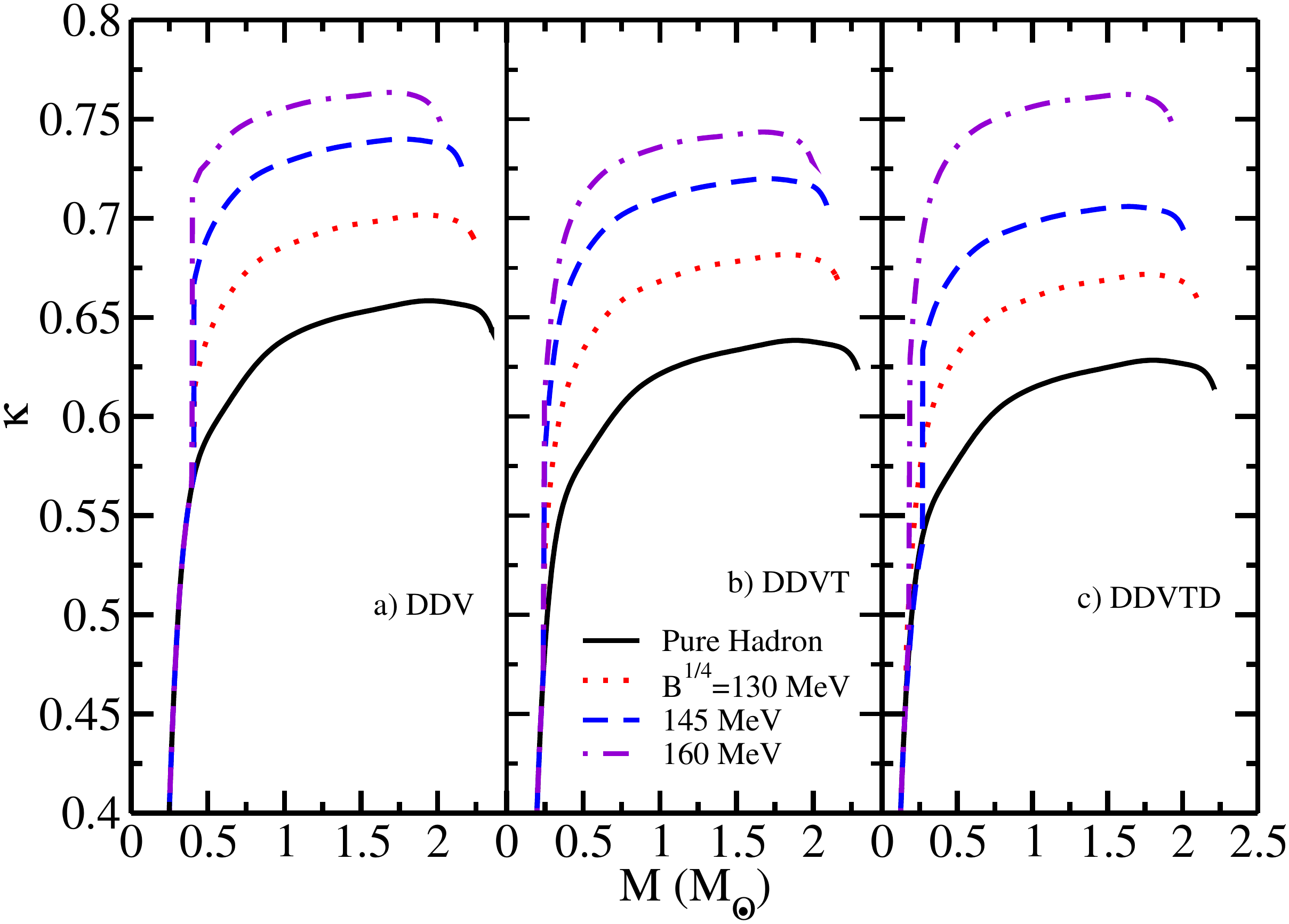}
	\caption{(color online) Same as Fig. \ref{fig14} but for (a) DDV, (b) DDVT, and (c) DDVTD EoSs.}
	\label{fig15}
\end{figure}

 To constrain the Kerr parameter for NSs, we studied the dependence of the Kerr parameter on the NS gravitational mass as displayed in Figs. \ref{fig14} and \ref{fig15} for the given parameter sets. From Fig. \ref{fig14}, the Kerr parameter for pure hadronic DD-LZ1 and DD-MEX parameter sets is found to be 0.64 and 0.67 respectively. This parameter increases for the hybrid stars with a maximum value of 0.73 at $B^{1/4}$=160 MeV for the DD-LZ1 set. For the DD-MEX set, the maximum value of the Kerr parameter is 0.75 at 160 MeV bag constant. For the DD-LZ1 parameter sets, the Kerr parameter remains almost unchanged once the star reaches a mass of around 1.4$M_{\odot}$ for pure hadronic matter and around 1.2$M_{\odot}$ for hybrid configurations. For DDV, DDVT, and DDVTD parameter sets as shown in Fig. \ref{fig15}, the Kerr parameter value for pure hadronic stars at the maximum mass is 0.64, 0.62, and 0.61 respectively. For hybrid star configurations, the value increases to 0.75 for all parameter sets at bag constant $B^{1/4}$=160 MeV. The Kerr parameter for HS configurations remains almost identical to the hadron star up to almost 0.4$M_{\odot}$. Therefore, by definition, the gravitational collapse of a RNS cannot form a Kerr black hole.\par 

\begin{figure}
	\includegraphics[width=0.48\textwidth]{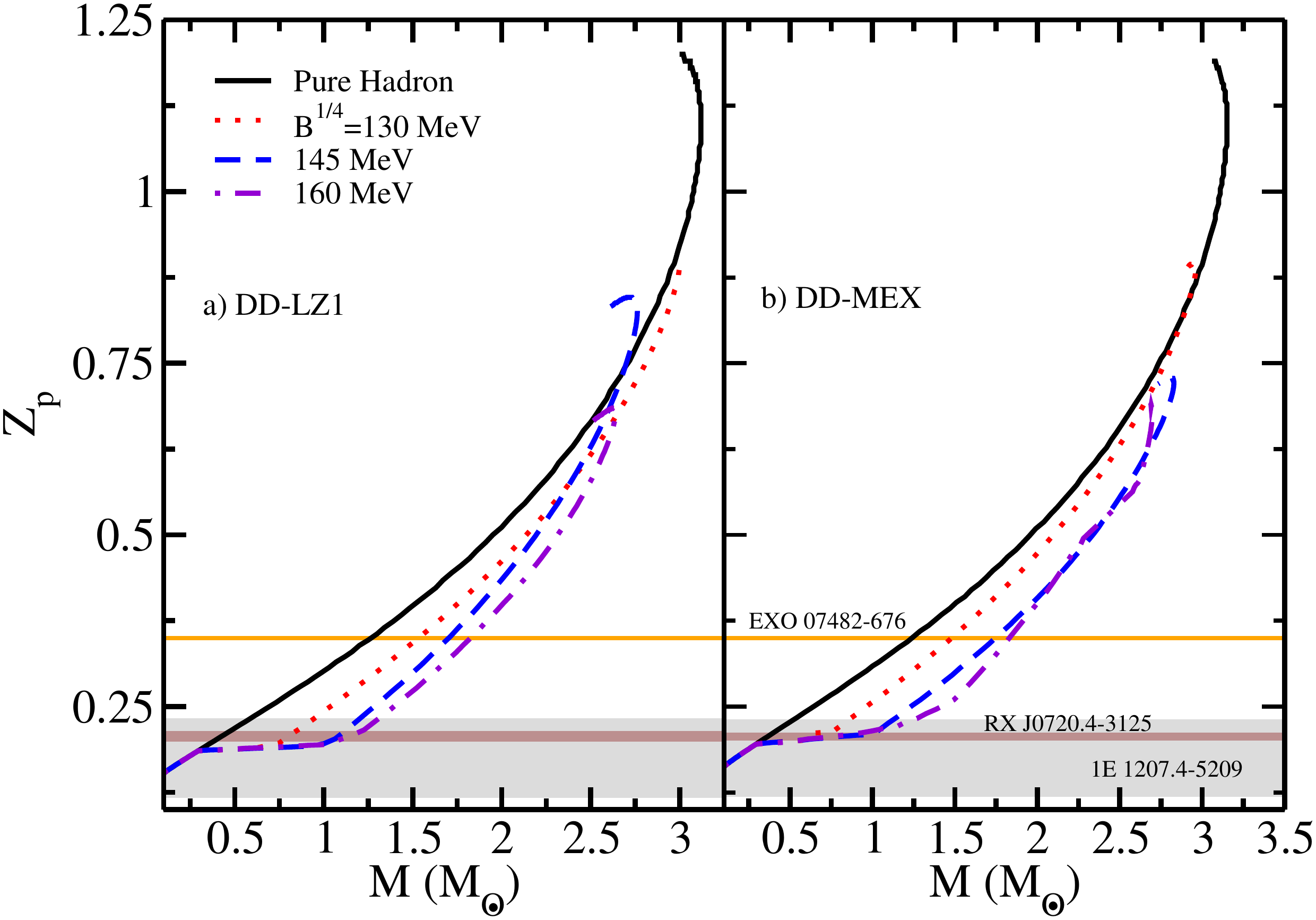}
	\caption{(color online) Polar redshift vs gravitational mass for pure hadron stars and hybrid star configurations for (a) DD-LZ1 and (b) DD-MEX EoSs. The observational limits imposed on the polar redshift from 1E 1207.4-5209 (grey band)\cite{Sanwal_2002}, RX J0720.4-3125 (brown band)\cite{refId0}, and EXO 07482-676 (orange horizontal line)\cite{cottam} are shown. }
	\label{fig16}
\end{figure}

Another important quantity related to the NSs is the redshift which has been investigated deeply \cite{Xia_2009,1994ApJ...424..823C,1986ApJ...304..115F}. The measurement of redshift can impose constraints on the compactness, and in turn, on the NS EoS. For a RNS, if the detector is placed in the direction of the polar plane of the star, the polar redshift, also called gravitational redshift, can be measured. For a detector directed tangentially, the forward and backward redshifts can be measured. The expression for the polar redshift is given as
\begin{equation}
Z_P(\Omega)=e^{-2\nu(\Omega)}-1
\end{equation}
where $\nu$ is the metric function. The variation of the polar redshift with the gravitational mass is depicted in Fig. \ref{fig16} for DD-LZ1 and DD-MEX EoSs. For pure hadronic stars, the polar redshift is found to be around 1.1 for both EoSs. With the QM present in the NSs, the polar redshift for DD-LZ1 decreases to a value 0.89, 0.84, and 0.64 for bag constants $B^{1/4}$=130, 145, and 160 MeV, respectively. Similarly for the DD-MEX set, the redshift decreases up to 0.68 for the 160 MeV bag constant. The observational limits imposed on the redshift from 1E 1207.4-5209 ($Z_P$=0.12-0.23) \cite{Sanwal_2002} , RX J0720.4-3125 ($Z_P$=0.205$_{-0.003}^{+0.006}$) \cite{refId0}, and EXO 07482-676 ($Z_P$=0.35) \cite{cottam} are also shown. The redshift prediction of $Z_P$=0.35 for EXO 07482-676 was based of the narrow absorption lines in the x-ray bursts. However, it was later seen that the spectral lines from EXO 07482-676 may be narrower than predicted \cite{Baub_ck_2013}. Therefore the estimates of the redshift from EXO 07482-676 are uncertain. \par 

\begin{figure}
	\includegraphics[width=0.48\textwidth]{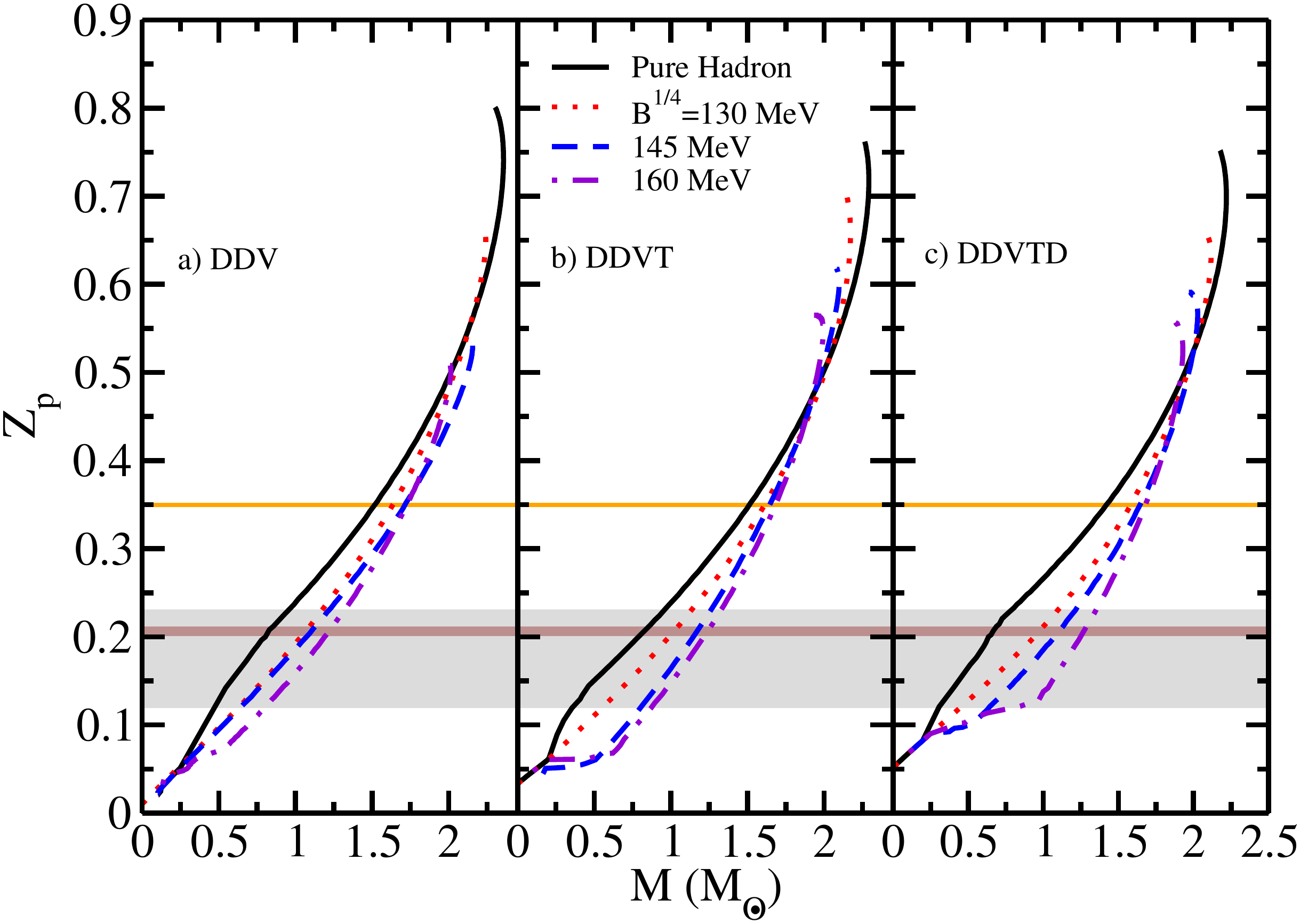}
	\caption{(color online) Same as Fig. \ref{fig16} but for (a) DDV, (b) DDVT, and (c) DDVTD EoSs. }
	\label{fig17}
\end{figure}
For the softer EoS group, the polar redshift variation with the gravitational mass is shown in Fig. \ref{fig17} for both pure HM and HS configurations. For the DDV set, the polar redshift is found to be 0.75 for the maximum mass of a pure hadronic star and decreases to 0.50 for the  hybrid star at a bag constant of 160 MeV. For DDVT and DDVTD EoSs, the redshift decreases from 0.72 and 0.70 for pure HM to 0.55 and 0.53 respectively for a hybrid star at $B^{1/4}$=160 MeV. The NS redshift provided by measuring the $\gamma$-ray burst annihilation lines has been interpreted as gravitationally redshifted 511 keV electron-positron pair annihilation from the NS surface \cite{Liang}. If this interpretation is correct, then it will support a NS with redshift in the range $0.2\le Z_P \le 0.5$ and thus will rule out almost every EoS studied in this work.\par

For the static NS, the phase transition to the QM for DD-LZ1 and DD-MEX parameter sets is studied in Ref. \cite{ddmex}. For DDV, DDVT, and DDVTD sets, the maximum mass obtained is around 2$M_{\odot}$ and hence the phase transition to QM will decrease the maximum mass to a value not satisfying any recent constraints on the mass and other NS properties. However, to study the properties of a pure hadronic EoS, the mass-radius profile for static stars is explained in Fig. \ref{fig3}. In addition to this, we study the tidal deformability of the given parameter sets. The equations describing the tidal deformation and its dependence on the star matter properties are described above.\par 

\begin{figure}
	\includegraphics[width=0.48\textwidth]{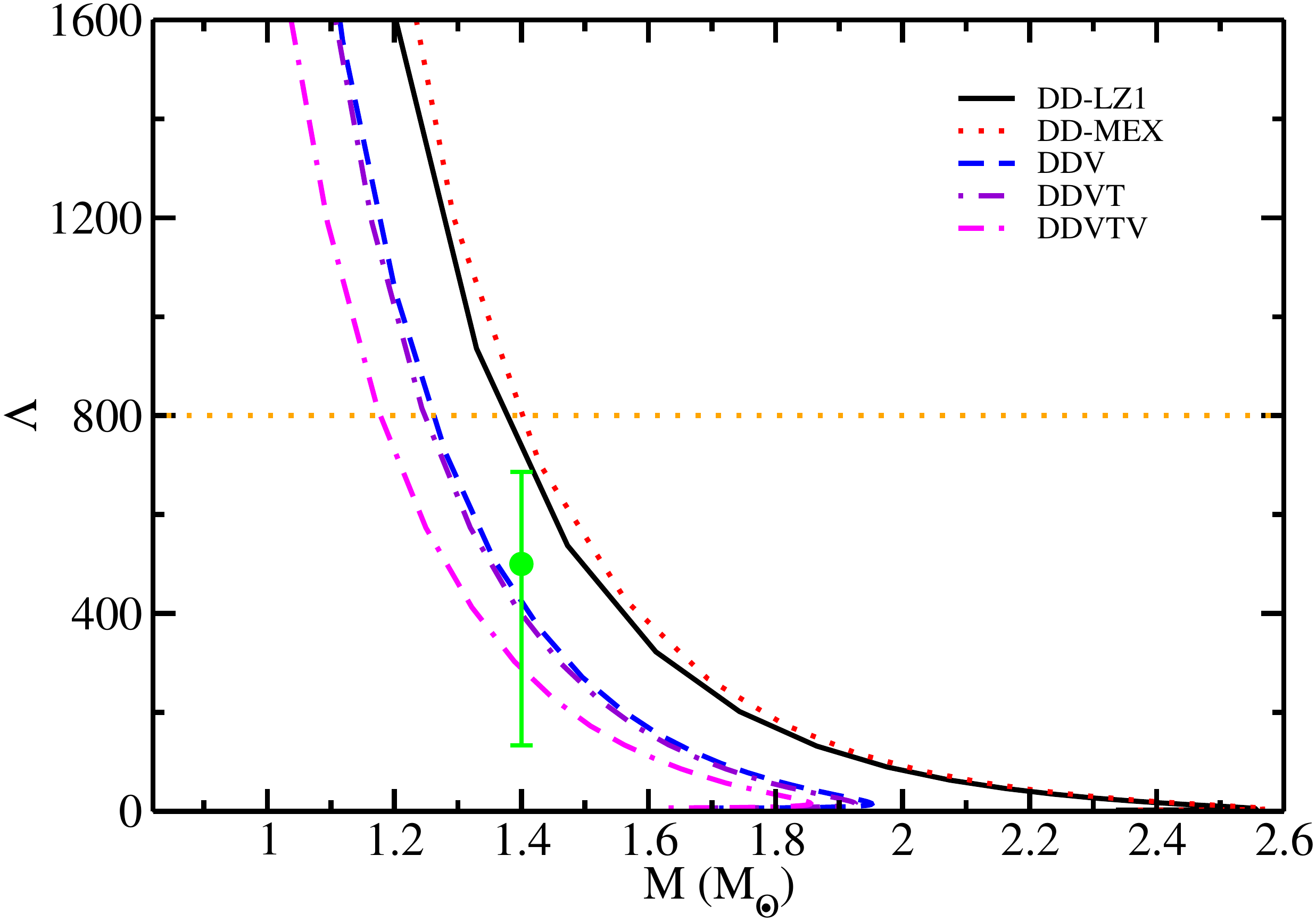}
	\caption{(color online) Dimensionless tidal deformability as a function of NS mass for DD-LZ1, DD-MEX, DDV, DDVT, and DDVTD EoSs.  The recent constraint using Bayesian analysis \cite{Li2021} and an upper limit on dimensionless tidal deformability set by measurement from GW170817 \cite{PhysRevLett.119.161101} are shown.}
	\label{fig18}
\end{figure}
 The dimensionless tidal deformability $\Lambda$ as a function of NS mass for the hadronic EoSs is shown in Fig. \ref{fig18}. The constraint on dimensionless tidal deformability obtained using Bayesian analysis is shown, $\Lambda_{1.4}=500_{-367}^{+186}$ \cite{Li2021}. The orange dotted curve represents an uper limit set on the tidal deformability from the measurement of GW170817 \cite{PhysRevLett.119.161101}. The tidal deformability depends upon the NS mass and the radius. The value decreases with increasing mass and becomes very small at the NS maximum mass. The dimensionless tidal deformability for DD-LZ1, DD-MEX, DDV, DDVT, and DDVTD EoSs at the canonical mass is found to be 727.17, 791.60, 391.23, 337.51, and 281.05 respectively. All these values lie well below the upper limit set by the GW170817 event. Using Bayesian analysis, Lim et al. \cite{PhysRevLett.121.062701} showed a 90\% and 65\% credibility interval on the dimensionless tidal deformability at 1.4$M_{\odot}$, 136 $< \Lambda <$ 519 and 256 $< \Lambda <$ 442, respectively. The DD-MEX set produces a little higher value of the tidal deformability. The value of $\Lambda_{1.4}$ for softer group EoS (DDV, DDVT, and DDVTD) is significantly lower than the stiffer group (DD-LZ1 and DD-MEX) because of the small maximum mass and the corresponding radius. However, the stiffer group EoSs cannot be neglected in comparison to the softer group.  The tidal deformability of softer group satisfies all the constraints imposed. The precise measurement of the tidal deformability for the BNS mergers with a maximum mass around 2$M_{\odot}$ by future gravitational wave detectors will lower the uncertainties in these values thereby constraining the EoSs.

\section{Summary and Conclusion}
\label{summary}
The properties of static and rotating NSs are studied with a hadron-quark phase transition. The hadronic matter is studied by employing the DD-RMF model. Recent parameter sets like DDV, DDVT, and DDVTD along with the DD-LZ1 and DD-MEX are used to study the hadronic EoS. The QM is studied using a modified version of the bag model, the vBag model. The vBag model includes the necessary repulsive vector interactions and D$\chi$SB. The vBag model coupling parameter $K_{\nu}$ controlling the stiffness of the EoS curve is held constant at 6 GeV$^{-2}$. The effective bag constant $B_{eff}^{1/4}$ is varied by taking the values 130, 145, and 160 MeV. The Gibbs technique is used to construct the mixed-phase between hadrons and quarks which accounts for the global charge neutrality of the system. The properties such as mass, radius, and the tidal deformability of static NS are studied. For RNS, the variation in the NS properties like maximum mass, radius, the  moment of inertia, rotational frequency, Kerr parameter, etc are studied in the presence of QM.\par

For static NSs, the maximum mass for DD-LZ1 and DD-MEX is found to be 2.55$M_{\odot}$ and 2.57$M_{\odot}$ respectively, forming a stiffer EoS group. For DDV, DDVT, and DDVTD EoSs, the maximum is found to be around 1.9$M_{\odot}$, thus lying in the softer EoS group. The phase-transition properties for SNS are not studied for the softer EoS group because it would result in a very low maximum mass not satisfying any mass constraints.\par 
For RNSs, the maximum mass is found to be 3.11$M_{\odot}$ for the DD-LZ1 set which in presence of QM reduces to 2.64$M_{\odot}$ satisfying the recent GW190814 possible maximum mass constraint. The DD-MEX set also predicts a maximum mass of 3.15$M_{\odot}$ decreasing to 2.69$M_{\odot}$ for $B_{eff}^{1/4}$=160 MeV bag constant. For the softer EoS group, the RNS mass lies in the range 2.2$M_{\odot}$-2.3$M_{\odot}$ which then reduces with increasing bag constant to satisfy the 2$M_{\odot}$ limit. The radius also decreases with increasing bag constant. The moment of inertia for the stiffer group lies in the range (2.2-2.3)$\times$10$^{45}$g.cm$^2$ for pure hadron EoSs. The phase transition to QM reduces the value to 1.7$\times$10$^{45}$g.cm$^2$ satisfying the recent constraints. For the softer group of EoSs, the moment of inertia is lowered in the  presence of QM to satisfy the constraints from GW170817 with universal relations. \par 
 The variation in the rotational frequency of a NS with the gravitational mass is also studied. The pure hadronic EoSs produce NSs with high rotational frequencies. For DD-LZ1 and DD-MEX, the rotational frequency at the maximum mass is 1525 and 1503 Hz, respectively. For DDV, DDVT, and DDVTD EoSs, the frequency obtained is in the range 1400-1500 Hz. The quarks produce the hybrid star configurations with larger rotational frequencies  as the quark star are more compact than hadron stars. Initially, for HS configuration at $B^{1/4}$=130 MeV, the rotating with frequency smaller than a pure hadronic star is formed. As the bag constant increases, the maximum mass decreases, and the corresponding frequency increases. All the pure hadronic and hybrid star configurations produce NSs with a frequency higher than the highest measured frequency of $\nu$=1122 Hz. \par 
  The ratio of rotational kinetic energy to the gravitational potential energy $\beta=T/W$ is studied to determine the dynamical stability of the RNS. For $\beta > \beta_d$ (=0.14-0.27), the star is considered to be dynamically unstable and hence emits gravitational radiation. The $T/W$ ratio for rotating pure hadronic stars is found to be 0.147 and 0.145 for DD-LZ1 and DD-MEX EoSs. The QM phase transition tends to increase the $T/W$ ratio with decreasing mass. For a bag constant of 160 MeV, the ratio is found to be 0.153 and 0.151 for DD-LZ1 and DD-MEX EoSs, respectively. For a softer EoS group, this ratio lies below the critical limit for pure hadronic stars, but increases to a value well within the critical limit. \par 
   The Kerr parameter is calculated for the RNSs whose measurement allows us to constrain the compactness of a star and hence the EoS. The precise value of the Kerr parameter for a NS is not known yet, but a maximum value of 0.75 is seen in most of the theoretical works. For the given parametrization sets, the Kerr parameter value lies around 0.65 for the stiffer group and 0.6 for the softer group. Following the inverse relationship with the gravitational mass, the Kerr parameter increases in the presence of quarks. For both stiffer and softer EoS groups, the value attains a maximum value of 0.75, which remains almost unchanged as the mass increases beyond 1$M{\odot}$. The dependence of polar redshift on the NS mass is also calculated. It is seen that the polar redshift decreases in presence of quarks. The redshift parameter measured for all hybrid star configurations lies well above the predicted value from EXO 07482-676, $Z_P$=0.35. \par 
   
   For static, spherically symmetric stars, we have also calculated the dimensionless tidal deformability. It is seen that all the parameter sets predict a value of tidal deformability satisfying the constraints from various measurements.\par 
   
   Thus, it is clear that the presence of quarks inside the NS affects both static and rotating NS properties. Eliminating the uncertainties present in the values of these quantities will allow us to rule out very stiff and very soft EoSs. The measurement of tidal deformability for RNS will help us to constraint its properties and hence determine a proper EoS in the near future. Additional gravitational-wave observations of binary NS mergers and more accurate measurements of other NS properties like mass, radius, tidal deformability will allow the universal relation-based bounds on canonical deformability to be further refined. The theoretical study of a uniformly RNS, along with the accurate measurements, may offer new information about the equation of state in high density regime. Besides, NSs through their evolution may provide us with a criterion to determine the final fate of a rotating compact star.



\begin{thebibliography}{142}%
	\makeatletter
	\providecommand \@ifxundefined [1]{%
		\@ifx{#1\undefined}
	}%
	\providecommand \@ifnum [1]{%
		\ifnum #1\expandafter \@firstoftwo
		\else \expandafter \@secondoftwo
		\fi
	}%
	\providecommand \@ifx [1]{%
		\ifx #1\expandafter \@firstoftwo
		\else \expandafter \@secondoftwo
		\fi
	}%
	\providecommand \natexlab [1]{#1}%
	\providecommand \enquote  [1]{``#1''}%
	\providecommand \bibnamefont  [1]{#1}%
	\providecommand \bibfnamefont [1]{#1}%
	\providecommand \citenamefont [1]{#1}%
	\providecommand \href@noop [0]{\@secondoftwo}%
	\providecommand \href [0]{\begingroup \@sanitize@url \@href}%
	\providecommand \@href[1]{\@@startlink{#1}\@@href}%
	\providecommand \@@href[1]{\endgroup#1\@@endlink}%
	\providecommand \@sanitize@url [0]{\catcode `\\12\catcode `\$12\catcode
		`\&12\catcode `\#12\catcode `\^12\catcode `\_12\catcode `\%12\relax}%
	\providecommand \@@startlink[1]{}%
	\providecommand \@@endlink[0]{}%
	\providecommand \url  [0]{\begingroup\@sanitize@url \@url }%
	\providecommand \@url [1]{\endgroup\@href {#1}{\urlprefix }}%
	\providecommand \urlprefix  [0]{URL }%
	\providecommand \Eprint [0]{\href }%
	\providecommand \doibase [0]{http://dx.doi.org/}%
	\providecommand \selectlanguage [0]{\@gobble}%
	\providecommand \bibinfo  [0]{\@secondoftwo}%
	\providecommand \bibfield  [0]{\@secondoftwo}%
	\providecommand \translation [1]{[#1]}%
	\providecommand \BibitemOpen [0]{}%
	\providecommand \bibitemStop [0]{}%
	\providecommand \bibitemNoStop [0]{.\EOS\space}%
	\providecommand \EOS [0]{\spacefactor3000\relax}%
	\providecommand \BibitemShut  [1]{\csname bibitem#1\endcsname}%
	\let\auto@bib@innerbib\@empty
	\bibitem [{\citenamefont {Abbott}\ \emph {et~al.}(2017)\citenamefont {Abbott},
		\citenamefont {Abbott} \emph {et~al.}}]{PhysRevLett.119.161101}%
	\BibitemOpen
	\bibfield  {author} {\bibinfo {author} {\bibfnamefont {B.~P.}\ \bibnamefont
			{Abbott}}, \bibinfo {author} {\bibfnamefont {R.}~\bibnamefont {Abbott}},
		\emph {et~al.} (\bibinfo {collaboration} {LIGO Scientific Collaboration and
			Virgo Collaboration}),\ }\href {\doibase 10.1103/PhysRevLett.119.161101}
	{\bibfield  {journal} {\bibinfo  {journal} {Phys. Rev. Lett.}\ }\textbf
		{\bibinfo {volume} {119}},\ \bibinfo {pages} {161101} (\bibinfo {year}
		{2017})}\BibitemShut {NoStop}%
	\bibitem [{\citenamefont {Abbott}\ \emph {et~al.}(2018)\citenamefont {Abbott},
		\citenamefont {Abbott} \emph {et~al.}}]{PhysRevLett.121.161101}%
	\BibitemOpen
	\bibfield  {author} {\bibinfo {author} {\bibfnamefont {B.~P.}\ \bibnamefont
			{Abbott}}, \bibinfo {author} {\bibfnamefont {R.}~\bibnamefont {Abbott}},
		\emph {et~al.} (\bibinfo {collaboration} {The LIGO Scientific Collaboration
			and the Virgo Collaboration}),\ }\href {\doibase
		10.1103/PhysRevLett.121.161101} {\bibfield  {journal} {\bibinfo  {journal}
			{Phys. Rev. Lett.}\ }\textbf {\bibinfo {volume} {121}},\ \bibinfo {pages}
		{161101} (\bibinfo {year} {2018})}\BibitemShut {NoStop}%
	\bibitem [{\citenamefont {Abbott}\ \emph {et~al.}(2019)\citenamefont {Abbott},
		\citenamefont {Abbott} \emph {et~al.}}]{PhysRevX.9.011001}%
	\BibitemOpen
	\bibfield  {author} {\bibinfo {author} {\bibfnamefont {B.~P.}\ \bibnamefont
			{Abbott}}, \bibinfo {author} {\bibfnamefont {R.}~\bibnamefont {Abbott}},
		\emph {et~al.} (\bibinfo {collaboration} {LIGO Scientific Collaboration and
			Virgo Collaboration}),\ }\href {\doibase 10.1103/PhysRevX.9.011001}
	{\bibfield  {journal} {\bibinfo  {journal} {Phys. Rev. X}\ }\textbf {\bibinfo
			{volume} {9}},\ \bibinfo {pages} {011001} (\bibinfo {year}
		{2019})}\BibitemShut {NoStop}%
	\bibitem [{\citenamefont {Abbott}\ \emph
		{et~al.}(2020{\natexlab{a}})\citenamefont {Abbott}, \citenamefont {Abbott}
		\emph {et~al.}}]{Abbott_2020a}%
	\BibitemOpen
	\bibfield  {author} {\bibinfo {author} {\bibfnamefont {R.}~\bibnamefont
			{Abbott}}, \bibinfo {author} {\bibfnamefont {T.~D.}\ \bibnamefont {Abbott}},
		\emph {et~al.},\ }\href {\doibase 10.3847/2041-8213/ab960f} {\bibfield
		{journal} {\bibinfo  {journal} {The Astrophys. Jour.}\ }\textbf {\bibinfo
			{volume} {896}},\ \bibinfo {pages} {L44} (\bibinfo {year}
		{2020}{\natexlab{a}})}\BibitemShut {NoStop}%
	\bibitem [{\citenamefont {Dexheimer}\ \emph {et~al.}(2020)\citenamefont
		{Dexheimer}, \citenamefont {Gomes}, \citenamefont {Klähn}, \citenamefont
		{Han},\ and\ \citenamefont {Salinas}}]{dexheimer2020gw190814}%
	\BibitemOpen
	\bibfield  {author} {\bibinfo {author} {\bibfnamefont {V.}~\bibnamefont
			{Dexheimer}}, \bibinfo {author} {\bibfnamefont {R.~O.}\ \bibnamefont
			{Gomes}}, \bibinfo {author} {\bibfnamefont {T.}~\bibnamefont {Klähn}},
		\bibinfo {author} {\bibfnamefont {S.}~\bibnamefont {Han}}, \ and\ \bibinfo
		{author} {\bibfnamefont {M.}~\bibnamefont {Salinas}},\ }\href@noop {} {}
	(\bibinfo {year} {2020}),\ \Eprint {http://arxiv.org/abs/2007.08493}
	{arXiv:2007.08493 [astro-ph.HE]} \BibitemShut {NoStop}%
	\bibitem [{\citenamefont {Tan}\ \emph {et~al.}(2020)\citenamefont {Tan},
		\citenamefont {Noronha-Hostler},\ and\ \citenamefont
		{Yunes}}]{tan2020neutron}%
	\BibitemOpen
	\bibfield  {author} {\bibinfo {author} {\bibfnamefont {H.}~\bibnamefont
			{Tan}}, \bibinfo {author} {\bibfnamefont {J.}~\bibnamefont
			{Noronha-Hostler}}, \ and\ \bibinfo {author} {\bibfnamefont {N.}~\bibnamefont
			{Yunes}},\ }\href@noop {} {} (\bibinfo {year} {2020}),\ \Eprint
	{http://arxiv.org/abs/2006.16296} {arXiv:2006.16296 [astro-ph.HE]}
	\BibitemShut {NoStop}%
	\bibitem [{\citenamefont {Fishbach}\ \emph {et~al.}(2020)\citenamefont
		{Fishbach}, \citenamefont {Essick},\ and\ \citenamefont
		{Holz}}]{Fishbach_2020}%
	\BibitemOpen
	\bibfield  {author} {\bibinfo {author} {\bibfnamefont {M.}~\bibnamefont
			{Fishbach}}, \bibinfo {author} {\bibfnamefont {R.}~\bibnamefont {Essick}}, \
		and\ \bibinfo {author} {\bibfnamefont {D.~E.}\ \bibnamefont {Holz}},\ }\href
	{\doibase 10.3847/2041-8213/aba7b6} {\bibfield  {journal} {\bibinfo
			{journal} {Astrophys. J.}\ }\textbf {\bibinfo {volume} {899}},\ \bibinfo
		{pages} {L8} (\bibinfo {year} {2020})}\BibitemShut {NoStop}%
	\bibitem [{\citenamefont {Rather}\ \emph
		{et~al.}(2020{\natexlab{a}})\citenamefont {Rather}, \citenamefont {Usmani},\
		and\ \citenamefont {Patra}}]{rather2020hadronquark}%
	\BibitemOpen
	\bibfield  {author} {\bibinfo {author} {\bibfnamefont {I.~A.}\ \bibnamefont
			{Rather}}, \bibinfo {author} {\bibfnamefont {A.~A.}\ \bibnamefont {Usmani}},
		\ and\ \bibinfo {author} {\bibfnamefont {S.~K.}\ \bibnamefont {Patra}},\
	}\href@noop {} {} (\bibinfo {year} {2020}{\natexlab{a}}),\ \Eprint
	{http://arxiv.org/abs/2011.14077} {arXiv:2011.14077 [nucl-th]} \BibitemShut
	{NoStop}%
	\bibitem [{\citenamefont {Godzieba}\ \emph {et~al.}(2020)\citenamefont
		{Godzieba}, \citenamefont {Radice},\ and\ \citenamefont
		{Bernuzzi}}]{godzieba2020maximum}%
	\BibitemOpen
	\bibfield  {author} {\bibinfo {author} {\bibfnamefont {D.~A.}\ \bibnamefont
			{Godzieba}}, \bibinfo {author} {\bibfnamefont {D.}~\bibnamefont {Radice}}, \
		and\ \bibinfo {author} {\bibfnamefont {S.}~\bibnamefont {Bernuzzi}},\
	}\href@noop {} {} (\bibinfo {year} {2020}),\ \Eprint
	{http://arxiv.org/abs/2007.10999} {arXiv:2007.10999 [astro-ph.HE]}
	\BibitemShut {NoStop}%
	\bibitem [{\citenamefont {Most}\ \emph {et~al.}(2020)\citenamefont {Most},
		\citenamefont {Papenfort}, \citenamefont {Weih},\ and\ \citenamefont
		{Rezzolla}}]{10.1093/mnrasl/slaa168}%
	\BibitemOpen
	\bibfield  {author} {\bibinfo {author} {\bibfnamefont {E.~R.}\ \bibnamefont
			{Most}}, \bibinfo {author} {\bibfnamefont {L.~J.}\ \bibnamefont {Papenfort}},
		\bibinfo {author} {\bibfnamefont {L.~R.}\ \bibnamefont {Weih}}, \ and\
		\bibinfo {author} {\bibfnamefont {L.}~\bibnamefont {Rezzolla}},\ }\href
	{\doibase 10.1093/mnrasl/slaa168} {\bibfield  {journal} {\bibinfo  {journal}
			{MNRAS}\ }\textbf {\bibinfo {volume} {499}},\ \bibinfo {pages} {L82}
		(\bibinfo {year} {2020})}\BibitemShut {NoStop}%
	\bibitem [{\citenamefont {Zhang}\ and\ \citenamefont {Li}(2020)}]{Zhang_2020}%
	\BibitemOpen
	\bibfield  {author} {\bibinfo {author} {\bibfnamefont {N.-B.}\ \bibnamefont
			{Zhang}}\ and\ \bibinfo {author} {\bibfnamefont {B.-A.}\ \bibnamefont {Li}},\
	}\href {\doibase 10.3847/1538-4357/abb470} {\bibfield  {journal} {\bibinfo
			{journal} {The Astrophys. Jour.}\ }\textbf {\bibinfo {volume} {902}},\
		\bibinfo {pages} {38} (\bibinfo {year} {2020})}\BibitemShut {NoStop}%
	\bibitem [{\citenamefont {Tsokaros}\ \emph {et~al.}(2020)\citenamefont
		{Tsokaros}, \citenamefont {Ruiz},\ and\ \citenamefont
		{Shapiro}}]{tsokaros2020gw190814}%
	\BibitemOpen
	\bibfield  {author} {\bibinfo {author} {\bibfnamefont {A.}~\bibnamefont
			{Tsokaros}}, \bibinfo {author} {\bibfnamefont {M.}~\bibnamefont {Ruiz}}, \
		and\ \bibinfo {author} {\bibfnamefont {S.~L.}\ \bibnamefont {Shapiro}},\
	}\href@noop {} {} (\bibinfo {year} {2020}),\ \Eprint
	{http://arxiv.org/abs/2007.05526} {arXiv:2007.05526 [astro-ph.HE]}
	\BibitemShut {NoStop}%
	\bibitem [{\citenamefont {Fattoyev}\ \emph {et~al.}(2020)\citenamefont
		{Fattoyev}, \citenamefont {Horowitz}, \citenamefont {Piekarewicz},\ and\
		\citenamefont {Reed}}]{fattoyev2020gw190814}%
	\BibitemOpen
	\bibfield  {author} {\bibinfo {author} {\bibfnamefont {F.~J.}\ \bibnamefont
			{Fattoyev}}, \bibinfo {author} {\bibfnamefont {C.~J.}\ \bibnamefont
			{Horowitz}}, \bibinfo {author} {\bibfnamefont {J.}~\bibnamefont
			{Piekarewicz}}, \ and\ \bibinfo {author} {\bibfnamefont {B.}~\bibnamefont
			{Reed}},\ }\href@noop {} {} (\bibinfo {year} {2020}),\ \Eprint
	{http://arxiv.org/abs/2007.03799} {arXiv:2007.03799 [nucl-th]} \BibitemShut
	{NoStop}%
	\bibitem [{\citenamefont {Lim}\ \emph {et~al.}(2020)\citenamefont {Lim},
		\citenamefont {Bhattacharya}, \citenamefont {Holt},\ and\ \citenamefont
		{Pati}}]{lim2020revisiting}%
	\BibitemOpen
	\bibfield  {author} {\bibinfo {author} {\bibfnamefont {Y.}~\bibnamefont
			{Lim}}, \bibinfo {author} {\bibfnamefont {A.}~\bibnamefont {Bhattacharya}},
		\bibinfo {author} {\bibfnamefont {J.~W.}\ \bibnamefont {Holt}}, \ and\
		\bibinfo {author} {\bibfnamefont {D.}~\bibnamefont {Pati}},\ }\href@noop {}
	{} (\bibinfo {year} {2020}),\ \Eprint {http://arxiv.org/abs/2007.06526}
	{arXiv:2007.06526 [nucl-th]} \BibitemShut {NoStop}%
	\bibitem [{\citenamefont {Tews}\ \emph {et~al.}(2020)\citenamefont {Tews},
		\citenamefont {Pang}, \citenamefont {Dietrich}, \citenamefont {Coughlin},
		\citenamefont {Antier}, \citenamefont {Bulla}, \citenamefont {Heinzel},\ and\
		\citenamefont {Issa}}]{tews2020nature}%
	\BibitemOpen
	\bibfield  {author} {\bibinfo {author} {\bibfnamefont {I.}~\bibnamefont
			{Tews}}, \bibinfo {author} {\bibfnamefont {P.~T.~H.}\ \bibnamefont {Pang}},
		\bibinfo {author} {\bibfnamefont {T.}~\bibnamefont {Dietrich}}, \bibinfo
		{author} {\bibfnamefont {M.~W.}\ \bibnamefont {Coughlin}}, \bibinfo {author}
		{\bibfnamefont {S.}~\bibnamefont {Antier}}, \bibinfo {author} {\bibfnamefont
			{M.}~\bibnamefont {Bulla}}, \bibinfo {author} {\bibfnamefont
			{J.}~\bibnamefont {Heinzel}}, \ and\ \bibinfo {author} {\bibfnamefont
			{L.}~\bibnamefont {Issa}},\ }\href@noop {} {} (\bibinfo {year} {2020}),\
	\Eprint {http://arxiv.org/abs/2007.06057} {arXiv:2007.06057 [astro-ph.HE]}
	\BibitemShut {NoStop}%
	\bibitem [{\citenamefont {Shibata}\ and\ \citenamefont
		{Taniguchi}(2006)}]{PhysRevD.73.064027}%
	\BibitemOpen
	\bibfield  {author} {\bibinfo {author} {\bibfnamefont {M.}~\bibnamefont
			{Shibata}}\ and\ \bibinfo {author} {\bibfnamefont {K.}~\bibnamefont
			{Taniguchi}},\ }\href {\doibase 10.1103/PhysRevD.73.064027} {\bibfield
		{journal} {\bibinfo  {journal} {Phys. Rev. D}\ }\textbf {\bibinfo {volume}
			{73}},\ \bibinfo {pages} {064027} (\bibinfo {year} {2006})}\BibitemShut
	{NoStop}%
	\bibitem [{\citenamefont {Sekiguchi}\ \emph {et~al.}(2011)\citenamefont
		{Sekiguchi}, \citenamefont {Kiuchi}, \citenamefont {Kyutoku},\ and\
		\citenamefont {Shibata}}]{PhysRevLett.107.051102}%
	\BibitemOpen
	\bibfield  {author} {\bibinfo {author} {\bibfnamefont {Y.}~\bibnamefont
			{Sekiguchi}}, \bibinfo {author} {\bibfnamefont {K.}~\bibnamefont {Kiuchi}},
		\bibinfo {author} {\bibfnamefont {K.}~\bibnamefont {Kyutoku}}, \ and\
		\bibinfo {author} {\bibfnamefont {M.}~\bibnamefont {Shibata}},\ }\href
	{\doibase 10.1103/PhysRevLett.107.051102} {\bibfield  {journal} {\bibinfo
			{journal} {Phys. Rev. Lett.}\ }\textbf {\bibinfo {volume} {107}},\ \bibinfo
		{pages} {051102} (\bibinfo {year} {2011})}\BibitemShut {NoStop}%
	\bibitem [{\citenamefont {Hotokezaka}\ \emph {et~al.}(2013)\citenamefont
		{Hotokezaka}, \citenamefont {Kiuchi}, \citenamefont {Kyutoku}, \citenamefont
		{Muranushi}, \citenamefont {Sekiguchi}, \citenamefont {Shibata},\ and\
		\citenamefont {Taniguchi}}]{PhysRevD.88.044026}%
	\BibitemOpen
	\bibfield  {author} {\bibinfo {author} {\bibfnamefont {K.}~\bibnamefont
			{Hotokezaka}}, \bibinfo {author} {\bibfnamefont {K.}~\bibnamefont {Kiuchi}},
		\bibinfo {author} {\bibfnamefont {K.}~\bibnamefont {Kyutoku}}, \bibinfo
		{author} {\bibfnamefont {T.}~\bibnamefont {Muranushi}}, \bibinfo {author}
		{\bibfnamefont {Y.-i.}\ \bibnamefont {Sekiguchi}}, \bibinfo {author}
		{\bibfnamefont {M.}~\bibnamefont {Shibata}}, \ and\ \bibinfo {author}
		{\bibfnamefont {K.}~\bibnamefont {Taniguchi}},\ }\href {\doibase
		10.1103/PhysRevD.88.044026} {\bibfield  {journal} {\bibinfo  {journal} {Phys.
				Rev. D}\ }\textbf {\bibinfo {volume} {88}},\ \bibinfo {pages} {044026}
		(\bibinfo {year} {2013})}\BibitemShut {NoStop}%
	\bibitem [{\citenamefont {Bauswein}\ \emph {et~al.}(2013)\citenamefont
		{Bauswein}, \citenamefont {Baumgarte},\ and\ \citenamefont
		{Janka}}]{PhysRevLett.111.131101}%
	\BibitemOpen
	\bibfield  {author} {\bibinfo {author} {\bibfnamefont {A.}~\bibnamefont
			{Bauswein}}, \bibinfo {author} {\bibfnamefont {T.~W.}\ \bibnamefont
			{Baumgarte}}, \ and\ \bibinfo {author} {\bibfnamefont {H.-T.}\ \bibnamefont
			{Janka}},\ }\href {\doibase 10.1103/PhysRevLett.111.131101} {\bibfield
		{journal} {\bibinfo  {journal} {Phys. Rev. Lett.}\ }\textbf {\bibinfo
			{volume} {111}},\ \bibinfo {pages} {131101} (\bibinfo {year}
		{2013})}\BibitemShut {NoStop}%
	\bibitem [{\citenamefont {Palenzuela}\ \emph {et~al.}(2015)\citenamefont
		{Palenzuela}, \citenamefont {Liebling}, \citenamefont {Neilsen},
		\citenamefont {Lehner}, \citenamefont {Caballero}, \citenamefont {O'Connor},\
		and\ \citenamefont {Anderson}}]{PhysRevD.92.044045}%
	\BibitemOpen
	\bibfield  {author} {\bibinfo {author} {\bibfnamefont {C.}~\bibnamefont
			{Palenzuela}}, \bibinfo {author} {\bibfnamefont {S.~L.}\ \bibnamefont
			{Liebling}}, \bibinfo {author} {\bibfnamefont {D.}~\bibnamefont {Neilsen}},
		\bibinfo {author} {\bibfnamefont {L.}~\bibnamefont {Lehner}}, \bibinfo
		{author} {\bibfnamefont {O.~L.}\ \bibnamefont {Caballero}}, \bibinfo {author}
		{\bibfnamefont {E.}~\bibnamefont {O'Connor}}, \ and\ \bibinfo {author}
		{\bibfnamefont {M.}~\bibnamefont {Anderson}},\ }\href {\doibase
		10.1103/PhysRevD.92.044045} {\bibfield  {journal} {\bibinfo  {journal} {Phys.
				Rev. D}\ }\textbf {\bibinfo {volume} {92}},\ \bibinfo {pages} {044045}
		(\bibinfo {year} {2015})}\BibitemShut {NoStop}%
	\bibitem [{\citenamefont {Bernuzzi}\ \emph {et~al.}(2016)\citenamefont
		{Bernuzzi}, \citenamefont {Radice}, \citenamefont {Ott}, \citenamefont
		{Roberts}, \citenamefont {M\"osta},\ and\ \citenamefont
		{Galeazzi}}]{PhysRevD.94.024023}%
	\BibitemOpen
	\bibfield  {author} {\bibinfo {author} {\bibfnamefont {S.}~\bibnamefont
			{Bernuzzi}}, \bibinfo {author} {\bibfnamefont {D.}~\bibnamefont {Radice}},
		\bibinfo {author} {\bibfnamefont {C.~D.}\ \bibnamefont {Ott}}, \bibinfo
		{author} {\bibfnamefont {L.~F.}\ \bibnamefont {Roberts}}, \bibinfo {author}
		{\bibfnamefont {P.}~\bibnamefont {M\"osta}}, \ and\ \bibinfo {author}
		{\bibfnamefont {F.}~\bibnamefont {Galeazzi}},\ }\href {\doibase
		10.1103/PhysRevD.94.024023} {\bibfield  {journal} {\bibinfo  {journal} {Phys.
				Rev. D}\ }\textbf {\bibinfo {volume} {94}},\ \bibinfo {pages} {024023}
		(\bibinfo {year} {2016})}\BibitemShut {NoStop}%
	\bibitem [{\citenamefont {Lehner}\ \emph {et~al.}(2016)\citenamefont {Lehner},
		\citenamefont {Liebling}, \citenamefont {Palenzuela}, \citenamefont
		{Caballero}, \citenamefont {O'Connor}, \citenamefont {Anderson},\ and\
		\citenamefont {Neilsen}}]{Lehner_2016}%
	\BibitemOpen
	\bibfield  {author} {\bibinfo {author} {\bibfnamefont {L.}~\bibnamefont
			{Lehner}}, \bibinfo {author} {\bibfnamefont {S.~L.}\ \bibnamefont
			{Liebling}}, \bibinfo {author} {\bibfnamefont {C.}~\bibnamefont
			{Palenzuela}}, \bibinfo {author} {\bibfnamefont {O.~L.}\ \bibnamefont
			{Caballero}}, \bibinfo {author} {\bibfnamefont {E.}~\bibnamefont {O'Connor}},
		\bibinfo {author} {\bibfnamefont {M.}~\bibnamefont {Anderson}}, \ and\
		\bibinfo {author} {\bibfnamefont {D.}~\bibnamefont {Neilsen}},\ }\href
	{\doibase 10.1088/0264-9381/33/18/184002} {\bibfield  {journal} {\bibinfo
			{journal} {Class. and Quan. Grav.}\ }\textbf {\bibinfo {volume} {33}},\
		\bibinfo {pages} {184002} (\bibinfo {year} {2016})}\BibitemShut {NoStop}%
	\bibitem [{\citenamefont {Radice}\ \emph {et~al.}(2018)\citenamefont {Radice},
		\citenamefont {Perego}, \citenamefont {Zappa},\ and\ \citenamefont
		{Bernuzzi}}]{Radice_2018}%
	\BibitemOpen
	\bibfield  {author} {\bibinfo {author} {\bibfnamefont {D.}~\bibnamefont
			{Radice}}, \bibinfo {author} {\bibfnamefont {A.}~\bibnamefont {Perego}},
		\bibinfo {author} {\bibfnamefont {F.}~\bibnamefont {Zappa}}, \ and\ \bibinfo
		{author} {\bibfnamefont {S.}~\bibnamefont {Bernuzzi}},\ }\href {\doibase
		10.3847/2041-8213/aaa402} {\bibfield  {journal} {\bibinfo  {journal} {The
				Astrophys. Jour.}\ }\textbf {\bibinfo {volume} {852}},\ \bibinfo {pages}
		{L29} (\bibinfo {year} {2018})}\BibitemShut {NoStop}%
	\bibitem [{\citenamefont {Köppel}\ \emph {et~al.}(2019)\citenamefont
		{Köppel}, \citenamefont {Bovard},\ and\ \citenamefont
		{Rezzolla}}]{K_ppel_2019}%
	\BibitemOpen
	\bibfield  {author} {\bibinfo {author} {\bibfnamefont {S.}~\bibnamefont
			{Köppel}}, \bibinfo {author} {\bibfnamefont {L.}~\bibnamefont {Bovard}}, \
		and\ \bibinfo {author} {\bibfnamefont {L.}~\bibnamefont {Rezzolla}},\ }\href
	{\doibase 10.3847/2041-8213/ab0210} {\bibfield  {journal} {\bibinfo
			{journal} {The Astrophys. Jour.}\ }\textbf {\bibinfo {volume} {872}},\
		\bibinfo {pages} {L16} (\bibinfo {year} {2019})}\BibitemShut {NoStop}%
	\bibitem [{\citenamefont {Hebeler}\ \emph {et~al.}(2010)\citenamefont
		{Hebeler}, \citenamefont {Lattimer}, \citenamefont {Pethick},\ and\
		\citenamefont {Schwenk}}]{PhysRevLett.105.161102}%
	\BibitemOpen
	\bibfield  {author} {\bibinfo {author} {\bibfnamefont {K.}~\bibnamefont
			{Hebeler}}, \bibinfo {author} {\bibfnamefont {J.~M.}\ \bibnamefont
			{Lattimer}}, \bibinfo {author} {\bibfnamefont {C.~J.}\ \bibnamefont
			{Pethick}}, \ and\ \bibinfo {author} {\bibfnamefont {A.}~\bibnamefont
			{Schwenk}},\ }\href {\doibase 10.1103/PhysRevLett.105.161102} {\bibfield
		{journal} {\bibinfo  {journal} {Phys. Rev. Lett.}\ }\textbf {\bibinfo
			{volume} {105}},\ \bibinfo {pages} {161102} (\bibinfo {year}
		{2010})}\BibitemShut {NoStop}%
	\bibitem [{\citenamefont {Hebeler}\ \emph {et~al.}(2013)\citenamefont
		{Hebeler}, \citenamefont {Lattimer}, \citenamefont {Pethick},\ and\
		\citenamefont {Schwenk}}]{Hebeler_2013}%
	\BibitemOpen
	\bibfield  {author} {\bibinfo {author} {\bibfnamefont {K.}~\bibnamefont
			{Hebeler}}, \bibinfo {author} {\bibfnamefont {J.~M.}\ \bibnamefont
			{Lattimer}}, \bibinfo {author} {\bibfnamefont {C.~J.}\ \bibnamefont
			{Pethick}}, \ and\ \bibinfo {author} {\bibfnamefont {A.}~\bibnamefont
			{Schwenk}},\ }\href {\doibase 10.1088/0004-637x/773/1/11} {\bibfield
		{journal} {\bibinfo  {journal} {The Astrophys. Jour.}\ }\textbf {\bibinfo
			{volume} {773}},\ \bibinfo {pages} {11} (\bibinfo {year} {2013})}\BibitemShut
	{NoStop}%
	\bibitem [{\citenamefont
		{Lattimer}(2012)}]{doi:10.1146/annurev-nucl-102711-095018}%
	\BibitemOpen
	\bibfield  {author} {\bibinfo {author} {\bibfnamefont {J.~M.}\ \bibnamefont
			{Lattimer}},\ }\href {\doibase 10.1146/annurev-nucl-102711-095018} {\bibfield
		{journal} {\bibinfo  {journal} {Annual Review of Nuclear and Particle
				Science}\ }\textbf {\bibinfo {volume} {62}},\ \bibinfo {pages} {485}
		(\bibinfo {year} {2012})}\BibitemShut {NoStop}%
	\bibitem [{\citenamefont {Miller}\ \emph {et~al.}(2019)\citenamefont {Miller},
		\citenamefont {Chirenti},\ and\ \citenamefont {Lamb}}]{Miller_2019}%
	\BibitemOpen
	\bibfield  {author} {\bibinfo {author} {\bibfnamefont {M.~C.}\ \bibnamefont
			{Miller}}, \bibinfo {author} {\bibfnamefont {C.}~\bibnamefont {Chirenti}}, \
		and\ \bibinfo {author} {\bibfnamefont {F.~K.}\ \bibnamefont {Lamb}},\ }\href
	{\doibase 10.3847/1538-4357/ab4ef9} {\bibfield  {journal} {\bibinfo
			{journal} {The Astrophys. Jour.}\ }\textbf {\bibinfo {volume} {888}},\
		\bibinfo {pages} {12} (\bibinfo {year} {2019})}\BibitemShut {NoStop}%
	\bibitem [{\citenamefont {Annala}\ \emph {et~al.}(2020)\citenamefont {Annala},
		\citenamefont {Gorda}, \citenamefont {Kurkela}, \citenamefont {Nättilä},\
		and\ \citenamefont {Vuorinen}}]{Annala2020}%
	\BibitemOpen
	\bibfield  {author} {\bibinfo {author} {\bibfnamefont {E.}~\bibnamefont
			{Annala}}, \bibinfo {author} {\bibfnamefont {T.}~\bibnamefont {Gorda}},
		\bibinfo {author} {\bibfnamefont {A.}~\bibnamefont {Kurkela}}, \bibinfo
		{author} {\bibfnamefont {J.}~\bibnamefont {Nättilä}}, \ and\ \bibinfo
		{author} {\bibfnamefont {A.}~\bibnamefont {Vuorinen}},\ }\href {\doibase
		10.1038/s41567-020-0914-9} {\bibfield  {journal} {\bibinfo  {journal} {Nature
				Phys.}\ }\textbf {\bibinfo {volume} {16}},\ \bibinfo {pages} {907} (\bibinfo
		{year} {2020})}\BibitemShut {NoStop}%
	\bibitem [{\citenamefont {Demorest}\ \emph {et~al.}(2010)\citenamefont
		{Demorest}, \citenamefont {Pennucci}, \citenamefont {Ransom}, \citenamefont
		{Roberts},\ and\ \citenamefont {Hessels}}]{Demorest2010}%
	\BibitemOpen
	\bibfield  {author} {\bibinfo {author} {\bibfnamefont {P.~B.}\ \bibnamefont
			{Demorest}}, \bibinfo {author} {\bibfnamefont {T.}~\bibnamefont {Pennucci}},
		\bibinfo {author} {\bibfnamefont {S.~M.}\ \bibnamefont {Ransom}}, \bibinfo
		{author} {\bibfnamefont {M.~S.~E.}\ \bibnamefont {Roberts}}, \ and\ \bibinfo
		{author} {\bibfnamefont {J.~W.~T.}\ \bibnamefont {Hessels}},\ }\href
	{\doibase 10.1038/nature09466} {\bibfield  {journal} {\bibinfo  {journal}
			{Nature}\ }\textbf {\bibinfo {volume} {467}},\ \bibinfo {pages} {1081}
		(\bibinfo {year} {2010})}\BibitemShut {NoStop}%
	\bibitem [{\citenamefont {Antoniadis}\ and\ \citenamefont {Freire~{\textit{et
					al.}}}(2013)}]{Antoniadis1233232}%
	\BibitemOpen
	\bibfield  {author} {\bibinfo {author} {\bibfnamefont {J.}~\bibnamefont
			{Antoniadis}}\ and\ \bibinfo {author} {\bibfnamefont {P.~C.~C.}\ \bibnamefont
			{Freire~{\textit{et al.}}}},\ }\href {\doibase 10.1126/science.1233232}
	{\bibfield  {journal} {\bibinfo  {journal} {Science}\ }\textbf {\bibinfo
			{volume} {340}} (\bibinfo {year} {2013}),\
		10.1126/science.1233232}\BibitemShut {NoStop}%
	\bibitem [{\citenamefont {Cromartie}\ and\ \citenamefont {Fonseca~{\textit{et
					al.}}}(2019)}]{Cromartie2020}%
	\BibitemOpen
	\bibfield  {author} {\bibinfo {author} {\bibfnamefont {H.~T.}\ \bibnamefont
			{Cromartie}}\ and\ \bibinfo {author} {\bibfnamefont {E.}~\bibnamefont
			{Fonseca~{\textit{et al.}}}},\ }\href {\doibase 10.1038/s41550-019-0880-2}
	{\bibfield  {journal} {\bibinfo  {journal} {Nature Astronomy}\ }\textbf
		{\bibinfo {volume} {4}},\ \bibinfo {pages} {72} (\bibinfo {year}
		{2019})}\BibitemShut {NoStop}%
	\bibitem [{\citenamefont {Rezzolla}\ \emph {et~al.}(2018)\citenamefont
		{Rezzolla}, \citenamefont {Most},\ and\ \citenamefont
		{Weih}}]{Rezzolla_2018}%
	\BibitemOpen
	\bibfield  {author} {\bibinfo {author} {\bibfnamefont {L.}~\bibnamefont
			{Rezzolla}}, \bibinfo {author} {\bibfnamefont {E.~R.}\ \bibnamefont {Most}},
		\ and\ \bibinfo {author} {\bibfnamefont {L.~R.}\ \bibnamefont {Weih}},\
	}\href {\doibase 10.3847/2041-8213/aaa401} {\bibfield  {journal} {\bibinfo
			{journal} {The Astrophys. Jour.}\ }\textbf {\bibinfo {volume} {852}},\
		\bibinfo {pages} {L25} (\bibinfo {year} {2018})}\BibitemShut {NoStop}%
	\bibitem [{\citenamefont {Margalit}\ and\ \citenamefont
		{Metzger}(2017)}]{Margalit_2017}%
	\BibitemOpen
	\bibfield  {author} {\bibinfo {author} {\bibfnamefont {B.}~\bibnamefont
			{Margalit}}\ and\ \bibinfo {author} {\bibfnamefont {B.~D.}\ \bibnamefont
			{Metzger}},\ }\href {\doibase 10.3847/2041-8213/aa991c} {\bibfield  {journal}
		{\bibinfo  {journal} {The Astrophys. Jour.}\ }\textbf {\bibinfo {volume}
			{850}},\ \bibinfo {pages} {L19} (\bibinfo {year} {2017})}\BibitemShut
	{NoStop}%
	\bibitem [{\citenamefont {Shibata}\ \emph {et~al.}(2019)\citenamefont
		{Shibata}, \citenamefont {Zhou}, \citenamefont {Kiuchi},\ and\ \citenamefont
		{Fujibayashi}}]{PhysRevD.100.023015}%
	\BibitemOpen
	\bibfield  {author} {\bibinfo {author} {\bibfnamefont {M.}~\bibnamefont
			{Shibata}}, \bibinfo {author} {\bibfnamefont {E.}~\bibnamefont {Zhou}},
		\bibinfo {author} {\bibfnamefont {K.}~\bibnamefont {Kiuchi}}, \ and\ \bibinfo
		{author} {\bibfnamefont {S.}~\bibnamefont {Fujibayashi}},\ }\href {\doibase
		10.1103/PhysRevD.100.023015} {\bibfield  {journal} {\bibinfo  {journal}
			{Phys. Rev. D}\ }\textbf {\bibinfo {volume} {100}},\ \bibinfo {pages}
		{023015} (\bibinfo {year} {2019})},\ \Eprint
	{http://arxiv.org/abs/1905.03656} {arXiv:1905.03656 [astro-ph.HE]}
	\BibitemShut {NoStop}%
	\bibitem [{\citenamefont {{Cook}}\ \emph {et~al.}(1994)\citenamefont {{Cook}},
		\citenamefont {{Shapiro}},\ and\ \citenamefont
		{{Teukolsky}}}]{1994ApJ...424..823C}%
	\BibitemOpen
	\bibfield  {author} {\bibinfo {author} {\bibfnamefont {G.~B.}\ \bibnamefont
			{{Cook}}}, \bibinfo {author} {\bibfnamefont {S.~L.}\ \bibnamefont
			{{Shapiro}}}, \ and\ \bibinfo {author} {\bibfnamefont {S.~A.}\ \bibnamefont
			{{Teukolsky}}},\ }\href {\doibase 10.1086/173934} {\bibfield  {journal}
		{\bibinfo  {journal} {Astrophys. J.}\ }\textbf {\bibinfo {volume} {424}},\
		\bibinfo {pages} {823} (\bibinfo {year} {1994})}\BibitemShut {NoStop}%
	\bibitem [{\citenamefont {Stergioulas}(1998)}]{Stergioulas1998}%
	\BibitemOpen
	\bibfield  {author} {\bibinfo {author} {\bibfnamefont {N.}~\bibnamefont
			{Stergioulas}},\ }\href {\doibase 10.12942/lrr-1998-8} {\bibfield  {journal}
		{\bibinfo  {journal} {L. Rev. Relativ.}\ }\textbf {\bibinfo {volume} {1}},\
		\bibinfo {pages} {L19} (\bibinfo {year} {1998})}\BibitemShut {NoStop}%
	\bibitem [{\citenamefont {Paschalidis}\ and\ \citenamefont
		{Stergioulas}(2017)}]{Paschalidis2017}%
	\BibitemOpen
	\bibfield  {author} {\bibinfo {author} {\bibfnamefont {V.}~\bibnamefont
			{Paschalidis}}\ and\ \bibinfo {author} {\bibfnamefont {N.}~\bibnamefont
			{Stergioulas}},\ }\href {\doibase 10.1007/s41114-017-0008-x} {\bibfield
		{journal} {\bibinfo  {journal} {L. Rev. Relativ.}\ }\textbf {\bibinfo
			{volume} {20}},\ \bibinfo {pages} {L19} (\bibinfo {year} {2017})}\BibitemShut
	{NoStop}%
	\bibitem [{\citenamefont {Rather}\ \emph
		{et~al.}(2020{\natexlab{b}})\citenamefont {Rather}, \citenamefont {Usmani},\
		and\ \citenamefont {Patra}}]{Rathernm}%
	\BibitemOpen
	\bibfield  {author} {\bibinfo {author} {\bibfnamefont {I.~A.}\ \bibnamefont
			{Rather}}, \bibinfo {author} {\bibfnamefont {A.~A.}\ \bibnamefont {Usmani}},
		\ and\ \bibinfo {author} {\bibfnamefont {S.~K.}\ \bibnamefont {Patra}},\
	}\href {\doibase 10.1088/1361-6471/aba116} {\bibfield  {journal} {\bibinfo
			{journal} {J. Phys. G: Nucl. and Part. Phys.}\ }\textbf {\bibinfo {volume}
			{47}},\ \bibinfo {pages} {105104} (\bibinfo {year}
		{2020}{\natexlab{b}})}\BibitemShut {NoStop}%
	\bibitem [{\citenamefont {Rather}\ \emph
		{et~al.}(2020{\natexlab{c}})\citenamefont {Rather}, \citenamefont {Kumar},
		\citenamefont {Das}, \citenamefont {Imran}, \citenamefont {Usmani},\ and\
		\citenamefont {Patra}}]{Ratherbag}%
	\BibitemOpen
	\bibfield  {author} {\bibinfo {author} {\bibfnamefont {I.~A.}\ \bibnamefont
			{Rather}}, \bibinfo {author} {\bibfnamefont {A.}~\bibnamefont {Kumar}},
		\bibinfo {author} {\bibfnamefont {H.~C.}\ \bibnamefont {Das}}, \bibinfo
		{author} {\bibfnamefont {M.}~\bibnamefont {Imran}}, \bibinfo {author}
		{\bibfnamefont {A.~A.}\ \bibnamefont {Usmani}}, \ and\ \bibinfo {author}
		{\bibfnamefont {S.~K.}\ \bibnamefont {Patra}},\ }\href {\doibase
		10.1142/S0218301320500445} {\bibfield  {journal} {\bibinfo  {journal} {Int.
				J. Mod. Phys. E}\ }\textbf {\bibinfo {volume} {29}},\ \bibinfo {pages}
		{2050044} (\bibinfo {year} {2020}{\natexlab{c}})}\BibitemShut {NoStop}%
	\bibitem [{\citenamefont {Ofengeim}\ \emph {et~al.}(2019)\citenamefont
		{Ofengeim}, \citenamefont {Gusakov}, \citenamefont {Haensel},\ and\
		\citenamefont {Fortin}}]{PhysRevD.100.103017}%
	\BibitemOpen
	\bibfield  {author} {\bibinfo {author} {\bibfnamefont {D.~D.}\ \bibnamefont
			{Ofengeim}}, \bibinfo {author} {\bibfnamefont {M.~E.}\ \bibnamefont
			{Gusakov}}, \bibinfo {author} {\bibfnamefont {P.}~\bibnamefont {Haensel}}, \
		and\ \bibinfo {author} {\bibfnamefont {M.}~\bibnamefont {Fortin}},\ }\href
	{\doibase 10.1103/PhysRevD.100.103017} {\bibfield  {journal} {\bibinfo
			{journal} {Phys. Rev. D}\ }\textbf {\bibinfo {volume} {100}},\ \bibinfo
		{pages} {103017} (\bibinfo {year} {2019})}\BibitemShut {NoStop}%
	\bibitem [{\citenamefont {Sulaksono}(2015)}]{doi:10.1142/S021830131550007X}%
	\BibitemOpen
	\bibfield  {author} {\bibinfo {author} {\bibfnamefont {A.}~\bibnamefont
			{Sulaksono}},\ }\href {\doibase 10.1142/S021830131550007X} {\bibfield
		{journal} {\bibinfo  {journal} {International Journal of Modern Physics E}\
		}\textbf {\bibinfo {volume} {24}},\ \bibinfo {pages} {1550007} (\bibinfo
		{year} {2015})},\ \Eprint
	{http://arxiv.org/abs/https://doi.org/10.1142/S021830131550007X}
	{https://doi.org/10.1142/S021830131550007X} \BibitemShut {NoStop}%
	\bibitem [{\citenamefont {Vautherin}\ and\ \citenamefont
		{Brink}(1972)}]{PhysRevC.5.626}%
	\BibitemOpen
	\bibfield  {author} {\bibinfo {author} {\bibfnamefont {D.}~\bibnamefont
			{Vautherin}}\ and\ \bibinfo {author} {\bibfnamefont {D.~M.}\ \bibnamefont
			{Brink}},\ }\href {\doibase 10.1103/PhysRevC.5.626} {\bibfield  {journal}
		{\bibinfo  {journal} {Phys. Rev. C}\ }\textbf {\bibinfo {volume} {5}},\
		\bibinfo {pages} {626} (\bibinfo {year} {1972})}\BibitemShut {NoStop}%
	\bibitem [{\citenamefont {Shen}\ \emph {et~al.}(1998)\citenamefont {Shen},
		\citenamefont {Toki}, \citenamefont {Oyamatsu},\ and\ \citenamefont
		{Sumiyoshi}}]{SHEN1998435}%
	\BibitemOpen
	\bibfield  {author} {\bibinfo {author} {\bibfnamefont {H.}~\bibnamefont
			{Shen}}, \bibinfo {author} {\bibfnamefont {H.}~\bibnamefont {Toki}}, \bibinfo
		{author} {\bibfnamefont {K.}~\bibnamefont {Oyamatsu}}, \ and\ \bibinfo
		{author} {\bibfnamefont {K.}~\bibnamefont {Sumiyoshi}},\ }\href {\doibase
		https://doi.org/10.1016/S0375-9474(98)00236-X} {\bibfield  {journal}
		{\bibinfo  {journal} {Nuclear Physics A}\ }\textbf {\bibinfo {volume}
			{637}},\ \bibinfo {pages} {435 } (\bibinfo {year} {1998})}\BibitemShut
	{NoStop}%
	\bibitem [{\citenamefont {Shen}(2002)}]{PhysRevC.65.035802}%
	\BibitemOpen
	\bibfield  {author} {\bibinfo {author} {\bibfnamefont {H.}~\bibnamefont
			{Shen}},\ }\href {\doibase 10.1103/PhysRevC.65.035802} {\bibfield  {journal}
		{\bibinfo  {journal} {Phys. Rev. C}\ }\textbf {\bibinfo {volume} {65}},\
		\bibinfo {pages} {035802} (\bibinfo {year} {2002})}\BibitemShut {NoStop}%
	\bibitem [{\citenamefont {{Douchin, F.}}\ and\ \citenamefont {{Haensel,
				P.}}(2001)}]{refId0}%
	\BibitemOpen
	\bibfield  {author} {\bibinfo {author} {\bibnamefont {{Douchin, F.}}}\ and\
		\bibinfo {author} {\bibnamefont {{Haensel, P.}}},\ }\href {\doibase
		10.1051/0004-6361:20011402} {\bibfield  {journal} {\bibinfo  {journal}
			{A\&A}\ }\textbf {\bibinfo {volume} {380}},\ \bibinfo {pages} {151} (\bibinfo
		{year} {2001})}\BibitemShut {NoStop}%
	\bibitem [{\citenamefont {Bao}\ and\ \citenamefont
		{Shen}(2014)}]{PhysRevC.89.045807}%
	\BibitemOpen
	\bibfield  {author} {\bibinfo {author} {\bibfnamefont {S.~S.}\ \bibnamefont
			{Bao}}\ and\ \bibinfo {author} {\bibfnamefont {H.}~\bibnamefont {Shen}},\
	}\href {\doibase 10.1103/PhysRevC.89.045807} {\bibfield  {journal} {\bibinfo
			{journal} {Phys. Rev. C}\ }\textbf {\bibinfo {volume} {89}},\ \bibinfo
		{pages} {045807} (\bibinfo {year} {2014})}\BibitemShut {NoStop}%
	\bibitem [{\citenamefont {Bao}\ \emph {et~al.}(2014)\citenamefont {Bao},
		\citenamefont {Hu}, \citenamefont {Zhang},\ and\ \citenamefont
		{Shen}}]{PhysRevC.90.045802}%
	\BibitemOpen
	\bibfield  {author} {\bibinfo {author} {\bibfnamefont {S.~S.}\ \bibnamefont
			{Bao}}, \bibinfo {author} {\bibfnamefont {J.~N.}\ \bibnamefont {Hu}},
		\bibinfo {author} {\bibfnamefont {Z.~W.}\ \bibnamefont {Zhang}}, \ and\
		\bibinfo {author} {\bibfnamefont {H.}~\bibnamefont {Shen}},\ }\href {\doibase
		10.1103/PhysRevC.90.045802} {\bibfield  {journal} {\bibinfo  {journal} {Phys.
				Rev. C}\ }\textbf {\bibinfo {volume} {90}},\ \bibinfo {pages} {045802}
		(\bibinfo {year} {2014})}\BibitemShut {NoStop}%
	\bibitem [{\citenamefont {Walecka}(1974)}]{Walecka:1974qa}%
	\BibitemOpen
	\bibfield  {author} {\bibinfo {author} {\bibfnamefont {J.~D.}\ \bibnamefont
			{Walecka}},\ }\href {\doibase 10.1016/0003-4916(74)90208-5} {\bibfield
		{journal} {\bibinfo  {journal} {Ann. Phys.}\ }\textbf {\bibinfo {volume}
			{\textbf{83}}},\ \bibinfo {pages} {491} (\bibinfo {year} {1974})}\BibitemShut
	{NoStop}%
	\bibitem [{\citenamefont {Horowitz}\ and\ \citenamefont
		{Piekarewicz}(2001)}]{PhysRevLett.86.5647}%
	\BibitemOpen
	\bibfield  {author} {\bibinfo {author} {\bibfnamefont {C.~J.}\ \bibnamefont
			{Horowitz}}\ and\ \bibinfo {author} {\bibfnamefont {J.}~\bibnamefont
			{Piekarewicz}},\ }\href {\doibase 10.1103/PhysRevLett.86.5647} {\bibfield
		{journal} {\bibinfo  {journal} {Phys. Rev. Lett.}\ }\textbf {\bibinfo
			{volume} {86}},\ \bibinfo {pages} {5647} (\bibinfo {year}
		{2001})}\BibitemShut {NoStop}%
	\bibitem [{\citenamefont {Sugahara}\ and\ \citenamefont
		{Toki}(1994)}]{SUGAHARA1994557}%
	\BibitemOpen
	\bibfield  {author} {\bibinfo {author} {\bibfnamefont {Y.}~\bibnamefont
			{Sugahara}}\ and\ \bibinfo {author} {\bibfnamefont {H.}~\bibnamefont
			{Toki}},\ }\href {\doibase https://doi.org/10.1016/0375-9474(94)90923-7}
	{\bibfield  {journal} {\bibinfo  {journal} {Nuclear Physics A}\ }\textbf
		{\bibinfo {volume} {579}},\ \bibinfo {pages} {557 } (\bibinfo {year}
		{1994})}\BibitemShut {NoStop}%
	\bibitem [{\citenamefont {Boguta}\ and\ \citenamefont
		{Bodmer}(1977)}]{BOGUTA1977413}%
	\BibitemOpen
	\bibfield  {author} {\bibinfo {author} {\bibfnamefont {J.}~\bibnamefont
			{Boguta}}\ and\ \bibinfo {author} {\bibfnamefont {A.}~\bibnamefont
			{Bodmer}},\ }\href {\doibase https://doi.org/10.1016/0375-9474(77)90626-1}
	{\bibfield  {journal} {\bibinfo  {journal} {Nuclear Physics A}\ }\textbf
		{\bibinfo {volume} {292}},\ \bibinfo {pages} {413 } (\bibinfo {year}
		{1977})}\BibitemShut {NoStop}%
	\bibitem [{\citenamefont {Serot}(1979)}]{SEROT1979146}%
	\BibitemOpen
	\bibfield  {author} {\bibinfo {author} {\bibfnamefont {B.~D.}\ \bibnamefont
			{Serot}},\ }\href {\doibase https://doi.org/10.1016/0370-2693(79)90804-9}
	{\bibfield  {journal} {\bibinfo  {journal} {Physics Letters B}\ }\textbf
		{\bibinfo {volume} {86}},\ \bibinfo {pages} {146 } (\bibinfo {year}
		{1979})}\BibitemShut {NoStop}%
	\bibitem [{\citenamefont {Singh}\ \emph {et~al.}(2014)\citenamefont {Singh},
		\citenamefont {Biswal}, \citenamefont {Bhuyan},\ and\ \citenamefont
		{Patra}}]{PhysRevC.89.044001}%
	\BibitemOpen
	\bibfield  {author} {\bibinfo {author} {\bibfnamefont {S.~K.}\ \bibnamefont
			{Singh}}, \bibinfo {author} {\bibfnamefont {S.~K.}\ \bibnamefont {Biswal}},
		\bibinfo {author} {\bibfnamefont {M.}~\bibnamefont {Bhuyan}}, \ and\ \bibinfo
		{author} {\bibfnamefont {S.~K.}\ \bibnamefont {Patra}},\ }\href {\doibase
		10.1103/PhysRevC.89.044001} {\bibfield  {journal} {\bibinfo  {journal} {Phys.
				Rev. C}\ }\textbf {\bibinfo {volume} {\textbf{89}}},\ \bibinfo {pages}
		{044001} (\bibinfo {year} {2014})}\BibitemShut {NoStop}%
	\bibitem [{\citenamefont {Kumar}\ \emph
		{et~al.}(2017{\natexlab{a}})\citenamefont {Kumar}, \citenamefont {Singh},
		\citenamefont {Agrawal},\ and\ \citenamefont {Patra}}]{Kumara:2017bti}%
	\BibitemOpen
	\bibfield  {author} {\bibinfo {author} {\bibfnamefont {B.}~\bibnamefont
			{Kumar}}, \bibinfo {author} {\bibfnamefont {S.~K.}\ \bibnamefont {Singh}},
		\bibinfo {author} {\bibfnamefont {B.~K.}\ \bibnamefont {Agrawal}}, \ and\
		\bibinfo {author} {\bibfnamefont {S.~K.}\ \bibnamefont {Patra}},\ }\href
	{\doibase 10.1016/j.nuclphysa.2017.07.001} {\bibfield  {journal} {\bibinfo
			{journal} {Nucl. Phys. A}\ }\textbf {\bibinfo {volume} {966}},\ \bibinfo
		{pages} {197} (\bibinfo {year} {2017}{\natexlab{a}})}\BibitemShut {NoStop}%
	\bibitem [{\citenamefont {Kumar}\ \emph {et~al.}(2018)\citenamefont {Kumar},
		\citenamefont {Patra},\ and\ \citenamefont {Agrawal}}]{PhysRevC.97.045806}%
	\BibitemOpen
	\bibfield  {author} {\bibinfo {author} {\bibfnamefont {B.}~\bibnamefont
			{Kumar}}, \bibinfo {author} {\bibfnamefont {S.~K.}\ \bibnamefont {Patra}}, \
		and\ \bibinfo {author} {\bibfnamefont {B.~K.}\ \bibnamefont {Agrawal}},\
	}\href {\doibase 10.1103/PhysRevC.97.045806} {\bibfield  {journal} {\bibinfo
			{journal} {Phys. Rev. C}\ }\textbf {\bibinfo {volume} {\textbf{97}}},\
		\bibinfo {pages} {045806} (\bibinfo {year} {2018})}\BibitemShut {NoStop}%
	\bibitem [{\citenamefont {Lalazissis}\ \emph {et~al.}(1997)\citenamefont
		{Lalazissis}, \citenamefont {K\"onig},\ and\ \citenamefont
		{Ring}}]{PhysRevC.55.540}%
	\BibitemOpen
	\bibfield  {author} {\bibinfo {author} {\bibfnamefont {G.~A.}\ \bibnamefont
			{Lalazissis}}, \bibinfo {author} {\bibfnamefont {J.}~\bibnamefont {K\"onig}},
		\ and\ \bibinfo {author} {\bibfnamefont {P.}~\bibnamefont {Ring}},\ }\href
	{\doibase 10.1103/PhysRevC.55.540} {\bibfield  {journal} {\bibinfo  {journal}
			{Phys. Rev. C}\ }\textbf {\bibinfo {volume} {55}},\ \bibinfo {pages} {540}
		(\bibinfo {year} {1997})}\BibitemShut {NoStop}%
	\bibitem [{\citenamefont {Das}\ \emph {et~al.}(2020)\citenamefont {Das},
		\citenamefont {Kumar}, \citenamefont {Kumar}, \citenamefont {Biswal},\ and\
		\citenamefont {Patra}}]{das2020bigapple}%
	\BibitemOpen
	\bibfield  {author} {\bibinfo {author} {\bibfnamefont {H.~C.}\ \bibnamefont
			{Das}}, \bibinfo {author} {\bibfnamefont {A.}~\bibnamefont {Kumar}}, \bibinfo
		{author} {\bibfnamefont {B.}~\bibnamefont {Kumar}}, \bibinfo {author}
		{\bibfnamefont {S.~K.}\ \bibnamefont {Biswal}}, \ and\ \bibinfo {author}
		{\bibfnamefont {S.~K.}\ \bibnamefont {Patra}},\ }\href@noop {} {} (\bibinfo
	{year} {2020}),\ \Eprint {http://arxiv.org/abs/2009.10690} {arXiv:2009.10690
		[nucl-th]} \BibitemShut {NoStop}%
	\bibitem [{\citenamefont {Brockmann}\ and\ \citenamefont
		{Toki}(1992)}]{PhysRevLett.68.3408}%
	\BibitemOpen
	\bibfield  {author} {\bibinfo {author} {\bibfnamefont {R.}~\bibnamefont
			{Brockmann}}\ and\ \bibinfo {author} {\bibfnamefont {H.}~\bibnamefont
			{Toki}},\ }\href {\doibase 10.1103/PhysRevLett.68.3408} {\bibfield  {journal}
		{\bibinfo  {journal} {Phys. Rev. Lett.}\ }\textbf {\bibinfo {volume} {68}},\
		\bibinfo {pages} {3408} (\bibinfo {year} {1992})}\BibitemShut {NoStop}%
	\bibitem [{\citenamefont {Nik\ifmmode \check{s}\else
			\v{s}\fi{}i\ifmmode~\acute{c}\else \'{c}\fi{}}\ \emph
		{et~al.}(2002)\citenamefont {Nik\ifmmode \check{s}\else
			\v{s}\fi{}i\ifmmode~\acute{c}\else \'{c}\fi{}}, \citenamefont {Vretenar},
		\citenamefont {Finelli},\ and\ \citenamefont {Ring}}]{PhysRevC.66.024306}%
	\BibitemOpen
	\bibfield  {author} {\bibinfo {author} {\bibfnamefont {T.}~\bibnamefont
			{Nik\ifmmode \check{s}\else \v{s}\fi{}i\ifmmode~\acute{c}\else \'{c}\fi{}}},
		\bibinfo {author} {\bibfnamefont {D.}~\bibnamefont {Vretenar}}, \bibinfo
		{author} {\bibfnamefont {P.}~\bibnamefont {Finelli}}, \ and\ \bibinfo
		{author} {\bibfnamefont {P.}~\bibnamefont {Ring}},\ }\href {\doibase
		10.1103/PhysRevC.66.024306} {\bibfield  {journal} {\bibinfo  {journal} {Phys.
				Rev. C}\ }\textbf {\bibinfo {volume} {66}},\ \bibinfo {pages} {024306}
		(\bibinfo {year} {2002})}\BibitemShut {NoStop}%
	\bibitem [{\citenamefont {Lalazissis}\ \emph {et~al.}(2005)\citenamefont
		{Lalazissis}, \citenamefont {Nik\ifmmode \check{s}\else
			\v{s}\fi{}i\ifmmode~\acute{c}\else \'{c}\fi{}}, \citenamefont {Vretenar},\
		and\ \citenamefont {Ring}}]{PhysRevC.71.024312}%
	\BibitemOpen
	\bibfield  {author} {\bibinfo {author} {\bibfnamefont {G.~A.}\ \bibnamefont
			{Lalazissis}}, \bibinfo {author} {\bibfnamefont {T.}~\bibnamefont
			{Nik\ifmmode \check{s}\else \v{s}\fi{}i\ifmmode~\acute{c}\else \'{c}\fi{}}},
		\bibinfo {author} {\bibfnamefont {D.}~\bibnamefont {Vretenar}}, \ and\
		\bibinfo {author} {\bibfnamefont {P.}~\bibnamefont {Ring}},\ }\href {\doibase
		10.1103/PhysRevC.71.024312} {\bibfield  {journal} {\bibinfo  {journal} {Phys.
				Rev. C}\ }\textbf {\bibinfo {volume} {71}},\ \bibinfo {pages} {024312}
		(\bibinfo {year} {2005})}\BibitemShut {NoStop}%
	\bibitem [{\citenamefont {Wei}\ \emph {et~al.}(2020)\citenamefont {Wei},
		\citenamefont {Zhao}, \citenamefont {Wang}, \citenamefont {Geng},
		\citenamefont {Sun}, \citenamefont {Niu},\ and\ \citenamefont
		{Long}}]{ddmex}%
	\BibitemOpen
	\bibfield  {author} {\bibinfo {author} {\bibfnamefont {B.}~\bibnamefont
			{Wei}}, \bibinfo {author} {\bibfnamefont {Q.}~\bibnamefont {Zhao}}, \bibinfo
		{author} {\bibfnamefont {Z.-H.}\ \bibnamefont {Wang}}, \bibinfo {author}
		{\bibfnamefont {J.}~\bibnamefont {Geng}}, \bibinfo {author} {\bibfnamefont
			{B.-Y.}\ \bibnamefont {Sun}}, \bibinfo {author} {\bibfnamefont {Y.-F.}\
			\bibnamefont {Niu}}, \ and\ \bibinfo {author} {\bibfnamefont {W.-H.}\
			\bibnamefont {Long}},\ }\href {\doibase 10.1088/1674-1137/44/7/074107}
	{\bibfield  {journal} {\bibinfo  {journal} {Ch. Phys. C}\ }\textbf {\bibinfo
			{volume} {44}},\ \bibinfo {pages} {074107} (\bibinfo {year}
		{2020})}\BibitemShut {NoStop}%
	\bibitem [{\citenamefont {Taninah}\ \emph {et~al.}(2020)\citenamefont
		{Taninah}, \citenamefont {Agbemava}, \citenamefont {Afanasjev},\ and\
		\citenamefont {Ring}}]{TANINAH2020135065}%
	\BibitemOpen
	\bibfield  {author} {\bibinfo {author} {\bibfnamefont {A.}~\bibnamefont
			{Taninah}}, \bibinfo {author} {\bibfnamefont {S.}~\bibnamefont {Agbemava}},
		\bibinfo {author} {\bibfnamefont {A.}~\bibnamefont {Afanasjev}}, \ and\
		\bibinfo {author} {\bibfnamefont {P.}~\bibnamefont {Ring}},\ }\href {\doibase
		https://doi.org/10.1016/j.physletb.2019.135065} {\bibfield  {journal}
		{\bibinfo  {journal} {Phys. Lett. B}\ }\textbf {\bibinfo {volume} {800}},\
		\bibinfo {pages} {135065} (\bibinfo {year} {2020})}\BibitemShut {NoStop}%
	\bibitem [{\citenamefont {Typel}\ and\ \citenamefont
		{Alvear~Terrero}(2020)}]{typel}%
	\BibitemOpen
	\bibfield  {author} {\bibinfo {author} {\bibfnamefont {S.}~\bibnamefont
			{Typel}}\ and\ \bibinfo {author} {\bibfnamefont {D.}~\bibnamefont
			{Alvear~Terrero}},\ }\href {\doibase 10.1140/epja/s10050-020-00172-2}
	{\bibfield  {journal} {\bibinfo  {journal} {Eur. Phys. Jour. A}\ }\textbf
		{\bibinfo {volume} {56}},\ \bibinfo {pages} {160} (\bibinfo {year}
		{2020})}\BibitemShut {NoStop}%
	\bibitem [{\citenamefont {Witten}(1984)}]{PhysRevD.30.272}%
	\BibitemOpen
	\bibfield  {author} {\bibinfo {author} {\bibfnamefont {E.}~\bibnamefont
			{Witten}},\ }\href {\doibase 10.1103/PhysRevD.30.272} {\bibfield  {journal}
		{\bibinfo  {journal} {Phys. Rev. D}\ }\textbf {\bibinfo {volume}
			{\textbf{30}}},\ \bibinfo {pages} {272} (\bibinfo {year} {1984})}\BibitemShut
	{NoStop}%
	\bibitem [{\citenamefont {Farhi}\ and\ \citenamefont
		{Jaffe}(1984)}]{PhysRevD.30.2379}%
	\BibitemOpen
	\bibfield  {author} {\bibinfo {author} {\bibfnamefont {E.}~\bibnamefont
			{Farhi}}\ and\ \bibinfo {author} {\bibfnamefont {R.~L.}\ \bibnamefont
			{Jaffe}},\ }\href {\doibase 10.1103/PhysRevD.30.2379} {\bibfield  {journal}
		{\bibinfo  {journal} {Phys. Rev. D}\ }\textbf {\bibinfo {volume}
			{\textbf{30}}},\ \bibinfo {pages} {2379} (\bibinfo {year}
		{1984})}\BibitemShut {NoStop}%
	\bibitem [{\citenamefont {Glendenning}(1992)}]{PhysRevD.46.1274}%
	\BibitemOpen
	\bibfield  {author} {\bibinfo {author} {\bibfnamefont {N.~K.}\ \bibnamefont
			{Glendenning}},\ }\href {\doibase 10.1103/PhysRevD.46.1274} {\bibfield
		{journal} {\bibinfo  {journal} {Phys. Rev. D}\ }\textbf {\bibinfo {volume}
			{\textbf{46}}},\ \bibinfo {pages} {1274} (\bibinfo {year}
		{1992})}\BibitemShut {NoStop}%
	\bibitem [{\citenamefont {Özel}\ \emph {et~al.}(2010)\citenamefont {Özel},
		\citenamefont {Psaltis}, \citenamefont {Ransom}, \citenamefont {Demorest},\
		and\ \citenamefont {Alford}}]{zel_2010}%
	\BibitemOpen
	\bibfield  {author} {\bibinfo {author} {\bibfnamefont {F.}~\bibnamefont
			{Özel}}, \bibinfo {author} {\bibfnamefont {D.}~\bibnamefont {Psaltis}},
		\bibinfo {author} {\bibfnamefont {S.}~\bibnamefont {Ransom}}, \bibinfo
		{author} {\bibfnamefont {P.}~\bibnamefont {Demorest}}, \ and\ \bibinfo
		{author} {\bibfnamefont {M.}~\bibnamefont {Alford}},\ }\href {\doibase
		10.1088/2041-8205/724/2/l199} {\bibfield  {journal} {\bibinfo  {journal} {The
				Astrophys. Jour.}\ }\textbf {\bibinfo {volume} {724}},\ \bibinfo {pages}
		{L199} (\bibinfo {year} {2010})}\BibitemShut {NoStop}%
	\bibitem [{\citenamefont {Kl\"ahn}\ \emph {et~al.}(2013)\citenamefont
		{Kl\"ahn}, \citenamefont {\L{}astowiecki},\ and\ \citenamefont
		{Blaschke}}]{PhysRevD.88.085001}%
	\BibitemOpen
	\bibfield  {author} {\bibinfo {author} {\bibfnamefont {T.}~\bibnamefont
			{Kl\"ahn}}, \bibinfo {author} {\bibfnamefont {R.}~\bibnamefont
			{\L{}astowiecki}}, \ and\ \bibinfo {author} {\bibfnamefont {D.}~\bibnamefont
			{Blaschke}},\ }\href {\doibase 10.1103/PhysRevD.88.085001} {\bibfield
		{journal} {\bibinfo  {journal} {Phys. Rev. D}\ }\textbf {\bibinfo {volume}
			{88}},\ \bibinfo {pages} {085001} (\bibinfo {year} {2013})}\BibitemShut
	{NoStop}%
	\bibitem [{\citenamefont {Bombaci}\ \emph {et~al.}(2016)\citenamefont
		{Bombaci}, \citenamefont {Logoteta}, \citenamefont {Vidaña},\ and\
		\citenamefont {Providência}}]{Bombaci2016}%
	\BibitemOpen
	\bibfield  {author} {\bibinfo {author} {\bibfnamefont {I.}~\bibnamefont
			{Bombaci}}, \bibinfo {author} {\bibfnamefont {D.}~\bibnamefont {Logoteta}},
		\bibinfo {author} {\bibfnamefont {I.}~\bibnamefont {Vidaña}}, \ and\
		\bibinfo {author} {\bibfnamefont {C.}~\bibnamefont {Providência}},\ }\href
	{\doibase 10.1140/epja/i2016-16058-5} {\bibfield  {journal} {\bibinfo
			{journal} {Eur. Phys. Jour. A}\ }\textbf {\bibinfo {volume} {52}},\ \bibinfo
		{pages} {58} (\bibinfo {year} {2016})}\BibitemShut {NoStop}%
	\bibitem [{\citenamefont {Chodos}\ \emph {et~al.}(1974)\citenamefont {Chodos},
		\citenamefont {Jaffe}, \citenamefont {Johnson}, \citenamefont {Thorn},\ and\
		\citenamefont {Weisskopf}}]{PhysRevD.9.3471}%
	\BibitemOpen
	\bibfield  {author} {\bibinfo {author} {\bibfnamefont {A.}~\bibnamefont
			{Chodos}}, \bibinfo {author} {\bibfnamefont {R.~L.}\ \bibnamefont {Jaffe}},
		\bibinfo {author} {\bibfnamefont {K.}~\bibnamefont {Johnson}}, \bibinfo
		{author} {\bibfnamefont {C.~B.}\ \bibnamefont {Thorn}}, \ and\ \bibinfo
		{author} {\bibfnamefont {V.~F.}\ \bibnamefont {Weisskopf}},\ }\href {\doibase
		10.1103/PhysRevD.9.3471} {\bibfield  {journal} {\bibinfo  {journal} {Phys.
				Rev. D}\ }\textbf {\bibinfo {volume} {\textbf{9}}},\ \bibinfo {pages} {3471}
		(\bibinfo {year} {1974})}\BibitemShut {NoStop}%
	\bibitem [{\citenamefont {Freedman}\ and\ \citenamefont
		{McLerran}(1978)}]{PhysRevD.17.1109}%
	\BibitemOpen
	\bibfield  {author} {\bibinfo {author} {\bibfnamefont {B.}~\bibnamefont
			{Freedman}}\ and\ \bibinfo {author} {\bibfnamefont {L.}~\bibnamefont
			{McLerran}},\ }\href {\doibase 10.1103/PhysRevD.17.1109} {\bibfield
		{journal} {\bibinfo  {journal} {Phys. Rev. D}\ }\textbf {\bibinfo {volume}
			{\textbf{17}}},\ \bibinfo {pages} {1109} (\bibinfo {year}
		{1978})}\BibitemShut {NoStop}%
	\bibitem [{\citenamefont {Kubis}\ and\ \citenamefont
		{Kutschera}(1997)}]{KUBIS1997191}%
	\BibitemOpen
	\bibfield  {author} {\bibinfo {author} {\bibfnamefont {S.}~\bibnamefont
			{Kubis}}\ and\ \bibinfo {author} {\bibfnamefont {M.}~\bibnamefont
			{Kutschera}},\ }\href {\doibase 10.1016/S0370-2693(97)00306-7} {\bibfield
		{journal} {\bibinfo  {journal} {Phys. Lett. B}\ }\textbf {\bibinfo {volume}
			{\textbf{399}}},\ \bibinfo {pages} {191} (\bibinfo {year}
		{1997})}\BibitemShut {NoStop}%
	\bibitem [{\citenamefont {Nambu}\ and\ \citenamefont
		{Jona-Lasinio}(1961{\natexlab{a}})}]{PhysRev.122.345}%
	\BibitemOpen
	\bibfield  {author} {\bibinfo {author} {\bibfnamefont {Y.}~\bibnamefont
			{Nambu}}\ and\ \bibinfo {author} {\bibfnamefont {G.}~\bibnamefont
			{Jona-Lasinio}},\ }\href {\doibase 10.1103/PhysRev.122.345} {\bibfield
		{journal} {\bibinfo  {journal} {Phys. Rev.}\ }\textbf {\bibinfo {volume}
			{122}},\ \bibinfo {pages} {345} (\bibinfo {year}
		{1961}{\natexlab{a}})}\BibitemShut {NoStop}%
	\bibitem [{\citenamefont {Nambu}\ and\ \citenamefont
		{Jona-Lasinio}(1961{\natexlab{b}})}]{PhysRev.124.246}%
	\BibitemOpen
	\bibfield  {author} {\bibinfo {author} {\bibfnamefont {Y.}~\bibnamefont
			{Nambu}}\ and\ \bibinfo {author} {\bibfnamefont {G.}~\bibnamefont
			{Jona-Lasinio}},\ }\href {\doibase 10.1103/PhysRev.124.246} {\bibfield
		{journal} {\bibinfo  {journal} {Phys. Rev.}\ }\textbf {\bibinfo {volume}
			{124}},\ \bibinfo {pages} {246} (\bibinfo {year}
		{1961}{\natexlab{b}})}\BibitemShut {NoStop}%
	\bibitem [{\citenamefont {Klevansky}(1992)}]{RevModPhys.64.649}%
	\BibitemOpen
	\bibfield  {author} {\bibinfo {author} {\bibfnamefont {S.~P.}\ \bibnamefont
			{Klevansky}},\ }\href {\doibase 10.1103/RevModPhys.64.649} {\bibfield
		{journal} {\bibinfo  {journal} {Rev. Mod. Phys.}\ }\textbf {\bibinfo {volume}
			{64}},\ \bibinfo {pages} {649} (\bibinfo {year} {1992})}\BibitemShut
	{NoStop}%
	\bibitem [{\citenamefont {Buballa}(2005)}]{BUBALLA2005205}%
	\BibitemOpen
	\bibfield  {author} {\bibinfo {author} {\bibfnamefont {M.}~\bibnamefont
			{Buballa}},\ }\href {\doibase https://doi.org/10.1016/j.physrep.2004.11.004}
	{\bibfield  {journal} {\bibinfo  {journal} {Physics Reports}\ }\textbf
		{\bibinfo {volume} {407}},\ \bibinfo {pages} {205 } (\bibinfo {year}
		{2005})}\BibitemShut {NoStop}%
	\bibitem [{\citenamefont {Li}\ \emph {et~al.}(2017)\citenamefont {Li},
		\citenamefont {Zhang}, \citenamefont {Zhao}, \citenamefont {Zhao},\ and\
		\citenamefont {Zong}}]{PhysRevD.95.056018}%
	\BibitemOpen
	\bibfield  {author} {\bibinfo {author} {\bibfnamefont {C.-M.}\ \bibnamefont
			{Li}}, \bibinfo {author} {\bibfnamefont {J.-L.}\ \bibnamefont {Zhang}},
		\bibinfo {author} {\bibfnamefont {T.}~\bibnamefont {Zhao}}, \bibinfo {author}
		{\bibfnamefont {Y.-P.}\ \bibnamefont {Zhao}}, \ and\ \bibinfo {author}
		{\bibfnamefont {H.-S.}\ \bibnamefont {Zong}},\ }\href {\doibase
		10.1103/PhysRevD.95.056018} {\bibfield  {journal} {\bibinfo  {journal} {Phys.
				Rev. D}\ }\textbf {\bibinfo {volume} {95}},\ \bibinfo {pages} {056018}
		(\bibinfo {year} {2017})}\BibitemShut {NoStop}%
	\bibitem [{\citenamefont {Li}\ \emph {et~al.}(2018)\citenamefont {Li},
		\citenamefont {Zhang}, \citenamefont {Yan}, \citenamefont {Huang},\ and\
		\citenamefont {Zong}}]{PhysRevD.97.103013}%
	\BibitemOpen
	\bibfield  {author} {\bibinfo {author} {\bibfnamefont {C.-M.}\ \bibnamefont
			{Li}}, \bibinfo {author} {\bibfnamefont {J.-L.}\ \bibnamefont {Zhang}},
		\bibinfo {author} {\bibfnamefont {Y.}~\bibnamefont {Yan}}, \bibinfo {author}
		{\bibfnamefont {Y.-F.}\ \bibnamefont {Huang}}, \ and\ \bibinfo {author}
		{\bibfnamefont {H.-S.}\ \bibnamefont {Zong}},\ }\href {\doibase
		10.1103/PhysRevD.97.103013} {\bibfield  {journal} {\bibinfo  {journal} {Phys.
				Rev. D}\ }\textbf {\bibinfo {volume} {97}},\ \bibinfo {pages} {103013}
		(\bibinfo {year} {2018})}\BibitemShut {NoStop}%
	\bibitem [{\citenamefont {Klähn}\ and\ \citenamefont
		{Fischer}(2015)}]{Kl_hn_2015}%
	\BibitemOpen
	\bibfield  {author} {\bibinfo {author} {\bibfnamefont {T.}~\bibnamefont
			{Klähn}}\ and\ \bibinfo {author} {\bibfnamefont {T.}~\bibnamefont
			{Fischer}},\ }\href {\doibase 10.1088/0004-637x/810/2/134} {\bibfield
		{journal} {\bibinfo  {journal} {The Astrophys. J.}\ }\textbf {\bibinfo
			{volume} {810}},\ \bibinfo {pages} {134} (\bibinfo {year}
		{2015})}\BibitemShut {NoStop}%
	\bibitem [{\citenamefont {Roupas}\ \emph {et~al.}(2020)\citenamefont {Roupas},
		\citenamefont {Panotopoulos},\ and\ \citenamefont {Lopes}}]{roupas2020qcd}%
	\BibitemOpen
	\bibfield  {author} {\bibinfo {author} {\bibfnamefont {Z.}~\bibnamefont
			{Roupas}}, \bibinfo {author} {\bibfnamefont {G.}~\bibnamefont
			{Panotopoulos}}, \ and\ \bibinfo {author} {\bibfnamefont {I.}~\bibnamefont
			{Lopes}},\ }\href@noop {} {} (\bibinfo {year} {2020}),\ \Eprint
	{http://arxiv.org/abs/2010.11020} {arXiv:2010.11020 [astro-ph.HE]}
	\BibitemShut {NoStop}%
	\bibitem [{\citenamefont {Horowitz}\ and\ \citenamefont
		{Serot}(1981)}]{walecka}%
	\BibitemOpen
	\bibfield  {author} {\bibinfo {author} {\bibfnamefont {C.}~\bibnamefont
			{Horowitz}}\ and\ \bibinfo {author} {\bibfnamefont {B.~D.}\ \bibnamefont
			{Serot}},\ }\href {\doibase https://doi.org/10.1016/0375-9474(81)90770-3}
	{\bibfield  {journal} {\bibinfo  {journal} {Nucl. Phys. A}\ }\textbf
		{\bibinfo {volume} {368}},\ \bibinfo {pages} {503 } (\bibinfo {year}
		{1981})}\BibitemShut {NoStop}%
	\bibitem [{\citenamefont {Furnstahl}\ \emph {et~al.}(1996)\citenamefont
		{Furnstahl}, \citenamefont {Serot},\ and\ \citenamefont
		{Tang}}]{FURNSTAHL1996539}%
	\BibitemOpen
	\bibfield  {author} {\bibinfo {author} {\bibfnamefont {R.}~\bibnamefont
			{Furnstahl}}, \bibinfo {author} {\bibfnamefont {B.~D.}\ \bibnamefont
			{Serot}}, \ and\ \bibinfo {author} {\bibfnamefont {H.~B.}\ \bibnamefont
			{Tang}},\ }\href {\doibase 10.1016/0375-9474(95)00488-2} {\bibfield
		{journal} {\bibinfo  {journal} {Nucl. Phys. A}\ }\textbf {\bibinfo {volume}
			{\textbf{598}}},\ \bibinfo {pages} {539} (\bibinfo {year}
		{1996})}\BibitemShut {NoStop}%
	\bibitem [{\citenamefont {Furnstahl}\ \emph {et~al.}(1997)\citenamefont
		{Furnstahl}, \citenamefont {Serot},\ and\ \citenamefont
		{Tang}}]{FURNSTAHL1997441}%
	\BibitemOpen
	\bibfield  {author} {\bibinfo {author} {\bibfnamefont {R.}~\bibnamefont
			{Furnstahl}}, \bibinfo {author} {\bibfnamefont {B.~D.}\ \bibnamefont
			{Serot}}, \ and\ \bibinfo {author} {\bibfnamefont {H.-B.}\ \bibnamefont
			{Tang}},\ }\href {\doibase https://doi.org/10.1016/S0375-9474(96)00472-1}
	{\bibfield  {journal} {\bibinfo  {journal} {Nucl. Phys. A}\ }\textbf
		{\bibinfo {volume} {615}},\ \bibinfo {pages} {441 } (\bibinfo {year}
		{1997})}\BibitemShut {NoStop}%
	\bibitem [{\citenamefont {Bouyssy}\ \emph {et~al.}(1987)\citenamefont
		{Bouyssy}, \citenamefont {Mathiot}, \citenamefont {Van~Giai},\ and\
		\citenamefont {Marcos}}]{PhysRevC.36.380}%
	\BibitemOpen
	\bibfield  {author} {\bibinfo {author} {\bibfnamefont {A.}~\bibnamefont
			{Bouyssy}}, \bibinfo {author} {\bibfnamefont {J.-F.}\ \bibnamefont
			{Mathiot}}, \bibinfo {author} {\bibfnamefont {N.}~\bibnamefont {Van~Giai}}, \
		and\ \bibinfo {author} {\bibfnamefont {S.}~\bibnamefont {Marcos}},\ }\href
	{\doibase 10.1103/PhysRevC.36.380} {\bibfield  {journal} {\bibinfo  {journal}
			{Phys. Rev. C}\ }\textbf {\bibinfo {volume} {36}},\ \bibinfo {pages} {380}
		(\bibinfo {year} {1987})}\BibitemShut {NoStop}%
	\bibitem [{\citenamefont {Brockmann}(1978)}]{PhysRevC.18.1510}%
	\BibitemOpen
	\bibfield  {author} {\bibinfo {author} {\bibfnamefont {R.}~\bibnamefont
			{Brockmann}},\ }\href {\doibase 10.1103/PhysRevC.18.1510} {\bibfield
		{journal} {\bibinfo  {journal} {Phys. Rev. C}\ }\textbf {\bibinfo {volume}
			{18}},\ \bibinfo {pages} {1510} (\bibinfo {year} {1978})}\BibitemShut
	{NoStop}%
	\bibitem [{\citenamefont {Long}\ \emph {et~al.}(2006)\citenamefont {Long},
		\citenamefont {{Van Giai}},\ and\ \citenamefont {Meng}}]{LONG2006150}%
	\BibitemOpen
	\bibfield  {author} {\bibinfo {author} {\bibfnamefont {W.-H.}\ \bibnamefont
			{Long}}, \bibinfo {author} {\bibfnamefont {N.}~\bibnamefont {{Van Giai}}}, \
		and\ \bibinfo {author} {\bibfnamefont {J.}~\bibnamefont {Meng}},\ }\href
	{\doibase https://doi.org/10.1016/j.physletb.2006.07.064} {\bibfield
		{journal} {\bibinfo  {journal} {Physics Letters B}\ }\textbf {\bibinfo
			{volume} {640}},\ \bibinfo {pages} {150 } (\bibinfo {year}
		{2006})}\BibitemShut {NoStop}%
	\bibitem [{\citenamefont {Cierniak}(2018)}]{vbageos}%
	\BibitemOpen
	\bibfield  {author} {\bibinfo {author} {\bibfnamefont {T.~F. T. B.~N.}\
			\bibnamefont {Cierniak}, \bibfnamefont {M.and~Klähn}},\ }\href {\doibase
		10.3390/universe4020030} {\bibfield  {journal} {\bibinfo  {journal}
			{Universe}\ }\textbf {\bibinfo {volume} {4}},\ \bibinfo {pages} {30}
		(\bibinfo {year} {2018})}\BibitemShut {NoStop}%
	\bibitem [{\citenamefont {Wei}\ \emph {et~al.}(2019)\citenamefont {Wei},
		\citenamefont {Irving}, \citenamefont {Salinas}, \citenamefont {Klähn},\
		and\ \citenamefont {Jaikumar}}]{Wei_2019}%
	\BibitemOpen
	\bibfield  {author} {\bibinfo {author} {\bibfnamefont {W.}~\bibnamefont
			{Wei}}, \bibinfo {author} {\bibfnamefont {B.}~\bibnamefont {Irving}},
		\bibinfo {author} {\bibfnamefont {M.}~\bibnamefont {Salinas}}, \bibinfo
		{author} {\bibfnamefont {T.}~\bibnamefont {Klähn}}, \ and\ \bibinfo {author}
		{\bibfnamefont {P.}~\bibnamefont {Jaikumar}},\ }\href {\doibase
		10.3847/1538-4357/ab53ea} {\bibfield  {journal} {\bibinfo  {journal} {The
				Astrophys. J.}\ }\textbf {\bibinfo {volume} {887}},\ \bibinfo {pages} {151}
		(\bibinfo {year} {2019})}\BibitemShut {NoStop}%
	\bibitem [{\citenamefont {Schertler}\ \emph {et~al.}(1999)\citenamefont
		{Schertler}, \citenamefont {Leupold},\ and\ \citenamefont
		{Schaffner-Bielich}}]{PhysRevC.60.025801}%
	\BibitemOpen
	\bibfield  {author} {\bibinfo {author} {\bibfnamefont {K.}~\bibnamefont
			{Schertler}}, \bibinfo {author} {\bibfnamefont {S.}~\bibnamefont {Leupold}},
		\ and\ \bibinfo {author} {\bibfnamefont {J.}~\bibnamefont
			{Schaffner-Bielich}},\ }\href {\doibase 10.1103/PhysRevC.60.025801}
	{\bibfield  {journal} {\bibinfo  {journal} {Phys. Rev. C}\ }\textbf {\bibinfo
			{volume} {\textbf{60}}},\ \bibinfo {pages} {025801} (\bibinfo {year}
		{1999})}\BibitemShut {NoStop}%
	\bibitem [{\citenamefont {Sharma}\ \emph {et~al.}(2007)\citenamefont {Sharma},
		\citenamefont {Panda},\ and\ \citenamefont {Patra}}]{PhysRevC.75.035808}%
	\BibitemOpen
	\bibfield  {author} {\bibinfo {author} {\bibfnamefont {B.~K.}\ \bibnamefont
			{Sharma}}, \bibinfo {author} {\bibfnamefont {P.~K.}\ \bibnamefont {Panda}}, \
		and\ \bibinfo {author} {\bibfnamefont {S.~K.}\ \bibnamefont {Patra}},\ }\href
	{\doibase 10.1103/PhysRevC.75.035808} {\bibfield  {journal} {\bibinfo
			{journal} {Phys. Rev. C}\ }\textbf {\bibinfo {volume} {\textbf{75}}},\
		\bibinfo {pages} {035808} (\bibinfo {year} {2007})}\BibitemShut {NoStop}%
	\bibitem [{\citenamefont {Burgio}\ \emph {et~al.}(2002)\citenamefont {Burgio},
		\citenamefont {Baldo}, \citenamefont {Sahu},\ and\ \citenamefont
		{Schulze}}]{PhysRevC.66.025802}%
	\BibitemOpen
	\bibfield  {author} {\bibinfo {author} {\bibfnamefont {G.~F.}\ \bibnamefont
			{Burgio}}, \bibinfo {author} {\bibfnamefont {M.}~\bibnamefont {Baldo}},
		\bibinfo {author} {\bibfnamefont {P.~K.}\ \bibnamefont {Sahu}}, \ and\
		\bibinfo {author} {\bibfnamefont {H.-J.}\ \bibnamefont {Schulze}},\ }\href
	{\doibase 10.1103/PhysRevC.66.025802} {\bibfield  {journal} {\bibinfo
			{journal} {Phys. Rev. C}\ }\textbf {\bibinfo {volume} {66}},\ \bibinfo
		{pages} {025802} (\bibinfo {year} {2002})}\BibitemShut {NoStop}%
	\bibitem [{\citenamefont {Orsaria}\ \emph {et~al.}(2014)\citenamefont
		{Orsaria}, \citenamefont {Rodrigues}, \citenamefont {Weber},\ and\
		\citenamefont {Contrera}}]{PhysRevC.89.015806}%
	\BibitemOpen
	\bibfield  {author} {\bibinfo {author} {\bibfnamefont {M.}~\bibnamefont
			{Orsaria}}, \bibinfo {author} {\bibfnamefont {H.}~\bibnamefont {Rodrigues}},
		\bibinfo {author} {\bibfnamefont {F.}~\bibnamefont {Weber}}, \ and\ \bibinfo
		{author} {\bibfnamefont {G.~A.}\ \bibnamefont {Contrera}},\ }\href {\doibase
		10.1103/PhysRevC.89.015806} {\bibfield  {journal} {\bibinfo  {journal} {Phys.
				Rev. C}\ }\textbf {\bibinfo {volume} {\textbf{89}}},\ \bibinfo {pages}
		{015806} (\bibinfo {year} {2014})}\BibitemShut {NoStop}%
	\bibitem [{\citenamefont {Logoteta}\ and\ \citenamefont
		{Bombaci}(2013)}]{PhysRevD.88.063001}%
	\BibitemOpen
	\bibfield  {author} {\bibinfo {author} {\bibfnamefont {D.}~\bibnamefont
			{Logoteta}}\ and\ \bibinfo {author} {\bibfnamefont {I.}~\bibnamefont
			{Bombaci}},\ }\href {\doibase 10.1103/PhysRevD.88.063001} {\bibfield
		{journal} {\bibinfo  {journal} {Phys. Rev. D}\ }\textbf {\bibinfo {volume}
			{88}},\ \bibinfo {pages} {063001} (\bibinfo {year} {2013})}\BibitemShut
	{NoStop}%
	\bibitem [{\citenamefont {Tolman}(1939)}]{PhysRev.55.364}%
	\BibitemOpen
	\bibfield  {author} {\bibinfo {author} {\bibfnamefont {R.~C.}\ \bibnamefont
			{Tolman}},\ }\href {\doibase 10.1103/PhysRev.55.364} {\bibfield  {journal}
		{\bibinfo  {journal} {Phys. Rev.}\ }\textbf {\bibinfo {volume} {55}},\
		\bibinfo {pages} {364} (\bibinfo {year} {1939})}\BibitemShut {NoStop}%
	\bibitem [{\citenamefont {Oppenheimer}\ and\ \citenamefont
		{Volkoff}(1939)}]{PhysRev.55.374}%
	\BibitemOpen
	\bibfield  {author} {\bibinfo {author} {\bibfnamefont {J.~R.}\ \bibnamefont
			{Oppenheimer}}\ and\ \bibinfo {author} {\bibfnamefont {G.~M.}\ \bibnamefont
			{Volkoff}},\ }\href {\doibase 10.1103/PhysRev.55.374} {\bibfield  {journal}
		{\bibinfo  {journal} {Phys. Rev.}\ }\textbf {\bibinfo {volume} {55}},\
		\bibinfo {pages} {374} (\bibinfo {year} {1939})}\BibitemShut {NoStop}%
	\bibitem [{\citenamefont {Hinderer}\ \emph {et~al.}(2010)\citenamefont
		{Hinderer}, \citenamefont {Lackey}, \citenamefont {Lang},\ and\ \citenamefont
		{Read}}]{PhysRevD.81.123016}%
	\BibitemOpen
	\bibfield  {author} {\bibinfo {author} {\bibfnamefont {T.}~\bibnamefont
			{Hinderer}}, \bibinfo {author} {\bibfnamefont {B.~D.}\ \bibnamefont
			{Lackey}}, \bibinfo {author} {\bibfnamefont {R.~N.}\ \bibnamefont {Lang}}, \
		and\ \bibinfo {author} {\bibfnamefont {J.~S.}\ \bibnamefont {Read}},\ }\href
	{\doibase 10.1103/PhysRevD.81.123016} {\bibfield  {journal} {\bibinfo
			{journal} {Phys. Rev. D}\ }\textbf {\bibinfo {volume} {81}},\ \bibinfo
		{pages} {123016} (\bibinfo {year} {2010})}\BibitemShut {NoStop}%
	\bibitem [{\citenamefont {Kumar}\ \emph
		{et~al.}(2017{\natexlab{b}})\citenamefont {Kumar}, \citenamefont {Biswal},\
		and\ \citenamefont {Patra}}]{PhysRevC.95.015801}%
	\BibitemOpen
	\bibfield  {author} {\bibinfo {author} {\bibfnamefont {B.}~\bibnamefont
			{Kumar}}, \bibinfo {author} {\bibfnamefont {S.~K.}\ \bibnamefont {Biswal}}, \
		and\ \bibinfo {author} {\bibfnamefont {S.~K.}\ \bibnamefont {Patra}},\ }\href
	{\doibase 10.1103/PhysRevC.95.015801} {\bibfield  {journal} {\bibinfo
			{journal} {Phys. Rev. C}\ }\textbf {\bibinfo {volume} {95}},\ \bibinfo
		{pages} {015801} (\bibinfo {year} {2017}{\natexlab{b}})}\BibitemShut
	{NoStop}%
	\bibitem [{\citenamefont {Hinderer}(2008)}]{Hinderer_2008}%
	\BibitemOpen
	\bibfield  {author} {\bibinfo {author} {\bibfnamefont {T.}~\bibnamefont
			{Hinderer}},\ }\href {\doibase 10.1086/533487} {\bibfield  {journal}
		{\bibinfo  {journal} {The Astrophys. J.}\ }\textbf {\bibinfo {volume}
			{677}},\ \bibinfo {pages} {1216} (\bibinfo {year} {2008})}\BibitemShut
	{NoStop}%
	\bibitem [{\citenamefont {{Butterworth}}\ and\ \citenamefont
		{{Ipser}}(1976)}]{1976ApJ...204..200B}%
	\BibitemOpen
	\bibfield  {author} {\bibinfo {author} {\bibfnamefont {E.~M.}\ \bibnamefont
			{{Butterworth}}}\ and\ \bibinfo {author} {\bibfnamefont {J.~R.}\ \bibnamefont
			{{Ipser}}},\ }\href {\doibase 10.1086/154163} {\bibfield  {journal} {\bibinfo
			{journal} {Astrophys. Jour.}\ }\textbf {\bibinfo {volume} {204}},\ \bibinfo
		{pages} {200} (\bibinfo {year} {1976})}\BibitemShut {NoStop}%
	\bibitem [{\citenamefont {{Friedman}}\ \emph {et~al.}(1986)\citenamefont
		{{Friedman}}, \citenamefont {{Ipser}},\ and\ \citenamefont
		{{Parker}}}]{1986ApJ...304..115F}%
	\BibitemOpen
	\bibfield  {author} {\bibinfo {author} {\bibfnamefont {J.~L.}\ \bibnamefont
			{{Friedman}}}, \bibinfo {author} {\bibfnamefont {J.~R.}\ \bibnamefont
			{{Ipser}}}, \ and\ \bibinfo {author} {\bibfnamefont {L.}~\bibnamefont
			{{Parker}}},\ }\href {\doibase 10.1086/164149} {\bibfield  {journal}
		{\bibinfo  {journal} {Astrophys. Jour.}\ }\textbf {\bibinfo {volume} {304}},\
		\bibinfo {pages} {115} (\bibinfo {year} {1986})}\BibitemShut {NoStop}%
	\bibitem [{\citenamefont {Friedman}\ \emph {et~al.}(1989)\citenamefont
		{Friedman}, \citenamefont {Ipser},\ and\ \citenamefont
		{Parker}}]{PhysRevLett.62.3015}%
	\BibitemOpen
	\bibfield  {author} {\bibinfo {author} {\bibfnamefont {J.~L.}\ \bibnamefont
			{Friedman}}, \bibinfo {author} {\bibfnamefont {J.~R.}\ \bibnamefont {Ipser}},
		\ and\ \bibinfo {author} {\bibfnamefont {L.}~\bibnamefont {Parker}},\ }\href
	{\doibase 10.1103/PhysRevLett.62.3015} {\bibfield  {journal} {\bibinfo
			{journal} {Phys. Rev. Lett.}\ }\textbf {\bibinfo {volume} {62}},\ \bibinfo
		{pages} {3015} (\bibinfo {year} {1989})}\BibitemShut {NoStop}%
	\bibitem [{\citenamefont {Lattimer}\ and\ \citenamefont
		{Prakash}(2000)}]{LATTIMER2000121}%
	\BibitemOpen
	\bibfield  {author} {\bibinfo {author} {\bibfnamefont {J.~M.}\ \bibnamefont
			{Lattimer}}\ and\ \bibinfo {author} {\bibfnamefont {M.}~\bibnamefont
			{Prakash}},\ }\href {\doibase https://doi.org/10.1016/S0370-1573(00)00019-3}
	{\bibfield  {journal} {\bibinfo  {journal} {Physics Reports}\ }\textbf
		{\bibinfo {volume} {333-334}},\ \bibinfo {pages} {121 } (\bibinfo {year}
		{2000})}\BibitemShut {NoStop}%
	\bibitem [{\citenamefont {Worley}\ \emph {et~al.}(2008)\citenamefont {Worley},
		\citenamefont {Krastev},\ and\ \citenamefont {Li}}]{Worley_2008}%
	\BibitemOpen
	\bibfield  {author} {\bibinfo {author} {\bibfnamefont {A.}~\bibnamefont
			{Worley}}, \bibinfo {author} {\bibfnamefont {P.~G.}\ \bibnamefont {Krastev}},
		\ and\ \bibinfo {author} {\bibfnamefont {B.-A.}\ \bibnamefont {Li}},\ }\href
	{\doibase 10.1086/589823} {\bibfield  {journal} {\bibinfo  {journal} {The
				Astrophys. Jour.}\ }\textbf {\bibinfo {volume} {685}},\ \bibinfo {pages}
		{390} (\bibinfo {year} {2008})}\BibitemShut {NoStop}%
	\bibitem [{\citenamefont {Stergioulas}(2003)}]{Stergioulas2003}%
	\BibitemOpen
	\bibfield  {author} {\bibinfo {author} {\bibfnamefont {N.}~\bibnamefont
			{Stergioulas}},\ }\href {\doibase 10.12942/lrr-2003-3} {\bibfield  {journal}
		{\bibinfo  {journal} {Living Rev. Relativ.}\ }\textbf {\bibinfo {volume}
			{6}},\ \bibinfo {pages} {3} (\bibinfo {year} {2003})}\BibitemShut {NoStop}%
	\bibitem [{\citenamefont {Stergioulas}\ and\ \citenamefont
		{Friedman}(1995)}]{Stergioulas_1995}%
	\BibitemOpen
	\bibfield  {author} {\bibinfo {author} {\bibfnamefont {N.}~\bibnamefont
			{Stergioulas}}\ and\ \bibinfo {author} {\bibfnamefont {J.~L.}\ \bibnamefont
			{Friedman}},\ }\href {\doibase 10.1086/175605} {\bibfield  {journal}
		{\bibinfo  {journal} {`Astrophys. Jour.}\ }\textbf {\bibinfo {volume}
			{444}},\ \bibinfo {pages} {306} (\bibinfo {year} {1995})}\BibitemShut
	{NoStop}%
	\bibitem [{\citenamefont {Stergioulas}(1996)}]{rnscode}%
	\BibitemOpen
	\bibfield  {author} {\bibinfo {author} {\bibfnamefont {N.}~\bibnamefont
			{Stergioulas}},\ }\href {http://www.gravity.phys.uwm.edu/rns} {\bibfield
		{journal} {\bibinfo  {journal} {Rotating Neutron star (RNS) code:
				http://www.gravity.phys.uwm.edu/rns}\ } (\bibinfo {year} {1996})}\BibitemShut
	{NoStop}%
	\bibitem [{\citenamefont {Li}\ and\ \citenamefont {Han}(2013)}]{LI2013276}%
	\BibitemOpen
	\bibfield  {author} {\bibinfo {author} {\bibfnamefont {B.-A.}\ \bibnamefont
			{Li}}\ and\ \bibinfo {author} {\bibfnamefont {X.}~\bibnamefont {Han}},\
	}\href {\doibase https://doi.org/10.1016/j.physletb.2013.10.006} {\bibfield
		{journal} {\bibinfo  {journal} {Phys. Lett. B}\ }\textbf {\bibinfo {volume}
			{727}},\ \bibinfo {pages} {276 } (\bibinfo {year} {2013})}\BibitemShut
	{NoStop}%
	\bibitem [{\citenamefont {Zhang}\ \emph {et~al.}(2020)\citenamefont {Zhang},
		\citenamefont {Liu}, \citenamefont {Xia}, \citenamefont {Li},\ and\
		\citenamefont {Biswal}}]{PhysRevC.101.034303}%
	\BibitemOpen
	\bibfield  {author} {\bibinfo {author} {\bibfnamefont {Y.}~\bibnamefont
			{Zhang}}, \bibinfo {author} {\bibfnamefont {M.}~\bibnamefont {Liu}}, \bibinfo
		{author} {\bibfnamefont {C.-J.}\ \bibnamefont {Xia}}, \bibinfo {author}
		{\bibfnamefont {Z.}~\bibnamefont {Li}}, \ and\ \bibinfo {author}
		{\bibfnamefont {S.~K.}\ \bibnamefont {Biswal}},\ }\href {\doibase
		10.1103/PhysRevC.101.034303} {\bibfield  {journal} {\bibinfo  {journal}
			{Phys. Rev. C}\ }\textbf {\bibinfo {volume} {101}},\ \bibinfo {pages}
		{034303} (\bibinfo {year} {2020})}\BibitemShut {NoStop}%
	\bibitem [{\citenamefont {Danielewicz}\ and\ \citenamefont
		{Lee}(2014)}]{DANIELEWICZ20141}%
	\BibitemOpen
	\bibfield  {author} {\bibinfo {author} {\bibfnamefont {P.}~\bibnamefont
			{Danielewicz}}\ and\ \bibinfo {author} {\bibfnamefont {J.}~\bibnamefont
			{Lee}},\ }\href {\doibase https://doi.org/10.1016/j.nuclphysa.2013.11.005}
	{\bibfield  {journal} {\bibinfo  {journal} {Nucl. Phys. A}\ }\textbf
		{\bibinfo {volume} {922}},\ \bibinfo {pages} {1 } (\bibinfo {year}
		{2014})}\BibitemShut {NoStop}%
	\bibitem [{\citenamefont {Abbott}\ \emph
		{et~al.}(2020{\natexlab{b}})\citenamefont {Abbott}, \citenamefont {Abbott}
		\emph {et~al.}}]{Abbott_eos}%
	\BibitemOpen
	\bibfield  {author} {\bibinfo {author} {\bibfnamefont {R.}~\bibnamefont
			{Abbott}}, \bibinfo {author} {\bibfnamefont {T.~D.}\ \bibnamefont {Abbott}},
		\emph {et~al.},\ }\href {\doibase 10.3847/2041-8213/ab960f} {\bibfield
		{journal} {\bibinfo  {journal} {Astrophys. Jour.}\ }\textbf {\bibinfo
			{volume} {896}},\ \bibinfo {pages} {L44} (\bibinfo {year}
		{2020}{\natexlab{b}})}\BibitemShut {NoStop}%
	\bibitem [{\citenamefont {Baym}\ \emph {et~al.}(1971)\citenamefont {Baym},
		\citenamefont {Pethick},\ and\ \citenamefont {Sutherland}}]{Baym:1971pw}%
	\BibitemOpen
	\bibfield  {author} {\bibinfo {author} {\bibfnamefont {G.}~\bibnamefont
			{Baym}}, \bibinfo {author} {\bibfnamefont {C.}~\bibnamefont {Pethick}}, \
		and\ \bibinfo {author} {\bibfnamefont {P.}~\bibnamefont {Sutherland}},\
	}\href {\doibase 10.1086/151216} {\bibfield  {journal} {\bibinfo  {journal}
			{Astrophys. J.}\ }\textbf {\bibinfo {volume} {170}},\ \bibinfo {pages} {299}
		(\bibinfo {year} {1971})}\BibitemShut {NoStop}%
	\bibitem [{\citenamefont {Rather}\ \emph {et~al.}(2021)\citenamefont {Rather},
		\citenamefont {Usmani},\ and\ \citenamefont {Patra}}]{rather2020effect}%
	\BibitemOpen
	\bibfield  {author} {\bibinfo {author} {\bibfnamefont {I.~A.}\ \bibnamefont
			{Rather}}, \bibinfo {author} {\bibfnamefont {A.}~\bibnamefont {Usmani}}, \
		and\ \bibinfo {author} {\bibfnamefont {S.}~\bibnamefont {Patra}},\ }\href
	{\doibase https://doi.org/10.1016/j.nuclphysa.2021.122189} {\bibfield
		{journal} {\bibinfo  {journal} {Nucl. Phys. A}\ }\textbf {\bibinfo {volume}
			{1010}},\ \bibinfo {pages} {122189} (\bibinfo {year} {2021})}\BibitemShut
	{NoStop}%
	\bibitem [{\citenamefont {Pais}\ and\ \citenamefont
		{Provid\^encia}(2016)}]{PhysRevC.94.015808}%
	\BibitemOpen
	\bibfield  {author} {\bibinfo {author} {\bibfnamefont {H.}~\bibnamefont
			{Pais}}\ and\ \bibinfo {author} {\bibfnamefont {C.}~\bibnamefont
			{Provid\^encia}},\ }\href {\doibase 10.1103/PhysRevC.94.015808} {\bibfield
		{journal} {\bibinfo  {journal} {Phys. Rev. C}\ }\textbf {\bibinfo {volume}
			{94}},\ \bibinfo {pages} {015808} (\bibinfo {year} {2016})}\BibitemShut
	{NoStop}%
	\bibitem [{\citenamefont {Grill}\ \emph {et~al.}(2014)\citenamefont {Grill},
		\citenamefont {Pais}, \citenamefont {Provid\^encia}, \citenamefont
		{Vida\~na},\ and\ \citenamefont {Avancini}}]{PhysRevC.90.045803}%
	\BibitemOpen
	\bibfield  {author} {\bibinfo {author} {\bibfnamefont {F.}~\bibnamefont
			{Grill}}, \bibinfo {author} {\bibfnamefont {H.}~\bibnamefont {Pais}},
		\bibinfo {author} {\bibfnamefont {C.}~\bibnamefont {Provid\^encia}}, \bibinfo
		{author} {\bibfnamefont {I.}~\bibnamefont {Vida\~na}}, \ and\ \bibinfo
		{author} {\bibfnamefont {S.~S.}\ \bibnamefont {Avancini}},\ }\href {\doibase
		10.1103/PhysRevC.90.045803} {\bibfield  {journal} {\bibinfo  {journal} {Phys.
				Rev. C}\ }\textbf {\bibinfo {volume} {90}},\ \bibinfo {pages} {045803}
		(\bibinfo {year} {2014})}\BibitemShut {NoStop}%
	\bibitem [{\citenamefont {Riley}\ \emph {et~al.}(2019)\citenamefont {Riley},
		\citenamefont {Watts}, \citenamefont {Bogdanov}, \citenamefont {Ray},
		\citenamefont {Ludlam}, \citenamefont {Guillot}, \citenamefont {Arzoumanian},
		\citenamefont {Baker}, \citenamefont {Bilous}, \citenamefont {Chakrabarty},
		\citenamefont {Gendreau}, \citenamefont {Harding}, \citenamefont {Ho},
		\citenamefont {Lattimer}, \citenamefont {Morsink},\ and\ \citenamefont
		{Strohmayer}}]{Riley_2019}%
	\BibitemOpen
	\bibfield  {author} {\bibinfo {author} {\bibfnamefont {T.~E.}\ \bibnamefont
			{Riley}}, \bibinfo {author} {\bibfnamefont {A.~L.}\ \bibnamefont {Watts}},
		\bibinfo {author} {\bibfnamefont {S.}~\bibnamefont {Bogdanov}}, \bibinfo
		{author} {\bibfnamefont {P.~S.}\ \bibnamefont {Ray}}, \bibinfo {author}
		{\bibfnamefont {R.~M.}\ \bibnamefont {Ludlam}}, \bibinfo {author}
		{\bibfnamefont {S.}~\bibnamefont {Guillot}}, \bibinfo {author} {\bibfnamefont
			{Z.}~\bibnamefont {Arzoumanian}}, \bibinfo {author} {\bibfnamefont {C.~L.}\
			\bibnamefont {Baker}}, \bibinfo {author} {\bibfnamefont {A.~V.}\ \bibnamefont
			{Bilous}}, \bibinfo {author} {\bibfnamefont {D.}~\bibnamefont {Chakrabarty}},
		\bibinfo {author} {\bibfnamefont {K.~C.}\ \bibnamefont {Gendreau}}, \bibinfo
		{author} {\bibfnamefont {A.~K.}\ \bibnamefont {Harding}}, \bibinfo {author}
		{\bibfnamefont {W.~C.~G.}\ \bibnamefont {Ho}}, \bibinfo {author}
		{\bibfnamefont {J.~M.}\ \bibnamefont {Lattimer}}, \bibinfo {author}
		{\bibfnamefont {S.~M.}\ \bibnamefont {Morsink}}, \ and\ \bibinfo {author}
		{\bibfnamefont {T.~E.}\ \bibnamefont {Strohmayer}},\ }\href {\doibase
		10.3847/2041-8213/ab481c} {\bibfield  {journal} {\bibinfo  {journal}
			{Astrophys. Jour.}\ }\textbf {\bibinfo {volume} {887}},\ \bibinfo {pages}
		{L21} (\bibinfo {year} {2019})}\BibitemShut {NoStop}%
	\bibitem [{\citenamefont {Jiang}\ \emph {et~al.}(2020)\citenamefont {Jiang},
		\citenamefont {Tang}, \citenamefont {Wang}, \citenamefont {Fan},\ and\
		\citenamefont {Wei}}]{Jiang_2020}%
	\BibitemOpen
	\bibfield  {author} {\bibinfo {author} {\bibfnamefont {J.-L.}\ \bibnamefont
			{Jiang}}, \bibinfo {author} {\bibfnamefont {S.-P.}\ \bibnamefont {Tang}},
		\bibinfo {author} {\bibfnamefont {Y.-Z.}\ \bibnamefont {Wang}}, \bibinfo
		{author} {\bibfnamefont {Y.-Z.}\ \bibnamefont {Fan}}, \ and\ \bibinfo
		{author} {\bibfnamefont {D.-M.}\ \bibnamefont {Wei}},\ }\href {\doibase
		10.3847/1538-4357/ab77cf} {\bibfield  {journal} {\bibinfo  {journal}
			{Astrophys. Jour.}\ }\textbf {\bibinfo {volume} {892}},\ \bibinfo {pages}
		{55} (\bibinfo {year} {2020})}\BibitemShut {NoStop}%
	\bibitem [{\citenamefont {Landry}\ \emph {et~al.}(2020)\citenamefont {Landry},
		\citenamefont {Essick},\ and\ \citenamefont
		{Chatziioannou}}]{PhysRevD.101.123007}%
	\BibitemOpen
	\bibfield  {author} {\bibinfo {author} {\bibfnamefont {P.}~\bibnamefont
			{Landry}}, \bibinfo {author} {\bibfnamefont {R.}~\bibnamefont {Essick}}, \
		and\ \bibinfo {author} {\bibfnamefont {K.}~\bibnamefont {Chatziioannou}},\
	}\href {\doibase 10.1103/PhysRevD.101.123007} {\bibfield  {journal} {\bibinfo
			{journal} {Phys. Rev. D}\ }\textbf {\bibinfo {volume} {101}},\ \bibinfo
		{pages} {123007} (\bibinfo {year} {2020})}\BibitemShut {NoStop}%
	\bibitem [{\citenamefont {Lim}\ \emph {et~al.}(2019)\citenamefont {Lim},
		\citenamefont {Holt},\ and\ \citenamefont {Stahulak}}]{PhysRevC.100.035802}%
	\BibitemOpen
	\bibfield  {author} {\bibinfo {author} {\bibfnamefont {Y.}~\bibnamefont
			{Lim}}, \bibinfo {author} {\bibfnamefont {J.~W.}\ \bibnamefont {Holt}}, \
		and\ \bibinfo {author} {\bibfnamefont {R.~J.}\ \bibnamefont {Stahulak}},\
	}\href {\doibase 10.1103/PhysRevC.100.035802} {\bibfield  {journal} {\bibinfo
			{journal} {Phys. Rev. C}\ }\textbf {\bibinfo {volume} {100}},\ \bibinfo
		{pages} {035802} (\bibinfo {year} {2019})}\BibitemShut {NoStop}%
	\bibitem [{\citenamefont {Kumar}\ and\ \citenamefont
		{Landry}(2019)}]{PhysRevD.99.123026}%
	\BibitemOpen
	\bibfield  {author} {\bibinfo {author} {\bibfnamefont {B.}~\bibnamefont
			{Kumar}}\ and\ \bibinfo {author} {\bibfnamefont {P.}~\bibnamefont {Landry}},\
	}\href {\doibase 10.1103/PhysRevD.99.123026} {\bibfield  {journal} {\bibinfo
			{journal} {Phys. Rev. D}\ }\textbf {\bibinfo {volume} {99}},\ \bibinfo
		{pages} {123026} (\bibinfo {year} {2019})}\BibitemShut {NoStop}%
	\bibitem [{\citenamefont {{Friedman}}\ \emph {et~al.}(1988)\citenamefont
		{{Friedman}}, \citenamefont {{Ipser}},\ and\ \citenamefont
		{{Sorkin}}}]{1988ApJ...325..722F}%
	\BibitemOpen
	\bibfield  {author} {\bibinfo {author} {\bibfnamefont {J.~L.}\ \bibnamefont
			{{Friedman}}}, \bibinfo {author} {\bibfnamefont {J.~R.}\ \bibnamefont
			{{Ipser}}}, \ and\ \bibinfo {author} {\bibfnamefont {R.~D.}\ \bibnamefont
			{{Sorkin}}},\ }\href {\doibase 10.1086/166043} {\bibfield  {journal}
		{\bibinfo  {journal} {Astrophys. Jour.}\ }\textbf {\bibinfo {volume} {325}},\
		\bibinfo {pages} {722} (\bibinfo {year} {1988})}\BibitemShut {NoStop}%
	\bibitem [{\citenamefont {Sorkin}(1981)}]{Sorkin:1981jc}%
	\BibitemOpen
	\bibfield  {author} {\bibinfo {author} {\bibfnamefont {R.}~\bibnamefont
			{Sorkin}},\ }\href {\doibase 10.1086/159282} {\bibfield  {journal} {\bibinfo
			{journal} {Astrophys. J.}\ }\textbf {\bibinfo {volume} {249}},\ \bibinfo
		{pages} {254} (\bibinfo {year} {1981})}\BibitemShut {NoStop}%
	\bibitem [{\citenamefont {Sorkin}(1982)}]{Sorkin:1982ut}%
	\BibitemOpen
	\bibfield  {author} {\bibinfo {author} {\bibfnamefont {R.~D.}\ \bibnamefont
			{Sorkin}},\ }\href {\doibase 10.1086/160034} {\bibfield  {journal} {\bibinfo
			{journal} {Astrophys. J.}\ }\textbf {\bibinfo {volume} {257}},\ \bibinfo
		{pages} {847} (\bibinfo {year} {1982})}\BibitemShut {NoStop}%
	\bibitem [{\citenamefont {Takami}\ \emph {et~al.}(2011)\citenamefont {Takami},
		\citenamefont {Rezzolla},\ and\ \citenamefont {Yoshida}}]{takami}%
	\BibitemOpen
	\bibfield  {author} {\bibinfo {author} {\bibfnamefont {K.}~\bibnamefont
			{Takami}}, \bibinfo {author} {\bibfnamefont {L.}~\bibnamefont {Rezzolla}}, \
		and\ \bibinfo {author} {\bibfnamefont {S.}~\bibnamefont {Yoshida}},\ }\href
	{\doibase 10.1111/j.1745-3933.2011.01085.x} {\bibfield  {journal} {\bibinfo
			{journal} {MNRAS}\ }\textbf {\bibinfo {volume} {416}},\ \bibinfo {pages} {L1}
		(\bibinfo {year} {2011})}\BibitemShut {NoStop}%
	\bibitem [{\citenamefont {{Backer}}\ \emph {et~al.}(1982)\citenamefont
		{{Backer}}, \citenamefont {{Kulkarni}}, \citenamefont {{Heiles}},
		\citenamefont {{Davis}},\ and\ \citenamefont {{Goss}}}]{Backer1982}%
	\BibitemOpen
	\bibfield  {author} {\bibinfo {author} {\bibfnamefont {D.~C.}\ \bibnamefont
			{{Backer}}}, \bibinfo {author} {\bibfnamefont {S.~R.}\ \bibnamefont
			{{Kulkarni}}}, \bibinfo {author} {\bibfnamefont {C.}~\bibnamefont
			{{Heiles}}}, \bibinfo {author} {\bibfnamefont {M.~M.}\ \bibnamefont
			{{Davis}}}, \ and\ \bibinfo {author} {\bibfnamefont {W.~M.}\ \bibnamefont
			{{Goss}}},\ }\href {\doibase 10.1038/300615a0} {\bibfield  {journal}
		{\bibinfo  {journal} {Nature}\ }\textbf {\bibinfo {volume} {300}},\ \bibinfo
		{pages} {615} (\bibinfo {year} {1982})}\BibitemShut {NoStop}%
	\bibitem [{\citenamefont {Hessels}\ \emph {et~al.}(2006)\citenamefont
		{Hessels}, \citenamefont {Ransom}, \citenamefont {Stairs}, \citenamefont
		{Freire}, \citenamefont {Kaspi},\ and\ \citenamefont {Camilo}}]{Hessels1901}%
	\BibitemOpen
	\bibfield  {author} {\bibinfo {author} {\bibfnamefont {J.~W.~T.}\
			\bibnamefont {Hessels}}, \bibinfo {author} {\bibfnamefont {S.~M.}\
			\bibnamefont {Ransom}}, \bibinfo {author} {\bibfnamefont {I.~H.}\
			\bibnamefont {Stairs}}, \bibinfo {author} {\bibfnamefont {P.~C.~C.}\
			\bibnamefont {Freire}}, \bibinfo {author} {\bibfnamefont {V.~M.}\
			\bibnamefont {Kaspi}}, \ and\ \bibinfo {author} {\bibfnamefont
			{F.}~\bibnamefont {Camilo}},\ }\href {\doibase 10.1126/science.1123430}
	{\bibfield  {journal} {\bibinfo  {journal} {Science}\ }\textbf {\bibinfo
			{volume} {311}},\ \bibinfo {pages} {1901} (\bibinfo {year}
		{2006})}\BibitemShut {NoStop}%
	\bibitem [{\citenamefont {Kaaret}\ \emph {et~al.}(2007)\citenamefont {Kaaret},
		\citenamefont {Prieskorn}, \citenamefont {in~{\textquotesingle}t~Zand},
		\citenamefont {Brandt}, \citenamefont {Lund}, \citenamefont {Mereghetti},
		\citenamefont {Götz}, \citenamefont {Kuulkers},\ and\ \citenamefont
		{Tomsick}}]{Kaaret_2007}%
	\BibitemOpen
	\bibfield  {author} {\bibinfo {author} {\bibfnamefont {P.}~\bibnamefont
			{Kaaret}}, \bibinfo {author} {\bibfnamefont {Z.}~\bibnamefont {Prieskorn}},
		\bibinfo {author} {\bibfnamefont {J.~J.~M.}\ \bibnamefont
			{in~{\textquotesingle}t~Zand}}, \bibinfo {author} {\bibfnamefont
			{S.}~\bibnamefont {Brandt}}, \bibinfo {author} {\bibfnamefont
			{N.}~\bibnamefont {Lund}}, \bibinfo {author} {\bibfnamefont {S.}~\bibnamefont
			{Mereghetti}}, \bibinfo {author} {\bibfnamefont {D.}~\bibnamefont {Götz}},
		\bibinfo {author} {\bibfnamefont {E.}~\bibnamefont {Kuulkers}}, \ and\
		\bibinfo {author} {\bibfnamefont {J.~A.}\ \bibnamefont {Tomsick}},\ }\href
	{\doibase 10.1086/513270} {\bibfield  {journal} {\bibinfo  {journal}
			{Astrophys. Jour.}\ }\textbf {\bibinfo {volume} {657}},\ \bibinfo {pages}
		{L97} (\bibinfo {year} {2007})}\BibitemShut {NoStop}%
	\bibitem [{\citenamefont {McNall}(1970)}]{10.1093/gji/21.1.103-a}%
	\BibitemOpen
	\bibfield  {author} {\bibinfo {author} {\bibfnamefont {D.}~\bibnamefont
			{McNall}},\ }\href {\doibase 10.1093/gji/21.1.103-a} {\bibfield  {journal}
		{\bibinfo  {journal} {Geophysical Journal International}\ }\textbf {\bibinfo
			{volume} {21}},\ \bibinfo {pages} {103} (\bibinfo {year} {1970})}\BibitemShut
	{NoStop}%
	\bibitem [{\citenamefont {{Tohline}}\ \emph {et~al.}(1985)\citenamefont
		{{Tohline}}, \citenamefont {{Durisen}},\ and\ \citenamefont
		{{McCollough}}}]{1985ApJ...298..220T}%
	\BibitemOpen
	\bibfield  {author} {\bibinfo {author} {\bibfnamefont {J.~E.}\ \bibnamefont
			{{Tohline}}}, \bibinfo {author} {\bibfnamefont {R.~H.}\ \bibnamefont
			{{Durisen}}}, \ and\ \bibinfo {author} {\bibfnamefont {M.}~\bibnamefont
			{{McCollough}}},\ }\href {\doibase 10.1086/163600} {\bibfield  {journal}
		{\bibinfo  {journal} {Astrophys. J.}\ }\textbf {\bibinfo {volume} {298}},\
		\bibinfo {pages} {220} (\bibinfo {year} {1985})}\BibitemShut {NoStop}%
	\bibitem [{\citenamefont {{Pickett}}\ \emph {et~al.}(1996)\citenamefont
		{{Pickett}}, \citenamefont {{Durisen}},\ and\ \citenamefont
		{{Davis}}}]{1996ApJ...458..714P}%
	\BibitemOpen
	\bibfield  {author} {\bibinfo {author} {\bibfnamefont {B.~K.}\ \bibnamefont
			{{Pickett}}}, \bibinfo {author} {\bibfnamefont {R.~H.}\ \bibnamefont
			{{Durisen}}}, \ and\ \bibinfo {author} {\bibfnamefont {G.~A.}\ \bibnamefont
			{{Davis}}},\ }\href {\doibase 10.1086/176852} {\bibfield  {journal} {\bibinfo
			{journal} {Astrophys. J.}\ }\textbf {\bibinfo {volume} {458}},\ \bibinfo
		{pages} {714} (\bibinfo {year} {1996})}\BibitemShut {NoStop}%
	\bibitem [{\citenamefont {{Imamura}}\ \emph {et~al.}(1995)\citenamefont
		{{Imamura}}, \citenamefont {{Toman}}, \citenamefont {{Durisen}},
		\citenamefont {{Pickett}},\ and\ \citenamefont
		{{Yang}}}]{1995ApJ...444..363I}%
	\BibitemOpen
	\bibfield  {author} {\bibinfo {author} {\bibfnamefont {J.~N.}\ \bibnamefont
			{{Imamura}}}, \bibinfo {author} {\bibfnamefont {J.}~\bibnamefont {{Toman}}},
		\bibinfo {author} {\bibfnamefont {R.~H.}\ \bibnamefont {{Durisen}}}, \bibinfo
		{author} {\bibfnamefont {B.~K.}\ \bibnamefont {{Pickett}}}, \ and\ \bibinfo
		{author} {\bibfnamefont {S.}~\bibnamefont {{Yang}}},\ }\href {\doibase
		10.1086/175611} {\bibfield  {journal} {\bibinfo  {journal} {Astrophys. J.}\
		}\textbf {\bibinfo {volume} {444}},\ \bibinfo {pages} {363} (\bibinfo {year}
		{1995})}\BibitemShut {NoStop}%
	\bibitem [{\citenamefont {{Centrella}}\ \emph {et~al.}(2001)\citenamefont
		{{Centrella}}, \citenamefont {{New}}, \citenamefont {{Lowe}},\ and\
		\citenamefont {{Brown}}}]{2001ApJ...550L.193C}%
	\BibitemOpen
	\bibfield  {author} {\bibinfo {author} {\bibfnamefont {J.~M.}\ \bibnamefont
			{{Centrella}}}, \bibinfo {author} {\bibfnamefont {K.~C.~B.}\ \bibnamefont
			{{New}}}, \bibinfo {author} {\bibfnamefont {L.~L.}\ \bibnamefont {{Lowe}}}, \
		and\ \bibinfo {author} {\bibfnamefont {J.~D.}\ \bibnamefont {{Brown}}},\
	}\href {\doibase 10.1086/319634} {\bibfield  {journal} {\bibinfo  {journal}
			{Astrophys. J. Lett.}\ }\textbf {\bibinfo {volume} {550}},\ \bibinfo {pages}
		{L193} (\bibinfo {year} {2001})}\BibitemShut {NoStop}%
	\bibitem [{\citenamefont {Lo}\ and\ \citenamefont
		{Lin}(2011)}]{spin.parameter}%
	\BibitemOpen
	\bibfield  {author} {\bibinfo {author} {\bibfnamefont {K.-W.}\ \bibnamefont
			{Lo}}\ and\ \bibinfo {author} {\bibfnamefont {L.-M.}\ \bibnamefont {Lin}},\
	}\href {\doibase 10.1088/0004-637X/728/1/12} {\bibfield  {journal} {\bibinfo
			{journal} {Astrophys. J.}\ }\textbf {\bibinfo {volume} {728}},\ \bibinfo
		{pages} {12} (\bibinfo {year} {2011})}\BibitemShut {NoStop}%
	\bibitem [{\citenamefont {Cipolletta}\ \emph {et~al.}(2015)\citenamefont
		{Cipolletta}, \citenamefont {Cherubini}, \citenamefont {Filippi},
		\citenamefont {Rueda},\ and\ \citenamefont {Ruffini}}]{PhysRevD.92.023007}%
	\BibitemOpen
	\bibfield  {author} {\bibinfo {author} {\bibfnamefont {F.}~\bibnamefont
			{Cipolletta}}, \bibinfo {author} {\bibfnamefont {C.}~\bibnamefont
			{Cherubini}}, \bibinfo {author} {\bibfnamefont {S.}~\bibnamefont {Filippi}},
		\bibinfo {author} {\bibfnamefont {J.~A.}\ \bibnamefont {Rueda}}, \ and\
		\bibinfo {author} {\bibfnamefont {R.}~\bibnamefont {Ruffini}},\ }\href
	{\doibase 10.1103/PhysRevD.92.023007} {\bibfield  {journal} {\bibinfo
			{journal} {Phys. Rev. D}\ }\textbf {\bibinfo {volume} {92}},\ \bibinfo
		{pages} {023007} (\bibinfo {year} {2015})}\BibitemShut {NoStop}%
	\bibitem [{\citenamefont {Koliogiannia}\ and\ \citenamefont
		{Moustakidis}(2020)}]{PhysRevC.101.015805}%
	\BibitemOpen
	\bibfield  {author} {\bibinfo {author} {\bibfnamefont {P.~S.}\ \bibnamefont
			{Koliogiannia}}\ and\ \bibinfo {author} {\bibfnamefont {C.~C.}\ \bibnamefont
			{Moustakidis}},\ }\href {\doibase 10.1103/PhysRevC.101.015805} {\bibfield
		{journal} {\bibinfo  {journal} {Phys. Rev. C}\ }\textbf {\bibinfo {volume}
			{101}},\ \bibinfo {pages} {015805} (\bibinfo {year} {2020})}\BibitemShut
	{NoStop}%
	\bibitem [{\citenamefont {Xia}\ and\ \citenamefont
		{Yong-Jiu}(2009)}]{Xia_2009}%
	\BibitemOpen
	\bibfield  {author} {\bibinfo {author} {\bibfnamefont {C.}~\bibnamefont
			{Xia}}\ and\ \bibinfo {author} {\bibfnamefont {W.}~\bibnamefont {Yong-Jiu}},\
	}\href {\doibase 10.1088/0256-307x/26/7/070402} {\bibfield  {journal}
		{\bibinfo  {journal} {Ch. Phys. Lett.}\ }\textbf {\bibinfo {volume} {26}},\
		\bibinfo {pages} {070402} (\bibinfo {year} {2009})}\BibitemShut {NoStop}%
	\bibitem [{\citenamefont {Sanwal}\ \emph {et~al.}(2002)\citenamefont {Sanwal},
		\citenamefont {Pavlov}, \citenamefont {Zavlin},\ and\ \citenamefont
		{Teter}}]{Sanwal_2002}%
	\BibitemOpen
	\bibfield  {author} {\bibinfo {author} {\bibfnamefont {D.}~\bibnamefont
			{Sanwal}}, \bibinfo {author} {\bibfnamefont {G.~G.}\ \bibnamefont {Pavlov}},
		\bibinfo {author} {\bibfnamefont {V.~E.}\ \bibnamefont {Zavlin}}, \ and\
		\bibinfo {author} {\bibfnamefont {M.~A.}\ \bibnamefont {Teter}},\ }\href
	{\doibase 10.1086/342368} {\bibfield  {journal} {\bibinfo  {journal}
			{Astrophys. J.}\ }\textbf {\bibinfo {volume} {574}},\ \bibinfo {pages} {L61}
		(\bibinfo {year} {2002})}\BibitemShut {NoStop}%
	\bibitem [{\citenamefont {Cottam}\ \emph {et~al.}(2002)\citenamefont {Cottam},
		\citenamefont {Paerels},\ and\ \citenamefont {Mendez}}]{cottam}%
	\BibitemOpen
	\bibfield  {author} {\bibinfo {author} {\bibfnamefont {J.}~\bibnamefont
			{Cottam}}, \bibinfo {author} {\bibfnamefont {F.}~\bibnamefont {Paerels}}, \
		and\ \bibinfo {author} {\bibfnamefont {M.}~\bibnamefont {Mendez}},\ }\href
	{\doibase 10.1038/nature01159} {\bibfield  {journal} {\bibinfo  {journal}
			{Nature}\ }\textbf {\bibinfo {volume} {420}},\ \bibinfo {pages} {51}
		(\bibinfo {year} {2002})}\BibitemShut {NoStop}%
	\bibitem [{\citenamefont {Bauböck}\ \emph {et~al.}(2013)\citenamefont
		{Bauböck}, \citenamefont {Psaltis},\ and\ \citenamefont
		{Özel}}]{Baub_ck_2013}%
	\BibitemOpen
	\bibfield  {author} {\bibinfo {author} {\bibfnamefont {M.}~\bibnamefont
			{Bauböck}}, \bibinfo {author} {\bibfnamefont {D.}~\bibnamefont {Psaltis}}, \
		and\ \bibinfo {author} {\bibfnamefont {F.}~\bibnamefont {Özel}},\ }\href
	{\doibase 10.1088/0004-637x/766/2/87} {\bibfield  {journal} {\bibinfo
			{journal} {Astrophys. J.}\ }\textbf {\bibinfo {volume} {766}},\ \bibinfo
		{pages} {87} (\bibinfo {year} {2013})}\BibitemShut {NoStop}%
	\bibitem [{\citenamefont {{Liang}}(1986)}]{Liang}%
	\BibitemOpen
	\bibfield  {author} {\bibinfo {author} {\bibfnamefont {E.~P.}\ \bibnamefont
			{{Liang}}},\ }\href {\doibase 10.1086/164206} {\bibfield  {journal} {\bibinfo
			{journal} {\apj}\ }\textbf {\bibinfo {volume} {304}},\ \bibinfo {pages}
		{682} (\bibinfo {year} {1986})}\BibitemShut {NoStop}%
	\bibitem [{\citenamefont {Li}\ \emph {et~al.}(2021)\citenamefont {Li},
		\citenamefont {Chen}, \citenamefont {Wen},\ and\ \citenamefont
		{Zhang}}]{Li2021}%
	\BibitemOpen
	\bibfield  {author} {\bibinfo {author} {\bibfnamefont {Y.}~\bibnamefont
			{Li}}, \bibinfo {author} {\bibfnamefont {H.}~\bibnamefont {Chen}}, \bibinfo
		{author} {\bibfnamefont {D.}~\bibnamefont {Wen}}, \ and\ \bibinfo {author}
		{\bibfnamefont {J.}~\bibnamefont {Zhang}},\ }\href {\doibase
		10.1140/epja/s10050-021-00342-w} {\bibfield  {journal} {\bibinfo  {journal}
			{The European Physical Journal A}\ }\textbf {\bibinfo {volume} {57}},\
		\bibinfo {pages} {31} (\bibinfo {year} {2021})}\BibitemShut {NoStop}%
	\bibitem [{\citenamefont {Lim}\ and\ \citenamefont
		{Holt}(2018)}]{PhysRevLett.121.062701}%
	\BibitemOpen
	\bibfield  {author} {\bibinfo {author} {\bibfnamefont {Y.}~\bibnamefont
			{Lim}}\ and\ \bibinfo {author} {\bibfnamefont {J.~W.}\ \bibnamefont {Holt}},\
	}\href {\doibase 10.1103/PhysRevLett.121.062701} {\bibfield  {journal}
		{\bibinfo  {journal} {Phys. Rev. Lett.}\ }\textbf {\bibinfo {volume} {121}},\
		\bibinfo {pages} {062701} (\bibinfo {year} {2018})}\BibitemShut {NoStop}%
\end{thebibliography}

%

\end{document}